%% file: article2_AGP_effective_EPJPlus_v2.tex
\documentclass[11pt, a4paper]{article}
\pdfoutput=1
\usepackage{amsmath, amsfonts, amssymb}
\usepackage{float}
\usepackage[margin=2cm]{geometry}
\usepackage{pgf}
\usepackage{longtable}
\usepackage{color,url}
\usepackage{colortbl}
\usepackage{multirow}
\usepackage{verbatim}
\usepackage[utf8]{inputenc}
\usepackage[colorlinks=true
,urlcolor=blue
,anchorcolor=blue
,citecolor=blue
,filecolor=blue
,linkcolor=blue
,menucolor=blue
,linktocpage=true
,pdfproducer=medialab
,pdfa=true
]{hyperref}

\usepackage{epsfig,psfrag,rotating,soul}
\usepackage{rotfloat}
\usepackage{epsf}
\usepackage{graphics}
\usepackage{color}

\usepackage{cite}

\newcommand{\RD}{\ensuremath{ R_{D}} }
\newcommand{\RDp}{\ensuremath{ R_{D^{(*)}}} }
\newcommand{\RKp}{\ensuremath{ R_{K^{(*)}}} }

\newcommand{\Rks}{\ensuremath{R_{K^{*0}}}}
\newcommand{\GeV}{\ensuremath{\,{\rm GeV}}}
\newcommand{\Clq}{\ensuremath{ C_{\ell q} } }
\newcommand{\Clqt}{\ensuremath{ C_{\ell q(3)} } }
\newcommand{\Clqo}{\ensuremath{ C_{\ell q(1)} } }

\newcommand{\CneNP}{\ensuremath{C_9^{\rm{NP}\, e}}}
\newcommand{\CteNP}{\ensuremath{C_{10}^{\rm{NP}\, e}}}
\newcommand{\CnmuNP}{\ensuremath{C_9^{\rm{NP}\, \mu}}}
\newcommand{\CtmuNP}{\ensuremath{C_{10}^{\rm{NP}\, \mu}}}

\renewcommand{\arraystretch}{1.5}

\begin{document}
\thispagestyle{empty}

\vspace*{2.5cm}

\vspace{0.5cm}

\begin{center}

\begin{Large}
\textbf{\textsc{Anomalies in B mesons decays: A phenomenological approach.}}
\end{Large}

\vspace{1cm}

{\sc
J. Alda$^{a, b}$%
\footnote{\tt \href{mailto:jalda@unizar.es}{jalda@unizar.es}}%
, J.~Guasch$^{c}$%
\footnote{\tt \href{mailto:jaume.guasch@ub.edu}{jaume.guasch@ub.edu}}%
, S.~Pe{\~n}aranda$^{a, b}$%
\footnote{\tt \href{mailto:siannah@unizar.es}{siannah@unizar.es}}%
}

\vspace*{.7cm}

{\sl
$^a$Departamento de F{\'\i}sica Te{\'o}rica, Facultad de Ciencias,\\
Universidad de Zaragoza, Pedro Cerbuna 12,  E-50009 Zaragoza, Spain

\vspace*{0.1cm}

$^b$Centro de Astropart{\'\i}culas y F{\'\i}sica de Altas Energ{\'\i}as (CAPA), 
Universidad de Zaragoza, Zaragoza, Spain

\vspace*{0.1cm}

$^c$Deptartament de F{\'\i}sica Qu{\`a}ntica i Astrof{\'\i}sica and Institut de Ci{\`e}ncies del Cosmos (ICCUB),\\
Universitat de Barcelona, Mart{\'\i} i Franqu{\`e}s 1, E-08028 Barcelona, Catalonia, Spain

}

\end{center}

\vspace*{0.1cm}

\begin{abstract}
\noindent

The experimental measurements on flavour physics, in tension
with Standard Model predictions, exhibit large sources of 
Lepton Flavour Universality violation. 
We perform an analysis of the effects of the global fits on the Wilson
coefficients assuming the Standard Model Effective Field Theory with
semileptonic dimension six operators at $1 \,\mathrm{TeV}$, and
by including a set of different scenarios in which the
New Physics contributions to the Wilson coefficients are present in one, 
two or three of the Wilson coefficients at a time. 
We compare the results of the global fit with respect
to two cases: the Standard Model and the more general case in which New Physics modifies three
independent Wilson coefficients. The last mentioned scenario is the
favoured one for explaining the tension between Standard Model
predictions and B-physics anomalies, but a specific more restricted
scenario can provide similar goodness with a smaller set of free
parameters. A discussion of the implications of 
our analysis in leptoquark models is included.

\end{abstract}

\def\thefootnote{\arabic{footnote}}
\setcounter{page}{0}
\setcounter{footnote}{0}

\newpage

\section{Introduction}
\label{intro}

In the last few years, many interesting measurements on 
flavour physics have been performed at the 
LHC~\cite{Aaij:2014pli,Aaij:2014ora,Aaij:2015esa,Aaij:2015oid,Aaij:2016flj,CMS:2014xfa,Aaij:2015yra,Aaij:2016cbx,Aaij:2017vad,Aaij:2017xqt,Aaij:2017vbb,Aaij:2017uff,Aaij:2018jhg,Aaij:2019wad,Aaij:2021vac},
BaBar~\cite{Lees:2012xj} and Belle~\cite{Abdesselam:2017kjf,Belle:2018ezy,Abdesselam:2019wac,Abdesselam:2019dgh,Abdesselam:2019lab}.
Several experimental collaborations observed Lepton Flavour Universality Violating (LFUV) processes in
$B$ meson decays that would be a clear sign for physics beyond the Standard
Model (SM). Some of these decays allow us to build optimised observables, 
as ratios of these decays, that are theoretically clean observables and whose measurements 
are in tension with SM predictions. One example is the case of the \RDp ratios, 
\begin{equation}
\RDp^\ell = \frac{\mathrm{BR}(B \to D^{(*)} \tau \bar{\nu}_\tau ) }{[\mathrm{BR}(B \to D^{(*)} e \bar{\nu}_e) + \mathrm{BR}(B \to D^{(*)} \mu \bar{\nu}_\mu)]/2}\ ,
\end{equation}
and
\begin{equation}
\RDp^\mu = \frac{\mathrm{BR}(B \to D^{(*)} \tau \bar{\nu}_\tau ) }{ \mathrm{BR}(B \to D^{(*)} \mu \bar{\nu}_\mu)}\ .
\end{equation}
In the $b\to c \ell \nu$ transitions, signs of violation of lepton
universality have been observed only in the $e-\tau$ and $\mu-\tau$
cases, while the universality has been tested to great precision in the
$e-\mu$ case~\cite{Abdesselam:2017kjf,Belle:2018ezy,Jung:2018lfu}. As a consequence,
both $\RDp^\ell$ and $\RDp^\mu$ should have similar predictions and
measurements. Note that \RD and \RDp ratios should only have similar
predictions in the SM or any other Lepton Flavour Universality (LFU) theory. 
The measured values of these ratios at BaBar, Belle and 
LHCb experiments are larger than the SM prediction ($R_D^{\ell\ \mathrm{SM}} = 0.299
\pm 0.003$, $R_{D^*}^{\ell\ \mathrm{SM}} = R_{D^*}^{\mu\ \mathrm{SM}} =
0.258 \pm 0.005$  \cite{Amhis:2019ckw}). The first deviation was
found by BaBar in 2012\cite{Lees:2012xj} 
\begin{equation}
R_D^\ell = 0.440 \pm 0.058 \pm 0.042\ ,\qquad\qquad
R_{D^*}^\ell = 0.332 \pm 0.024 \pm 0.018\ .
\end{equation}
The latest experimental results
from Belle are \cite{Abdesselam:2019dgh}, 
\begin{equation}
R_D^\ell = 0.307 \pm 0.037 \pm 0.016\ , \qquad\qquad R_{D^*}^\ell = 0.283 \pm 0.018 \pm 0.014\ ,
\end{equation}
and from LHCb \cite{Aaij:2017uff},
\begin{equation}
R_{D^*}^\mu = 0.291 \pm 0.019 \pm 0.026 \pm 0.013\ .
\end{equation}

The combined result from the Belle measurements has a compatibility with 
the SM predictions of $1.2\,\sigma$, much better than previous
measurements of these observables (see, for example, a compatibility of 
$3.6\,\sigma$ as of 2016 \cite{Amhis:2016xyh}). The world average of the 
experimental values for the \RDp ratios, as obtained by the Heavy
Flavour Averaging Group (HFLAV), assuming universality in the lighter 
leptons, is \cite{Amhis:2019ckw}
\begin{equation}
R_D^\mathrm{ave} = 0.340 \pm 0.027 \pm 0.013,\qquad\qquad
R_{D^*}^\mathrm{ave} = 0.295 \pm 0.011 \pm 0.008.
\end{equation}
$R_D$ exceeds the SM value by $1.4\,\sigma$, and $R_{D^*}$ by $2.5\,\sigma$. When combined together, included their correlation, the excess is $3.08\,\sigma$.

Another class of $B$ meson observables showing signs of LFUV is related
to $b  \to s \ell^+ \ell^- $ processes, namely the optimised angular
observable $P_5'$ \cite{DescotesGenon:2012zf} and the \RKp ratios,
\begin{equation}
\RKp = \frac{\mathrm{BR}(B\to K^{(*)} \mu^+ \mu^- )}{\mathrm{BR}(B\to K^{(*)} e^+ e^- )}\ .
\end{equation}

The \RKp ratios are observables that have small theoretical
uncertainties, and in the SM, $R_K = R_{K^*} = 1$ with uncertainties of
the order of $1\%$ \cite{Hiller:2003js,Bordone:2016gaq} as a consequence
of LFU.  The latest experimental results from
LHCb, in the specified regions of $q^2$ di-lepton invariant mass, are:
\begin{align}
R_K^{[1.1, 6]} = 0.846^{+0.042}_{-0.039}{}^{+0.013}_{-0.012}\,, \qquad &\mbox{\cite{Aaij:2021vac} }\nonumber \\
R_{K^*}^{[0.045,1.1]} = 0.66^{+0.11}_{-0.07}\pm 0.03\,, \qquad\qquad
R_{K^*}^{[1.1,6]} = 0.69^{+0.11}_{-0.07}\pm 0.05\ . \qquad
&\mbox{\cite{Aaij:2017vbb} }
\end{align}

The compatibility of the individual measurements with respect to the SM
predictions is of $3.1\sigma$ for the $R_K$ ratio, $2.3\sigma$ for the
$R_{K^*}$ ratio in the low-$q^2$ region and $2.4\sigma$ in the
central-$q^2$ region. The Belle collaboration has also recently reported
experimental results for the \RKp ratios~\cite{Abdesselam:2019lab,Abdesselam:2019wac}, 
although with less precision than the LHCb measurements. 

A great theoretical effort has been devoted to the understanding of the deviations in the $\RKp$ observables~\cite{Altmannshofer:2017fio,Hiller:2014yaa,Hiller:2014ula,Crivellin:2015lwa,Crivellin:2015era,Hurth:2016fbr,Capdevila:2017ert,Chobanova:2017ghn,Altmannshofer:2017wqy,Altmannshofer:2017yso,Hiller:2017bzc,Geng:2017svp,Ciuchini:2017mik,Alda:2018mfy,Coy:2019rfr}, 
the deviations in the $\RDp$ observables~\cite{Celis:2016azn,Ivanov:2017mrj,Altmannshofer:2017poe,Bigi:2017jbd,Iguro:2017ysu,Alok:2017qsi,Jung:2018lfu,Azatov:2018knx,Bhattacharya:2018kig,Huang:2018nnq,Blanke:2018yud,Murgui:2019czp,Blanke:2019qrx,Hu:2019bdf,Mandal:2020htr,Iguro:2020keo,Becirevic:2020rzi}, 
and combined explanations for the deviations in $\RKp$ and $\RDp$ \cite{Bhattacharya:2014wla,Hiller:2016kry,Bhattacharya:2016mcc,Cai:2017wry,Alok:2017jaf,Feruglio:2017rjo,Buttazzo:2017ixm,DiLuzio:2017vat,Bordone:2017bld,Blanke:2018sro,Becirevic:2018afm,Kumar:2018kmr,Angelescu:2018tyl,Bifani:2018zmi,Saad:2020ucl,Babu:2020hun}
and references therein. Besides, the experimental data have been used to constrain New Physics (NP) models. 
Several global fits have been performed in the
literature~\cite{Capdevila:2017bsm,Celis:2017doq,Alok:2017sui,Camargo-Molina:2018cwu,Datta:2019zca,Aebischer:2019mlg,Aoude:2020dwv}. 

One of the most widely used tools to study any possible New Physics (NP)
contribution that explain the above experimental results is the
Effective Field Theories. The effective Hamiltonian approach allows us to
perform a model-independent analysis of NP effects. In this way, it is possible to
obtain constraints on NP contributions to the Wilson coefficients of the
Hamiltonian from the experimental results. 

In this work we investigate the effects of the global fits to the Wilson
coefficients assuming a model independent effective Hamiltonian approach
and including a discussion of the consequences of our analysis in
leptoquark models. We define different scenarios for the
phenomenological study by considering the NP contributions to the Wilson
coefficients in such a way that NP is present in one, two or three of the Wilson
coefficients simultaneously. These scenarios are used to study the
impact of the global fits on the Wilson coefficients and, therefore, 
to exhibit more clearly which combinations of Wilson coefficients are 
preferred and/or constrained by experimental data.

We begin in Sect.~\ref{sec:EFT} by presenting a brief summary of the Effective Field
Theory used to describe possible NP contributions to $B$ decays 
observables. Then,
Sect.~\ref{sec:fits} is devoted to the global fits to the Wilson
coefficients, presenting the set of scenarios that we are going to
analyse. As already explained, we will work in different scenarios that
arise by considering the presence of NP contributions in one, two or 
three of the Wilson coefficients. We will compare the results of the global
fit in each scenario with respect to two cases: the SM and the best fit
point of the three independent Wilson coefficients scenario
(the most general case). This particular choice of the Wilson
coefficients that will enter our analysis is the main difference 
with respect to previous global fits analysis in the recent literature. 
Sect.~\ref{sec:VII} is devoted to
discuss in more detail the most general proposed scenario, Scenario
VII, in which the prediction of the $\RDp$ and $\RKp$ observables is improved.
Finally, the phenomenological implications of our analysis in leptoquark models are included in
Sect.~\ref{sec:leptoq}. Conclusions are presented in Sect.~\ref{sec:conclu}.
Appendix~\ref{app:pulls} contains the list of observables that
contribute to the global fit with their prediction in
the most general scenario: the global fit to three independent Wilson coefficients
receiving NP contributions.

\section{Effective field theories for $B$ observables}\label{sec:EFT}

One of the most widely used tools to study any possible New Physics (NP)
contribution is the Effective Field Theories. The Standard Model
Effective Field Theory (SMEFT) is formulated at an energy scale 
$\mu_\mathrm{SMEFT} = \Lambda$ higher than the electroweak scale, and
the degrees of freedom are all the SM fields. The Weak Effective Theory
(WET) is formulated at an energy scale below the electroweak scale, 
for example $\mu_\mathrm{WET} = m_b$, and the top quark, Higgs, W and Z
bosons are integrated out.

In this work, all the numerical analyses will be performed using
  only the SMEFT operators, while the WET Lagrangian will be useful for
  the discussion of the results.

\subsection{Weak Effective Theory}

The relevant terms of the WET
Lagrangian~\cite{Buras:1998raa,Aebischer:2015fzz,Aebischer:2017gaw,Tanaka:2012nw}
are
\begin{equation}
\mathcal{L}_{\text{eff}} = -\frac{4 G_F}{\sqrt{2}}V_{cb}\sum_{\ell = e, \mu, \tau} (1 + C_{VL}^\ell) \mathcal{O}_{VL}^\ell + \frac{4G_F}{\sqrt{2}}V_{tb}V_{ts}^*\frac{e^2}{16\pi^2}\sum_{\ell=e,\mu} (C_9^\ell \mathcal{O}_9^\ell  + C_{10}^\ell \mathcal{O}_{10}^\ell) \ ,\label{eq:Lagr_WET}
\end{equation}
where $G_F$ is the Fermi constant, $e$ is the electromagnetic coupling, $V_{qq'}$
are the elements of the Cabibbo-Kobayashi-Maskawa (CKM) matrix and with the dimension six
operators defined as,  
\begin{equation}
\mathcal{O}_{VL}^\ell = (\bar{c}_L \gamma_\alpha b_L)(\bar{\ell}_L \gamma^\alpha \nu_\ell)\ ,\qquad \mathcal{O}_9^\ell = (\bar{s}_L \gamma_\alpha b_L)(\bar{\ell} \gamma^\alpha \ell)\ , \qquad \mathcal{O}_{10}^\ell = (\bar{s}_L \gamma_\alpha b_L)(\bar{\ell} \gamma^\alpha \gamma_5 \ell) \ ,
\end{equation}
and their corresponding Wilson coefficients $C_{VL}^\ell$, $C_9^\ell$
and $C_{10}^\ell$. The $C_9^\ell$ and $C_{10}^\ell$ Wilson coefficients
have contributions from the SM processes as well as any NP contribution, 
\begin{equation}
C_i^\ell = C_i^{\mathrm{SM}\, \ell} + C_i^{\mathrm{NP}\, \ell}\ ,\qquad\qquad i= 9,10\ ,
\end{equation}
whereas $C_{VL}^\ell$ only receives contributions from NP. 
In the present work we analyse the NP contributions.

The \RDp ratios obey the following expression \cite{Bhattacharya:2016mcc,Feruglio:2017rjo}:
\begin{align}
\RDp^\ell &= \RDp^{\ell, \mathrm{SM}} \frac{|1+ C_{VL}^\tau|^2}{ (|1+C_{VL}^e|^2 + |1+C_{VL}^\mu|^2)/2}\ , \nonumber\\
\RDp^\mu &= \RDp^{\mu, \mathrm{SM}} \frac{|1+ C_{VL}^\tau|^2}{ |1+C_{VL}^\mu|^2}\ . \label{eq:RD}
\end{align}

The dependence of the \RKp ratios on the Wilson coefficients has been
previously obtained in \cite{Alda:2018mfy}, where an analytic 
computation of $\Rks$ as a function of $\CnmuNP$, $\CtmuNP$ in the
region $1.1 \leq q^2 \leq 6.0\GeV^2$
was performed. The result is given by~\cite{Alda:2018mfy} 
\begin{equation}
R_{K^*}^{[1.1,6]} \simeq \frac{0.9875+0.1759\, \mathrm{Re}\,\CnmuNP - 0.2954\,  \mathrm{Re}\, \CtmuNP + 0.0212|\CnmuNP|^2 + 0.0350 |\CtmuNP|^2}{1\,\  \ \ \ \ +0.1760\, \mathrm{Re}\,\CneNP - 0.3013\, \mathrm{Re}\,  \CteNP + 0.0212|\CneNP|^2 + 0.0357 |\CteNP|^2}\ .
\end{equation}

\subsection{Standard Model Effective Field Theory}

We consider NP contributions at an energy scale $\Lambda$ ($\Lambda \sim
\mathcal{O}(\mathrm{TeV})$) described by the SMEFT
Lagrangian as given in~\cite{Grzadkowski:2010es}, where a complete list
of the independent dimension-six operators that are allowed by the SM
gauge symmetries is presented. 
The SMEFT Lagrangian is given by~\cite{Grzadkowski:2010es}
\begin{equation}
\mathcal{L}_\mathrm{SMEFT} = \frac{1}{\Lambda^2}\left(\Clqo^{ijkl}\, O_{\ell q(1)}^{ijkl} + \Clqt^{ijkl}\,  O^{ijkl}_{\ell q(3)}   \right) \ ,
\label{eq:Lagr_SMEFT}
\end{equation}
where the dimension six operators are defined as 
\begin{equation}
O_{\ell q(1)}^{ijkl} = (\bar{\ell}_i \gamma_\mu \ell_j)(\bar{q}_k \gamma^\mu  q_l),\qquad\qquad O_{\ell q(3)}^{ijkl}= (\bar{\ell}_i \gamma_\mu \tau^I \ell_j)(\bar{q}_k \gamma^\mu \tau^I q_l) ,
\label{eq:oper6}
\end{equation}
$\ell$ and $q$ are the lepton and quark $SU(2)_L$ doublets defined in
the mass basis\footnote{As given in~\cite{Dedes:2017zog}, and used
  on the package \texttt{wilson}~\cite{Aebischer:2018bkb}, in both
  the ``Warsaw'' basis and the ``Warsaw mass'' basis, the lepton and
  $d$-quark fields are defined so that their mass matrices are
  diagonal. Consequently, translating from one to another does not
  modify the Lagrangian in Eq.~\eqref{eq:Lagr_SMEFT}.},
$\tau^I$ the Pauli matrices, and ${i,j,k,l}$ denote
generation indices. The $O_{\ell q(1)}$ operator couples two
$SU(2)_L$-singlet currents, while the $O_{\ell q(3)}$ operator couples
two $SU(2)_L$-triplet currents. Consequently, $O_{\ell q(1)}$ only
mediates flavour-changing neutral processes and $O_{\ell q(3)}$
mediates both flavour-changing neutral and charged processes. 
We will restrict our analysis to operators including only third-generation quarks and 
same-generation leptons, and we will use the following notation for their Wilson coefficients:
\begin{equation}
\Clq^e \equiv \Clq^{1133}\ , \qquad\qquad \Clq^\mu \equiv \Clq^{2233}\ , \qquad\qquad \Clq^\tau \equiv \Clq^{3333}\ .
\label{eq:wcs}
\end{equation}
This particular choice of the Wilson coefficients that will enter our
analysis is motivated by the fact that the most prominent discrepancies
between SM predictions and experimental measurements, namely
$R_{K^{(*)}}$ and $R_{D^{(*)}}$, affect the third quark generation. From
a symmetry point of view, this would amount to imposing an 
$U(2)^3 = U(2)_q \times U(2)_u \times U(2)_d$ symmetry between the first
and second quark generations~\cite{Barbieri:2011ci,Barbieri:2012uh,AguilarSaavedra:2018nen}, 
that remain SM-like. No restriction is imposed on the third quark
generation. In the lepton sector we only consider diagonal entries in 
order to avoid Lepton Flavour Violating (LFV) decays. This flavour
structure for NP contributions has been presented
in~\cite{AguilarSaavedra:2018nen} as a minimal working set-up.

These operators generate the $C_{VL}^\ell$, $C_9^\ell$ and $C_{10}^\ell$
operators of the electroweak effective field theory when matched at the electroweak 
scale $\mu_\mathrm{EW}$. Using the package
\texttt{wilson}~\cite{Aebischer:2018bkb},
we define the $\Clq$ operators at
$\Lambda = 1 \,\mathrm{TeV}$, we calculate their running down to $\mu_\mathrm{EW} =
M_Z$, then match them with the WET operators and finally run the
down to $\mu = m_b$, where the $B$-physics observables are computed.
By taking the SMEFT Wilson coefficients at $\Lambda = 1 \,\mathrm{TeV}$,
we found the following relations between the Wilson coefficients at high
and low energies:  
\begin{align}
C_9^{\mathrm{NP}\ e, \mu}(m_b) = -0.583 \, \Clqo^{e, \mu} - 0.596 \, \Clqt^{e,
  \mu}\ , &\qquad C_{10}^{\mathrm{NP}\ e, \mu}(m_b) = 0.588 \, \Clqo^{e,
  \mu} + 0.591 \, \Clqt^{e, \mu}\ , \nonumber\\
  C_{VL}^{e, \mu}(m_b) = 0.0012 \, \Clqo^{e, \mu} - 0.0644\, \Clqt^{e, \mu}\ ,
          &\qquad C_{VL}^\tau(m_b) = -0.0598\, \Clqt^\tau\ .
\label{eq:running}
\end{align}

It is important to note that the Renormalization Group-induced SMEFT
operators shift the Fermi constant $G_F$~\cite{Jenkins:2017jig} in the
Lagrangian~\eqref{eq:Lagr_WET} from its SM value $G_F^0$, determined
experimentally from the muon lifetime. This shift is already included in
the matching conditions of Eq.~\eqref{eq:running}. Note that both the
contributions from the SMEFT operators entering in the redefinition of
the vacuum expectation value and the ones that are relevant for
the muon decay are included
in our analysis. These two contributions are implemented in the
package \texttt{wilson}~\cite{Aebischer:2018bkb}. 
The elements of the CKM matrix are also affected by SMEFT
contributions~\cite{Descotes-Genon:2018foz}. Those contributions
have not been included in the present work. 

The $\mathcal{O}_{\ell q}$ operators~(\ref{eq:oper6}) also produce unwanted
contributions to the $B \to K^{(*)} \nu \bar{\nu}$
decays~\cite{Feruglio:2017rjo,Aebischer:2018iyb}. In order to obey these
constraints, we will fix the relation at the scale $\Lambda = 1\,\mathrm{TeV}$ 
\begin{equation}
\Clqo^i = \Clqt^i \equiv \Clq^i\ .\label{eq:C1C3}
\end{equation}
While the above relation eliminates the tree-level contribution to the
$B \to K^{(*)} \nu \bar{\nu}$ decays, the Renormalization Group (RG) generates a one-loop
contribution proportional to the $C_{\ell q(3)}$ coefficients. However,
we have checked that this term is only a correction of $0.1\%$ of the SM
prediction. Relation~\eqref{eq:C1C3} also has the positive consequence
of a partial cancellation of loop-induced effects in $Z$-pole and LFV
observables. 

Finally, an important point to emphasise here is
that the dimension-six operators affect a large range of observables
because of the RG equations that give mixing between different particle
sectors. Therefore, 
any NP prediction based on Wilson coefficients has to be confronted not 
only with the \RKp an \RDp measurements, but also with additional
several measurements involving the decays of $B$ mesons. In the case of 
the SMEFT, the evolution of the RG produces a
mix of the low-energy effective operators. More concretely, the 
$\mathcal{O}_{\ell q}$ operators mix under RG evolution
with~\cite{Jenkins:2013wua,Alonso:2013hga,Buchalla:2013mpa} 
\begin{equation} 
\mathcal{O}_{\varphi\ell(1)}^{jk} = (\varphi^\dagger i \overleftrightarrow D_{\mu} \varphi)(\bar{\ell}_j \gamma^\mu \ell_k )\ ,\qquad \mathcal{O}_{\varphi\ell(3)}^{jk} = (\varphi^\dagger i \overleftrightarrow D_{\mu}^I \varphi)(\bar{\ell}_j \gamma^\mu \tau^I \ell_k )\ ,\qquad\mathcal{O}_{\varphi e}^{jk} = (\varphi^\dagger i \overleftrightarrow D_{\mu} \varphi)(\bar{e}_j \gamma^\mu e_k )\ ,
 \end{equation}
that modify the $W$ and $Z$ couplings to leptons. In consequence, NP in
the semileptonic couplings of third generation quarks will indirectly affect electroweak observables, such as
the mass of the $W$ boson, the hadronic cross section of the $Z$ boson 
$\sigma^0_\mathrm{had}$ or the branching ratios of the $Z$ to different
leptons. In order to keep the predictions consistent with this range of 
experimental test, global fits have proven to be a valuable
tool~\cite{Celis:2017doq,Camargo-Molina:2018cwu,Aebischer:2019mlg,Aoude:2020dwv}. 

\section{Global fits}\label{sec:fits}

We have performed global fits to the $\Clq$ Wilson coefficients using
the package \texttt{smelli} v1.3~\cite{Aebischer:2018iyb}. The global fit
includes the \RKp and \RDp observables, the electroweak precision
observables, $W$ and $Z$ decay widths and branching ratios to leptons,
the $b\to s \mu\mu$ observables (including $P_5'$ and the branching ratio of
$B_s \to \mu\mu$) and the $b \to s \nu \bar{\nu}$ observables. The SM input parameters
are presented in Table~\ref{tab:tabSMinputs}. These values are
taken from open-source code \texttt{flavio} v1.5~\cite{Straub:2018kue}; sources used by the
program are quoted when available\footnote{ We have supplemented the experimental
measurements of the \texttt{flavio} v1.5 database with updated 
values for $R_K$~\cite{Aaij:2021vac}, $\RDp$~\cite{Abdesselam:2019lab},
$B\to K^*\ell^+\ell^-$ differential
observables~\cite{Aaij:2020nrf,Aaij:2020umj}, $B_{(s)}\to\mu^+\mu^-$~\cite{LHCb:2020zud} 
and a re-analysis of the electroweak precision tests from
LEP~\cite{Janot:2019oyi}.}. Note that the experimental measurement of
the $\mu\to e\bar{\nu}\nu$ decay, used to determine the SM input
parameters, is not included in the fit in order to ensure the
consistency of the procedure. The parameters $V_{us}$, $V_{ub}$,
$V_{cb}$ and $\delta_{KM}$ of the CKM matrix are treated as nuisance
parameters of the fit, and the remaining elements are determined
implementing the unitarity of the matrix. A complete analysis would
require including the SMEFT corrections to the CKM matrix, which have
not been considered in this work. 

\begin{table}
\centering
\renewcommand{\arraystretch}{1.3}
\begin{tabular}{|c|c|c|}\hline
$G_F^0$ & $1.1663787(6)\times 10^{-5}~\mathrm{GeV}^{-2}$ & PDG 2014 \cite{Agashe:2014kda} \\\hline
$\alpha_e(M_Z)$ & $0.00781616(86)$ &\cite{Straub:2018kue} \\\hline
$\alpha_s(M_Z)$ & $0.1182(8)$ & FLAG 2019 \cite{Aoki:2019cca}
 \\\hline
$\sin^2 \hat{\theta}_W(M_Z)$, $\overline{\mathrm{MS}}$ & $0.23129(5)$ & PDG 2017 \cite{Patrignani:2016xqp} \\\hline
$V_{us}$ & 0.2248(8) &  FLAG 2017 $N_f=2+1+1$ \cite{Aoki:2016frl} \\\hline
$|V_{ub}|$ & $3.73(14)\times 10^{-3}$ & FLAG 2017 $N_f=2+1$ $B\to \pi\ell\nu$ \cite{Aoki:2016frl} \\\hline
$V_{cb}$ & $4.221(78)\times 10^{-2}$ & \cite{Straub:2018kue} \\\hline
$\delta_\mathrm{KM}$ & 1.27(12) & \cite{Straub:2018kue}\\\hline
$m_u (2~\mathrm{GeV})$, $\overline{\mathrm{MS}}$ & $2.130(41)~\mathrm{MeV}$ & \cite{Bazavov:2018omf} \\\hline
$m_d (2~\mathrm{GeV})$, $\overline{\mathrm{MS}}$ & $4.675(56)~\mathrm{MeV}$ & \cite{Bazavov:2018omf} \\\hline
$m_s (2~\mathrm{GeV})$, $\overline{\mathrm{MS}}$ & $92.47(69)~\mathrm{MeV}$ & \cite{Bazavov:2018omf} \\\hline
$m_c (m_c)$, $\overline{\mathrm{MS}}$ & $1.273(10)~\mathrm{GeV}$ & \cite{Bazavov:2018omf} \\\hline
$m_b (m_b)$, $\overline{\mathrm{MS}}$ & $4.195(14)~\mathrm{GeV}$ & \cite{Bazavov:2018omf} \\\hline
\end{tabular}
\caption{SM input parameters.}
\label{tab:tabSMinputs}
\end{table}
We proceed to study observables by defining some specific scenarios
for combinations of the $\Clq^i$ operators such that NP contributions
to the Wilson coefficients emerge in one, two or three of the Wilson 
coefficients simultaneously: in Scenarios I-III NP only modifies the
$\Clq$ operators in one lepton flavour at a time; in Scenarios IV-VI 
NP is present in two of the Wilson coefficients simultaneously; and finally
in Scenarios VII-IX we consider the more general case in which three of
the $\Clq^i$ operators receive NP contributions. The more general one of
these last three scenarios is Scenario VII, in which we consider three 
independent Wilson coefficients. This scenario is discussed in more 
detail in Sect.~\ref{sec:VII}.

The goodness of each fit is evaluated with its difference of $\chi^2$
with respect to the SM, $\Delta \chi^2_\mathrm{SM} = \chi^2_\mathrm{SM} -
\chi^2_\mathrm{fit}$. The package \texttt{smelli} actually computes the
differences of the logarithms of the likelihood function 
$\Delta \log L = -\frac{1}{2} \Delta \chi^2$. The $\chi^2_\mathrm{fit}$
includes the experimental and theoretical uncertainties and correlations
of the observables. In order to compare two 
fits $A$ and $B$, we use the pull between them in units of $\sigma$, 
defined as~\cite{Descotes-Genon:2015uva,Capdevila:2018jhy}
\begin{equation}
\mathrm{Pull}_{A \to B} = \sqrt{2} \mathrm{Erf}^{-1}[F(\Delta \chi^2_A - \Delta \chi^2_B; n_B - n_A )]\,,
\end{equation}
where $\mathrm{Erf}^{-1}$ is the inverse of the error function, $F$ is
the cumulative distribution function of the $\chi^2$ distribution and
$n$ is the number of degrees of freedom of each fit. We will compare
each scenario against two cases: the SM ($\Clq = 0$,
$n=0$) and the fit to three independent Wilson coefficients (Scenario
VII), which is the more general and descriptive case. The pull from the
SM quantifies how much each scenario is preferred over the SM to
describe the data. The larger the pull, the better the description of
the data of the preferred scenario. The pull of Scenario VII quantifies
how much the fit over the whole space of parameters is preferred over the simpler and
more constrained fits. From the analysis of this pull, we are able to discuss the
relevance of the proposed scenarios, the larger the pull means that the more
restricted scenario represents a worse description of the experimental data.

\begin{table}
\centering
\begin{tabular}{|c|c|c|c|c|c|c|c|}\hline
\multicolumn{2}{|c|}{\multirow{2}*{Scenario}}&\multirow{2}*{$C_{\ell
    q}^e$} &\multirow{2}*{$C_{\ell q}^\mu$}&\multirow{2}*{$C_{\ell
    q}^\tau$} &\multirow{2}*{$\Delta\chi^2_\mathrm{SM}$} & Pull & Pull
\\ \multicolumn{2}{|c|}{} & & & & & from SM & to VII\\\hline
I& $e$	& $-0.14 \pm 0.04$ & & & 8.84 &	2.97 $\sigma$ & 4.37 $\sigma$\\\hline
II& $\mu$ & & $0.10 \pm 0.04$ & & 5.47 & 2.34 $\sigma$ & 4.73 $\sigma$\\\hline
III& $\tau$ & & & $-0.38\pm0.19$ & 3.85 & 1.96	$\sigma$ & 4.89 $\sigma$\\\hline
IV& $e$ and $\mu$ & $-0.25 \pm 0.07$ &	$0.24 \pm 0.06$	& & 28.42 & 4.97
$\sigma$ & 1.75 $\sigma$\\\hline
V& $e$ and $\tau$ & $-0.14 \pm 0.06$ & & $-0.4 \pm 0.3$ & 12.98 & 3.17 $\sigma$ & 4.30 $\sigma$\\\hline
VI& $\mu$ and $\tau$ &	& $0.10 \pm 0.06$ & $-0.3 \pm 0.3$ & 8.73 & 2.49
$\sigma$ & 4.77 $\sigma$\\\hline
VII& $e$, $\mu$ and $\tau$ & $-0.25 \pm 0.02$ & $0.211 \pm 0.016$ &
$-0.3 \pm 0.4$ & 31.50 & 4.97 $\sigma$ &\\\hline
VIII& $e = \mu = \tau$	& $-0.0139 \pm 0.0003$ & $-0.0139 \pm 0.0003$ &
$-0.0139 \pm 0.0003$ & 0.30 & 0.55 $\sigma$ & 5.23 $\sigma$\\\hline
IX& $e = -\mu = \tau$ & $-0.232\pm0.001$ & $0.232\pm0.001$ &
$-0.232\pm0.001$ & 30.74 & 5.54 $\sigma$ & 0.41 $\sigma$\\\hline
\end{tabular}
\caption{Best-fit values with $1 \sigma$ uncertainties and pulls from the Standard Model and of
  Scenario VII for several combinations of $\Clq^i$ operators.}
\label{tab:Fits}
\end{table}
The results of the fits are summarised in Table~\ref{tab:Fits} for
several combinations of $\Clq^i$ operators, with one-, two- or three-lepton flavour
present simultaneously in the Wilson coefficients as defined below.
The best fit values at 1 $\sigma$ and pulls from the SM and to Scenario
VII for all scenarios are included in this table.

\begin{figure}
\begin{center}
\begin{tabular}{ccc}
\includegraphics[width=0.3\textwidth]{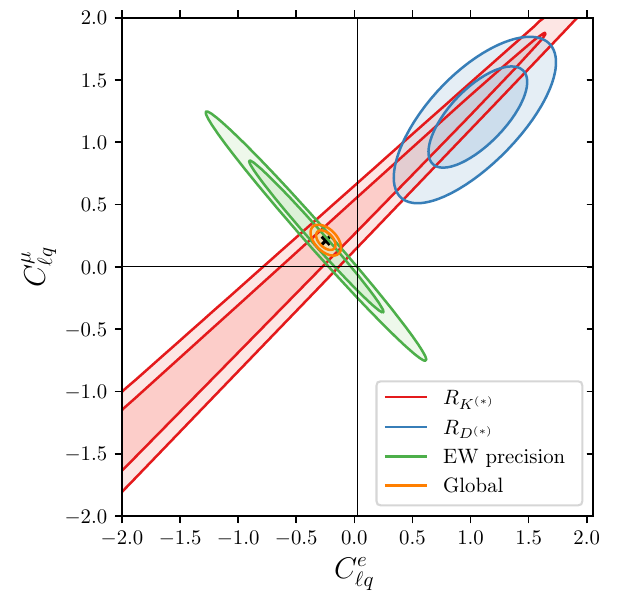}&
\includegraphics[width=0.3\textwidth]{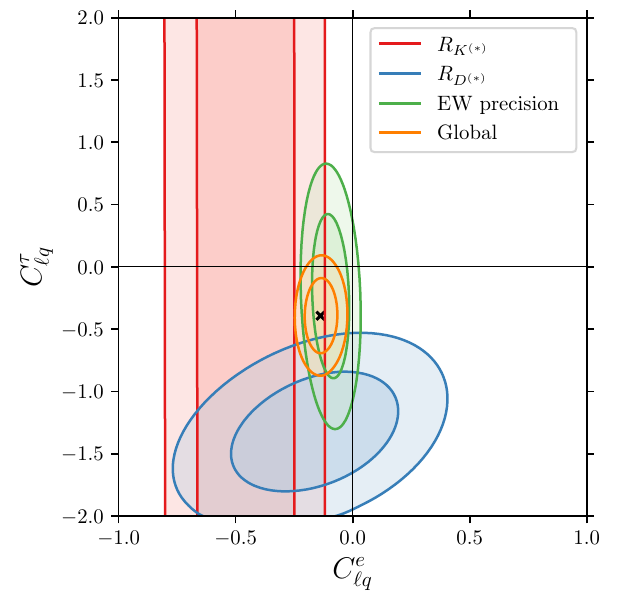}&
\includegraphics[width=0.3\textwidth]{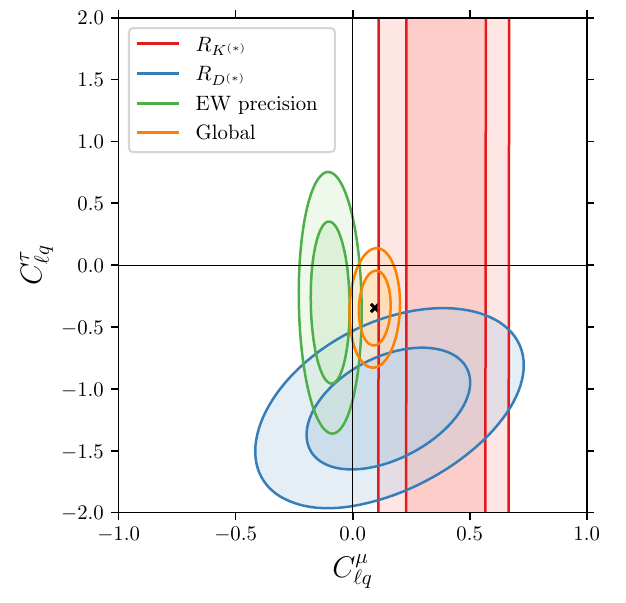}\\
(a)&(b)&(c)
\end{tabular}
\caption{$1\sigma$ and $2\sigma$ contours for scenarios with two
  lepton flavours present 
in the Wilson coefficients: (a) Scenario IV, (b) Scenario V, and (c)
Scenario VI. All available data is considered.}
\label{im:globalfits}
\end{center}
\end{figure}

\begin{figure}
\begin{center}
\begin{tabular}{cc}
\resizebox{0.45\textwidth}{!}{\input{RKplot_21.pgf}}&
\resizebox{0.45\textwidth}{!}{\input{RDplot_21.pgf}}\\
(a)&(b)
\end{tabular}
\caption{Central value and $1 \sigma$ uncertainty of the (a) $\RKp$
  observables, and (b) $\RDp$ observables (blue lines) in Scenarios 
IV, VII and IX, compared to the SM prediction (yellow) and experimental measurements (green).}
\label{im:RK}
\end{center}
\end{figure}
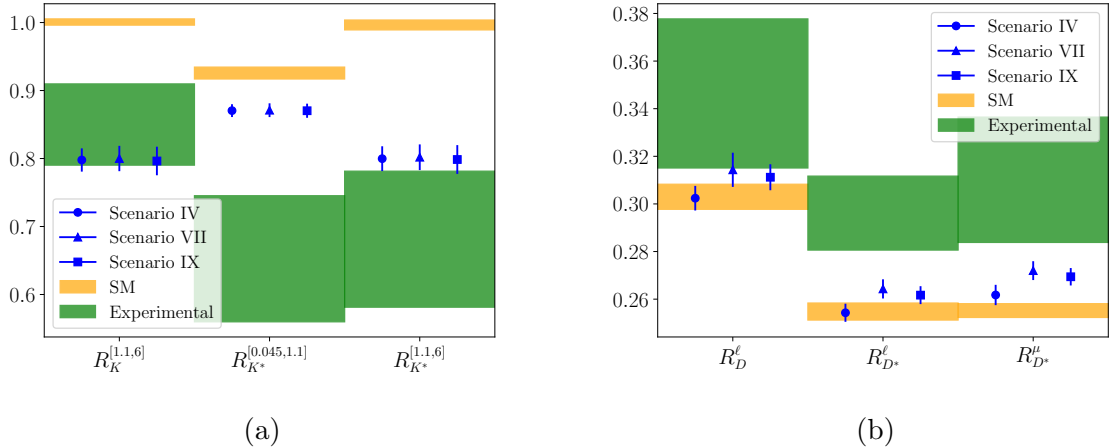

\begin{itemize}
\item \textbf{Scenarios I, II and III:} In these scenarios, NP only
  modifies the $\Clq^i$ operators in one lepton flavour at a time, i.e. 
  $\Clq^e$, $\Clq^\mu$ or $\Clq^\tau$. The largest pull from the SM
  prediction, almost $3\ \sigma$, is found
  in Scenario I when the coupling to electrons is added. This result 
is a reflection of the great impact of the electroweak precision 
observables in the global fit. The fit to only muons in Scenario
II displays only a pull from the SM of $2.34\ \sigma$;
if we restricted our fit to only $b \to s \ell^+ \ell^-$ observables this
fit would display a better pull, in line with the common wisdom about the 
  anomalies, explaining them through NP in the muon
  sector~\cite{Altmannshofer:2017yso,Ciuchini:2017mik,Capdevila:2017bsm,Descotes-Genon:2015uva,DAmico:2017mtc}. The
  worst pull is obtained in the fit to the tau coefficient, 
with $1.96\ \sigma$, as it does not modify the value of the \RKp
ratios. 
Scenarios I and II both produce SM-like predictions for the observables 
$R_D$ and $R_{D^*}$: $R^\ell_D = 0.3006$ and $R^\ell_{D^*}=0.2528$ for Scenario I
and $R^\ell_D = 0.3048$ and $R^\ell_{D^*}=0.2563$ for Scenario II.
Scenario III, with a larger value of its Wilson
coefficient, produces values closer to the average of the experimental measurements;
i.e. $R^\ell_D = 0.318$ and $R^\ell_{D^*}=0.268$. In order to fully
address the anomaly in these observables, a larger deviation from the 
SM would be needed; however such a deviation would be in conflict with 
the electroweak precision data, as we will see later in
Sect.~\ref{sec:VII}, and in agreement with~\cite{Capdevila:2017iqn}.

\item \textbf{Scenarios IV, V and VI:} In these scenarios NP is present
  in two of the Wilson coefficients. The best fit corresponds to
  Scenario IV, where the contributions to $\Clq^e$ and $\Clq^\mu$ are
  favoured with a pull of $4.97\ \sigma$ with respect to the
  SM. Fig.~\ref{im:globalfits} shows the allowed regions for these
  fits. In the fit to Scenario IV, the \RKp and \RDp observables
  constrain the $\Clq^e - \Clq^\mu$ combination, while the
  LFU-conserving electroweak precision observables tightly constrain the 
combination $\Clq^e + \Clq^\mu$. It is clear that EW precision
observables play an important role in the global fit and the preferred
values for the Wilson coefficients. The reason for this behaviour is
justified by deviations in Z-couplings to leptons, the $\tau$-leptonic
decays and the Z and W decays widths, as shown in~\cite{Feruglio:2018jnu}. 
The values of the \RKp and \RDp observables in this scenario are given in 
Table~\ref{tab:observables}. Together, these
  sets of observables constrain the fit to a narrow ellipse around the
  best fit point. In Scenarios V and VI, the $\Clq^\tau$ coefficient is
  determined by the electroweak precision observables, that are
  compatible with a SM-like coefficient, and by \RDp observables, that
  prefer a large negative value. All the experimental constraints for
  $\Clq^\tau$ show large uncertainties, which result in less statistical
  significance of these fits and $\Clq^\tau$ still being compatible with
  zero at $2\,\sigma$ level. The central values with 1 $\sigma$ uncertainties
  of the \RKp and \RDp observables for Scenario IV (the best-fit scenario
  in this subset) are shown in Table~\ref{tab:observables} and
  Fig.~\ref{im:RK}. Below we compare these results in various scenarios.

\item \textbf{Scenario VII:} In this fit, the three $\Clq$ operators
  receive independent NP contribution. The pull from the SM, $4.97\ \sigma$, is
  similar to that of Scenario IV, and the values of $\Clq^e$ and
  $\Clq^\mu$ are similar too; therefore, the predictions for the \RKp
  observables are very similar, as shown in Fig.~\ref{im:RK}a. The
  value of $\Clq^\tau$ is close to that of Scenarios III, V and VI, which
  allows a better fit to the \RDp observables, and especially to
  $R_D^\ell$, that is compatible at $1\,\sigma$ with its experimental
  value, as shown in Fig.~\ref{im:RK}b. Therefore, we conclude that the
  prediction of the \RDp and \RKp observables is improved in Scenario VII. 
We will discuss this scenario in more detail in Sect.~\ref{sec:VII}.

\item \textbf{Scenario VIII:} This scenario has \textit{universal}
  couplings; the three Wilson coefficients have the same
  \textit{universal} contribution and do not violate LFU. It has the 
smallest pull with respect to the SM ($0.30\,\sigma$). This shows that
LFU NP cannot explain experimental data, and LFU violation is needed 
to accommodate it.

\item \textbf{Scenario IX:} In this scenario, the three Wilson
  coefficients have the same absolute value, but $\Clq^\mu$ has 
the opposite sign. This particular arrangement of the coefficients 
was inspired by the similar absolute values of $\Clq^e$ and $\Clq^\mu$ 
in Scenario VII. This choice produces a good fit, with a pull of 
$5.54\,\sigma$. It is also the only scenario that remains compatible at $1\,\sigma$ with Scenario VII.
\end{itemize}

\begin{table}
\centering
\begin{tabular}{|c|c|c|c|c|}\hline
Observable & Scenario IV & Scenario VII & Scenario IX & Measurement \\\hline
$R_K^{[1.1, 6]}$ & $0.799 \pm 0.017$ & $0.800 \pm 0.018$ & $0.79 \pm 0.02$ & $0.85 \pm 0.04$ \\\hline
$R_{K^*}^{[0.045,\ 1.1]}$ & $0.870 \pm 0.009$ & $0.871 \pm 0.010$ & $0.870 \pm 0.010$ & $0.65 \pm 0.09$\\\hline
$R_{K^*}^{[1.1,\ 6]}$ & $0.800 \pm 0.018$ & $0.802 \pm 0.019$  & $0.80 \pm 0.02$ & $0.68 \pm 0.10$ \\\hline
$R_D^\ell$ & $0.302 \pm 0.005$ & $0.314 \pm 0.007$ & $0.311 \pm 0.005$ & $0.35 \pm 0.03$ \\\hline
$R_{D^*}^\ell$ & $0.254 \pm 0.004$ & $0.264 \pm 0.004$ & $0.261 \pm 0.004$ & $0.296 \pm 0.016$ \\\hline
$R_{D^*}^\mu$ & $0.261 \pm 0.004$ & $0.272 \pm 0.004$ & $0.269 \pm 0.004$ & $0.31 \pm 0.03$ \\\hline
\end{tabular}
\caption{Values of the $\RKp$ and $\RDp$ observables in the Scenarios with best pulls.}
\label{tab:observables}
\end{table}
The results for the \RKp and \RDp observables in the scenarios with
best pulls, Scenarios IV, VII and IX, are presented in
Table~\ref{tab:observables}. Fig.~\ref{im:RK} shows the results 
for the central value and $1\,\sigma$ uncertainty of these two
observables in the three scenarios, compared to the SM prediction 
(yellow area) and experimental measurements (green area). These 
three scenarios have similar fits for the Wilson
coefficients $\Clq^e$ and $\Clq^\mu$ and therefore reproduce the 
experimental value of $R_K^{[1.1,6]}$ and reduce the tension in 
$R_{K^*}^{[1.1,6]}$. The main difference between Scenarios IV, VII and
IX is the fit for $\Clq^\tau$: Scenario IV has no NP contribution in 
the $\tau$ sector and consequently predicts SM-like \RDp ratios;  
Scenario VII has a large contribution to $\Clq^\tau$ and is able 
to produce a prediction for $R_D^\ell$ compatible with the experimental 
results, and significantly improve the predictions for $R_{D^*}^\ell$ 
and $R_{D^*}^\mu$; Scenario IX has an intermediate value of $\Clq^\tau$, 
and consequently its predictions for the \RDp ratios are not as good as in Scenario VII. 

In addition to the observables included in our global fits, it is also
possible to constrain the NP contributions to Wilson coefficients using 
high-energy collision data from LHC. In particular, it is known that 
high $p_T$ tails in proton-proton collisions producing tau leptons
provide bounds that are competitive to those from the $\RDp$ ratios in 
$B$-physics~\cite{Faroughy:2016osc}. Reference~\cite{Faroughy:2016osc} finds the bound 
$|C_{\ell q(3)}^\tau|/\Lambda^2 < 2.6\ \mathrm{TeV}^{-2}$ by recasting the
$pp\to\tau^+\tau^-$ searches in ATLAS 13 TeV with 3.2 fb${}^{-1}$. 
The constraint $|C_{VL}^\tau| < 0.32$ is established~\cite{Greljo:2018tzh}
for mono-$\tau$ searches $pp\to \tau X + E_T\!\!\!\!\!\!\!/\ \ $, by combining 
the results from ATLAS with 36.1 fb${}^{-1}$ and CMS with 
35.9 fb${}^{-1}$, at 13 TeV. In order to compare this constraint in the 
WET with our fits in the SMEFT basis, we use the matching condition in
Eq.~\eqref{eq:running}, obtaining that 
$|C_{\ell q(3)}^\tau|< 5.35$. Therefore, we can conclude that all the results of
our fits are clearly compatible with the limits imposed by the high-$p_T$ phenomena.

\subsection{Scenario VII}\label{sec:VII}

Since the Scenario VII is the more general one and we found that the
prediction of the $\RDp$ and $\RKp$ observables is improved in this case,
we discuss in this section this Scenario in more detail.

The $\chi^2$ of the fit can be expressed as a series expansion around
its minimum~\cite{Capdevila:2018jhy} 
\begin{equation}
  \chi^2(\Clq^k) = \chi^2_\mathrm{fit} +  \delta\Clq^i\, \mathbb{H}_{ij}\, \delta\Clq^j
  + \mathcal{O}((\delta \Clq^k)^3)\ ,
\end{equation}
where $\delta \Clq^i = \Clq^i - {\Clq^i}|_\mathrm{BF}$ represent the deviation with respect to the
best fit (BF) and $\mathbb{H}$
is the Hessian matrix evaluated at the best fit. In Scenario VII, the
Hessian matrix takes the value: 
\begin{equation}
\mathbb{H} = \begin{pmatrix}
1.07524\times10^4 &  -1.11206\times10^4 & 4.75434\\
-1.11206\times10^4 &  1.33503\times10^4 & -8.39386\\
4.75434 &  -8.39386 &  26.9816\\
\end{pmatrix}\,.
\end{equation}

Within the quadratic approximation, the points with constant
$\Delta\chi^2$ (e.g. all the points that are $1\ \sigma$ away 
from the best fit) are located in the surface of an ellipsoid. 
The length and orientation of the ellipsoid can be found with 
the Singular Value Decomposition (SVD) of the Hessian,

\begin{equation}
\mathbb{H} = U \Sigma U^T\,,
\end{equation}
where $U$ is an orthogonal matrix whose columns are the directions of
the principal axes, and $\Sigma$ is a diagonal matrix. The lengths of
the semi-axes for a given value of $\Delta \chi^2$ are
\begin{equation}
a_j = \sqrt{\frac{\Delta \chi^2}{\Sigma_{jj}}}\,.
\end{equation}

In a $\chi^2$ distribution with $3$ degrees of freedom, the $1\ \sigma$
confidence region corresponds to $\Delta \chi^2 = 3.527$. The lengths of
the semi-axes, in decreasing order, are 
\begin{equation}
a_1 = 0.362,\qquad a_2 =0.064,\qquad a_3=0.0123\,.
\end{equation}

The orientation of the axes, also in decreasing order of $a_i$, is given by
\begin{equation}
U = \begin{pmatrix}
-0.001560 &  -0.7470 & -0.6648 \\
-0.01932 & -0.6648 &  0.7470 \\
-0.9999 &  -0.002450 &  -4.0615\times10^{-4}
\end{pmatrix}\,.
\end{equation}

The first direction (i.e. the one that is less constrained by the fit)
corresponds to the $\tau$ coefficient, while the second and third
directions contain an equal mix of the two other Wilson coefficients
that can be given as  
\begin{align}
\label{eq:Cellipsoid}
  C_1 &\sim -\Clq^\tau, \\
  C_2 &\sim \frac{1}{\sqrt{2}}(-\Clq^e - \Clq^\mu), \qquad C_3 	\sim \frac{1}{\sqrt{2}}(-\Clq^e + \Clq^\mu),\nonumber\\
\Clq^e &\sim \frac{1}{\sqrt{2}}(-C_2 - C_3), \qquad \Clq^\mu 	\sim \frac{1}{\sqrt{2}}(-C_2 + C_3)\,.
\end{align}

The physical interpretation of the orientation of the axes is pretty
clear from our analysis. We conclude that the NP effects in $\tau$ (axis 1)
are mostly uncorrelated with those
of the lighter leptons, and NP in $e$ and $\mu$ is better described as
a combination of LFU effects (axis 2) and LFUV effects (axis 3). The
coordinates of the best fit point (see Scenario VII in Table~\ref{tab:Fits}),
expressed in terms of this basis, are $C_1 = 0.336$, $C_2 = 0.043$ and $C_3 = 0.321$.
The value obtained for the coordinate 3 implies a simultaneous decrease
in the electronic part and an increase in the muonic part to describe
the LFUV observables; and the value of coordinate 2 so close to 0
indicates that the LFU processes are not changed with respect to the
SM. 

The extrema of the $1 \sigma$ confidence ellipsoid are located at
\begin{equation}
\left.\Clq^i\right|_{js} = \left.\Clq^i\right|_\mathrm{BF} + s\, U_{ik} A_{kj}\,,
\label{eq:axis123}
\end{equation}
where $j=1,2,3$, $s=\pm1$ and $A_{kj} = a_j \delta_{kj}$. 

Other notable points on the ellipsoid are found moving from the best fit
point in the direction of the $\Clq^e$, $\Clq^\mu$ and $\Clq^\tau$ axes
($j=e, \mu, \tau$). The distance from the best fit to the ellipsoid when
changing only one Wilson coefficient $j$ is 
\begin{equation}
a_j = \sqrt{\frac{\Delta \chi^2}{\mathbb{H}_{jj}}},\qquad\qquad j=e,\,\mu,\,\tau\,,
\end{equation}
and the points of the ellipsoid obtained when only one Wilson coefficient is changed from its BF value are given by
\begin{equation}
\left.\Clq^i\right|_{js} = \left.\Clq^i\right|_\mathrm{BF} + s\, a_j \delta^i_{j},\qquad\qquad j=e,\,\mu,\,\tau\,.
\label{eq:axisemutau}
\end{equation}

Finally, the points on the $1\,\sigma$ ellipsoid closest and furthest in
the direction connecting the best fit point and the SM benchmark are
given by 
\begin{equation}
\left.\Clq^i\right|_{\mathrm{SM}\,s} = \left.\Clq^i\right|_\mathrm{BF}(1
+ s\, a_\mathrm{SM})\,, 
\end{equation}
where the distance $a_\mathrm{SM}$ is given by
\begin{equation}
a_\mathrm{SM} = \sqrt{\frac{\Delta \chi^2}{\Clq^i|_\mathrm{BF}\
    \mathbb{H}_{ij}\ \Clq^j|_\mathrm{BF}  }  }\,. 
\end{equation}

The Wilson coefficients at these points of the ellipse, from the
corresponding best fit point to the ellipsoid, at $1 \sigma$ confidence
level, are given in Table~\ref{tab:extrema}. 
\begin{table}
\centering
\begin{tabular}{|c|c|c|c|c|c|}\hline
$j$ & $s$ & $\Clq^e$ & $\Clq^\mu$ & $\Clq^\tau$ & $\Delta \chi^2$\\\hline
BF & & -0.246 & 0.211 & -0.336 & \\\hline
1 & $+$ & -0.246 & 0.21 & -0.698 & 3.47\\\hline
1 & $-$ & -0.245 & 0.211 & 0.0251 & 3.65\\\hline
2 & $+$ & -0.294 & 0.168 & -0.336 & 3.25\\\hline
2 & $-$ & -0.198 & 0.253 & -0.337 & 3.22\\\hline
3 & $+$ & -0.323 & 0.297 & -0.336 & 3.84\\\hline
3 & $-$ & -0.168 & 0.124 & -0.336 & 3.57\\\hline
$e$ & $+$ & -0.159 & 0.211 & -0.336 & 3.62\\\hline
$e$ & $-$ & -0.332 & 0.211 & -0.336 & 3.74\\\hline
$\mu$ & $+$ & -0.246 & 0.292 & -0.336 & 3.71\\\hline
$\mu$ & $-$ & -0.246 & 0.129 & -0.336 & 3.62\\\hline
$\tau$ & $+$ & -0.246 & 0.211 & 0.0251 & 3.66\\\hline
$\tau$ & $-$ & -0.246 & 0.211 & -0.698 & 3.47\\\hline
SM & $+$ & -0.330 & 0.283 & -0.452 & 3.88 \\\hline
SM & $-$ & -0.161 & 0.138 & -0.221 & 3.69 \\\hline
\end{tabular}
\caption{Values of the Wilson coefficients at some points located at $1
  \sigma$ confidence ellipsoid around the best fit point in Scenario
  VII.} 
\label{tab:extrema}
\end{table}

\begin{figure}
\begin{center}
\resizebox{0.8\textwidth}{!}{\input{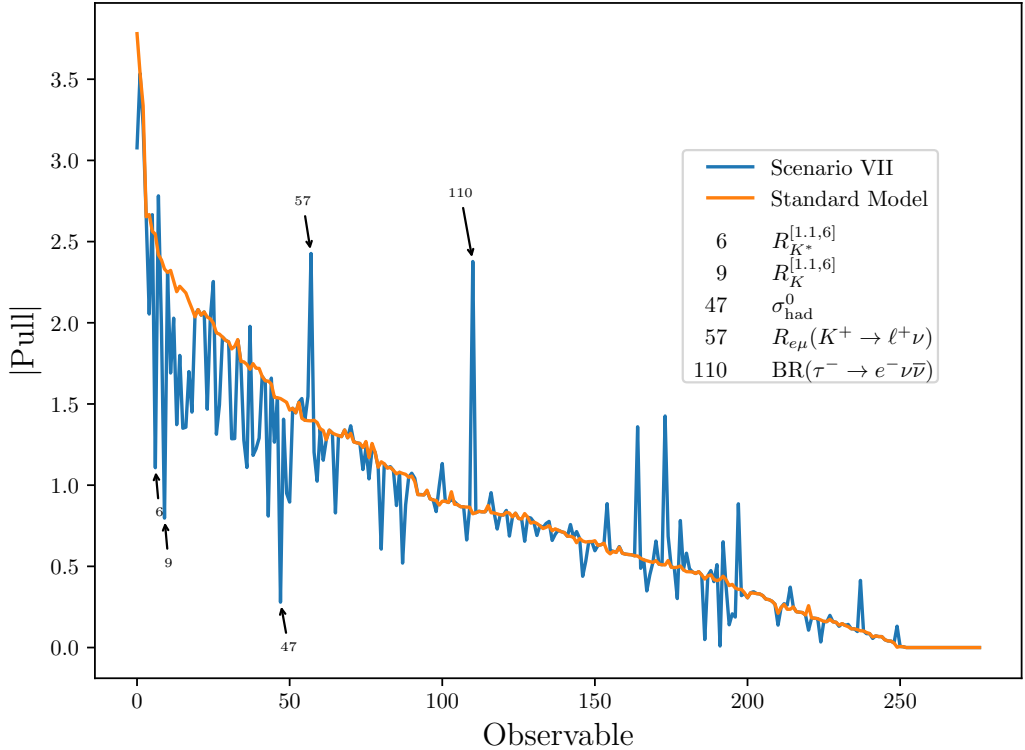}}
\caption{Pulls in the Standard Model (orange) and Scenario VII (blue) of the observables included in the global fit.}
\label{im:pulls}
\end{center}
\end{figure}

The pull for a single observable is defined as
\begin{equation}
\mathrm{Pull}_\mathcal{O}(\Clq) = \frac{\mathcal{O}(\Clq) - \mathcal{O}_\mathrm{exp}}{\sqrt{\sigma_\mathrm{exp}^2 + \sigma_\mathrm{th}^2(\Clq) }}\,.
\end{equation}
The theoretical uncertainties of the observables in general depend on
the SMEFT coefficients. The package
\texttt{smelli}~\cite{Aebischer:2018iyb} treats the theoretical
uncertainties in two different ways: in some observables, such as the
EW precision tests, the theoretical uncertainty is considered
negligible compared to the experimental uncertainty. In other cases,
like the $B$-physics observables, both theoretical and experimental
uncertainties are included, but they are assumed to be Gaussian.
The list of observables that contribute to the global fit with their
prediction in Scenario VII as well as the pulls that compare the
predictions against experimental measurements for NP models (NP pull)
and in the SM (SM pull) is presented in Appendix~\ref{app:pulls}.
Notice that the values of these pulls are approximate, as they do not
take in account the correlation between observables. 

Fig.~\ref{im:pulls} shows the pull of the observables included in the
global fit for Scenario VII with respect to their experimental
measurement (blue line), compared to the same pull in the SM (orange line). 
It is clear that, for most of the observables, the NP either improves their prediction,
especially for $R_K$, $R_{K^*}$ (observables 9 and 6 in the table presented in
Appendix~\ref{app:pulls}) and the hadronic $Z$ cross-section
$\sigma^0_\mathrm{had}$ (observable 47), as well as the differential
branching ratios of $B\to K^{(*)} \mu\mu$ in several low-$q^2$ 
bins;~\footnote{See for example observables 12, 15, 17, 23, 27, 31, 32,
  35, 36, 38, 40, 49, 50,  65, 80, 87 in Appendix~\ref{app:pulls}} or
leave the prediction mostly unchanged. Nevertheless, in the case of the
following observables, the pull of the Scenario VII is significantly
worse than that of the SM: 
\begin{align}
R_{e \mu}(K^+ \to \ell^+ \nu) = \frac{\mathrm{BR}(K^+ \to e^+
  \nu)}{\mathrm{BR}(K^+ \to \mu^+ \nu)}\ , &\qquad\qquad
\mathrm{BR}(\tau^- \to e^- \nu \bar{\nu})\ , \nonumber\\
R_{D^*}^{\mu/e}=R_{\mu e}(B \to D^* \ell^+ \nu) = \frac{\mathrm{BR}(B
  \to D^* \mu^+ \nu)}{\mathrm{BR}(B \to D^* e^+ \nu)}\ , &\qquad\qquad \mathrm{BR}(\pi^+ \to e^+ \nu )\ .
\label{eq:worseobs}
\end{align}
Those observables correspond to observables 57, 110, 164 and 173,
respectively, in the table given in Appendix~\ref{app:pulls}. Scenario
VII also produces worse predictions of the \RKp ratios in the low-recoil
bins $q^2 > 14~\mathrm{GeV}^2$ (observables 154 and 197 in
Appendix~\ref{app:pulls}). 

In order to identify which operators are constraining the fit in each
direction, we use the difference of the pulls, defined as~\cite{Capdevila:2018jhy}: 
\begin{equation}
\delta'_{js}(\mathcal{O} ) = \mathrm{Pull}_\mathcal{O} (\Clq|_\mathrm{BF}) - \mathrm{Pull}_\mathcal{O} (\Clq|_{js})\,,
\end{equation}
where $js$ represents the direction of the corresponding axis, as
described in Eqs.~(\ref{eq:axis123}) and (\ref{eq:axisemutau}). 
The observables with the largest values of the square of $\delta'$ for
each extreme of the ellipse are shown in Table~\ref{tab:deltas}. We can
see that the values of both $\Clq^e$ and $\Clq^\mu$ are constrained
mostly by electroweak precision tests: the $W-$mass,
the electron asymmetry in the $Z$ decay $A_e$,
the forward-backward asymmetry $A_\mathrm{FB}(Z \to \bar{b}b)$ and the
Z-decay width $\Gamma_Z$ (corresponding to observable 39-$m_W$,
observable 14-$A_e$, observable 7-$A_\mathrm{FB}$ and
observable 180-$\Gamma_Z$ as presented in Appendix~\ref{app:pulls}),
as well as by the $\RKp$
data (observable 9 is $R_K^{[1.1, 6]}$). The coefficient $\Clq^\tau$ is
constrained by $\tau$ observables: the
branching ratios of $\tau \to e \bar{\nu}\nu$ and $\tau \to \mu
\bar{\nu}\nu$ (observables 110 and 25) and the ratios $R_{D^*}^\ell$ and $R_{D^*}^\mu$
(observables 4 and 18). This result is in agreement
with~\cite{Feruglio:2016gvd}.
\begin{table}
\centering
\begin{tabular}{|c|c|c||c|c|c||c|c|c|}\hline
  \multicolumn{3}{|c||}{$\Clq^e$} & \multicolumn{3}{|c||}{$\Clq^\mu$}
  & \multicolumn{3}{|c|}{$\Clq^\tau$}\\\hline
No. & Observable & $\delta'^2$ & No. & Observable & $\delta'^2$ & No. & Observable & $\delta'^2$ \\\hline
39  & $m_W$ & 1.513 & 39 & $m_W$ & 1.312 & 110 & $\mathrm{BR}(\tau^-\to e^-\nu\overline{\nu})$ & 1.060 \\\hline
14 & $A_e$ & 0.418 & 9 & $R_K^{[1.1,6]}$ & 0.391 & 25 & $\mathrm{BR}(\tau^-\to \mu^-\nu\overline{\nu})$ & 1.026 \\\hline
9 & $R_K^{[1.1,6]}$ & 0.348 & 14 & $A_e$ & 0.290 & 47 & $\sigma^0_\mathrm{had}$ & 0.566 \\\hline
7 & $A_\mathrm{FB}$ & 0.306 & 180 & $\Gamma_Z$ & 0.232 & 4 & $R^\ell_{D^*}$ & 0.487 \\\hline
180 & $\Gamma_Z$ & 0.268 & 7 & $A_\mathrm{FB}$ & 0.213 & 18 & $R^\mu_{D^*}$ & 0.179 \\\hline
\end{tabular}\\[0.3em]
\begin{tabular}{|c|c|c||c|c|c||c|c|c|}\hline
\multicolumn{3}{|c||}{Axis 1} & \multicolumn{3}{|c||}{Axis 2} & \multicolumn{3}{|c|}{Axis 3}   \\\hline
No. & Observable & $\delta'^2$ & No. & Observable & $\delta'^2$ & No. & Observable & $\delta'^2$ \\\hline
110 & $\mathrm{BR}(\tau^-\to e^-\nu\overline{\nu})$ & 1.055& 39 & $m_W$ & 1.64 & 9 & $R_K^{[1.1,6]}$ & 1.419 \\\hline
25 & $\mathrm{BR}(\tau^-\to \mu^-\nu\overline{\nu})$ & 1.021  & 14 & $A_e$ & 0.410 & 173 & $\mathrm{BR}(\pi^+\to e\nu) $ & 0.475 \\\hline
47 & $\sigma^0_\mathrm{had}$ & 0.570 & 7 & $A_\mathrm{FB} $ & 0.301 & 164 & $R_{D^*}^{\mu/e}$ & 0.440 \\\hline
4 & $R^\ell_{D^*}$ & 0.495 & 180 & $\Gamma_Z$ & 0.291 & 6 & $R_{K^*}^{[1.1,6]}$ & 0.276 \\\hline
18 & $R^\mu_{D^*}$ & 0.182 & 100 & $A_\tau$ & 0.079 & 57 & $R_{e\mu}(K^+\to\ell^+ \nu)$ & 0.135 \\\hline
\end{tabular}\\[0.3em]
\begin{tabular}{|c|c|c|}\hline
 \multicolumn{3}{|c|}{SM direction}  \\\hline
No. & Observable & $\delta'^2$ \\\hline
9 & $R_K^{[1.1,6]}$ & 1.278\\\hline
173 & $\mathrm{BR}(\pi^+\to e^+ \nu)$ & 0.435\\\hline
164 & $R_{D^*}^{\mu/e}$ & 0.401\\\hline
110 & $\mathrm{BR}(\tau^-\to e^-\nu\overline{\nu})$ & 0.287\\\hline
6 & $R_{K^*}^{[1.1,6]}$ & 0.249\\\hline
\end{tabular}
\caption{Observables with the largest difference of pulls between the
  best fit and the extreme of the $1 \sigma$ confidence
  ellipsoid. The number of the observables corresponds to the ones given in
  Appendix~\ref{app:pulls}.} 
\label{tab:deltas}
\end{table}

If we focus instead on the principal directions of the uncertainty
ellipsoid, the picture is clearer: axis 1 is still dominated by $\tau$
observables.  Axis 2 is constrained by the electroweak precision
tests: $m_W$, $\Gamma_Z$, $A_\mathrm{FB}(Z \to \bar{b}b)$  and the
leptonic asymmetries $A_e$ and $A_\tau$ (observables 14 and 100).
Axis 3, on the other hand, is constrained by observables sensitive to lepton
universality violations in the $e$-$\mu$ sector: $\RKp$ observables
(observable 9 is $R_K^{[1.1, 6]}$ and observable 6 is  $R_{K^*}^{[1.1,
  6]}$), but also the equivalent $\RDp$ observable $R_{\mu e}(B \to D^*
\ell^+ \nu)$ (observable 164), the leptonic branching ratio of $\pi^+
\to e^+ \nu$ (observable 173) and the ratio $R_{e \mu}(K^+ \to \ell^+
\nu)$ (observable 78), all of them defined
in~\eqref{eq:worseobs}. Indeed,
this separation between electroweak and $\RKp$ observables
is already visible in Fig.~\ref{im:globalfits}a: the allowed
region by EW precision observables (green) is focused around
a constant value of $\Clq^e + \Clq^\mu$ approximately aligned with axis
3, while the allowed region of the $\RKp$ observables (red) is
focused around a constant value of $\Clq^e - \Clq^\mu$,
approximately aligned with axis 2.

\begin{figure}
\begin{center}
\begin{tabular}{cc}
\includegraphics[width=0.45\textwidth]{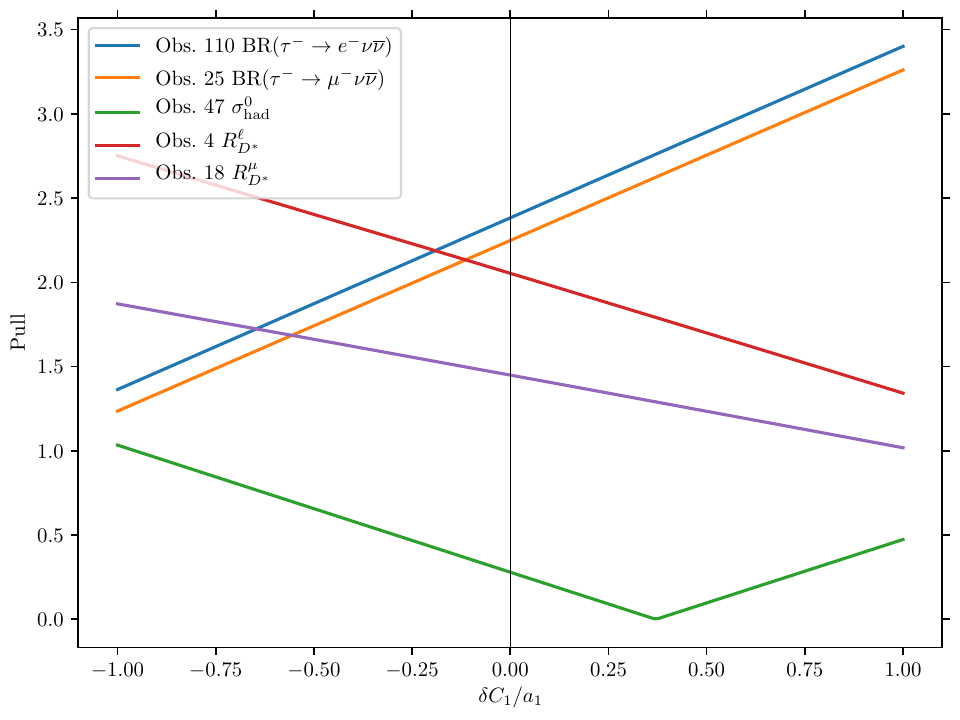} & \includegraphics[width=0.45\textwidth]{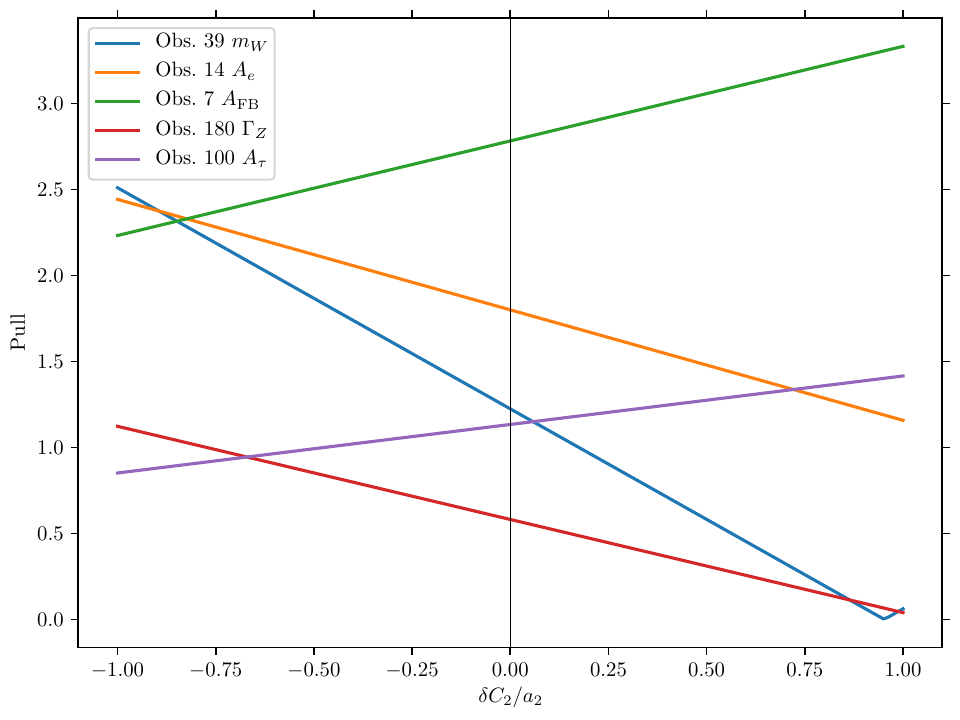}\\
(a)&(b)\\\\
\includegraphics[width=0.45\textwidth]{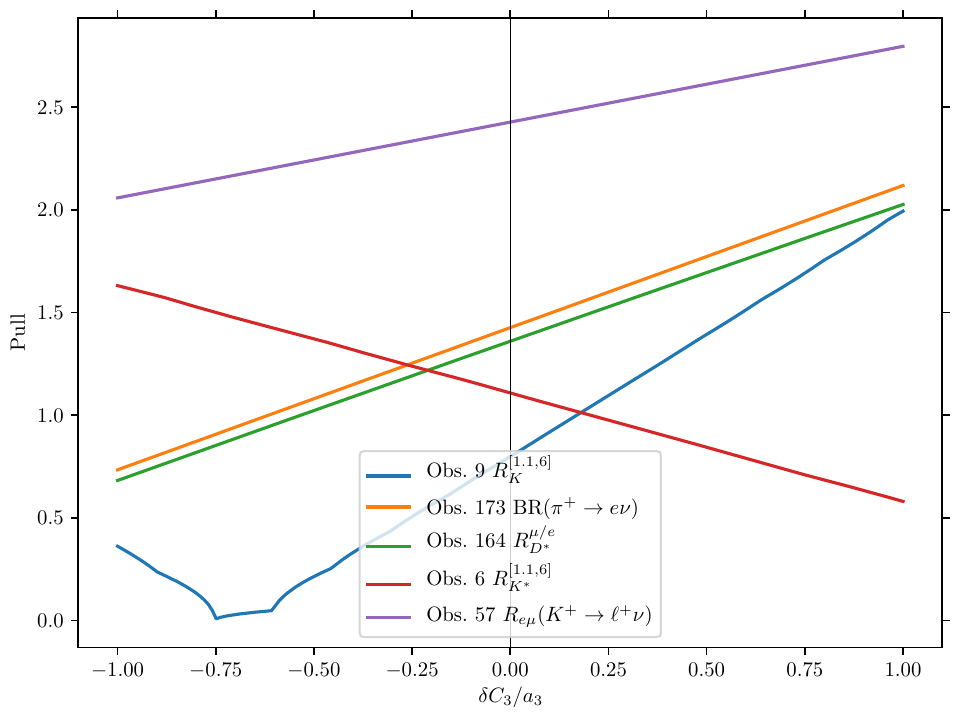} &\includegraphics[width=0.45\textwidth]{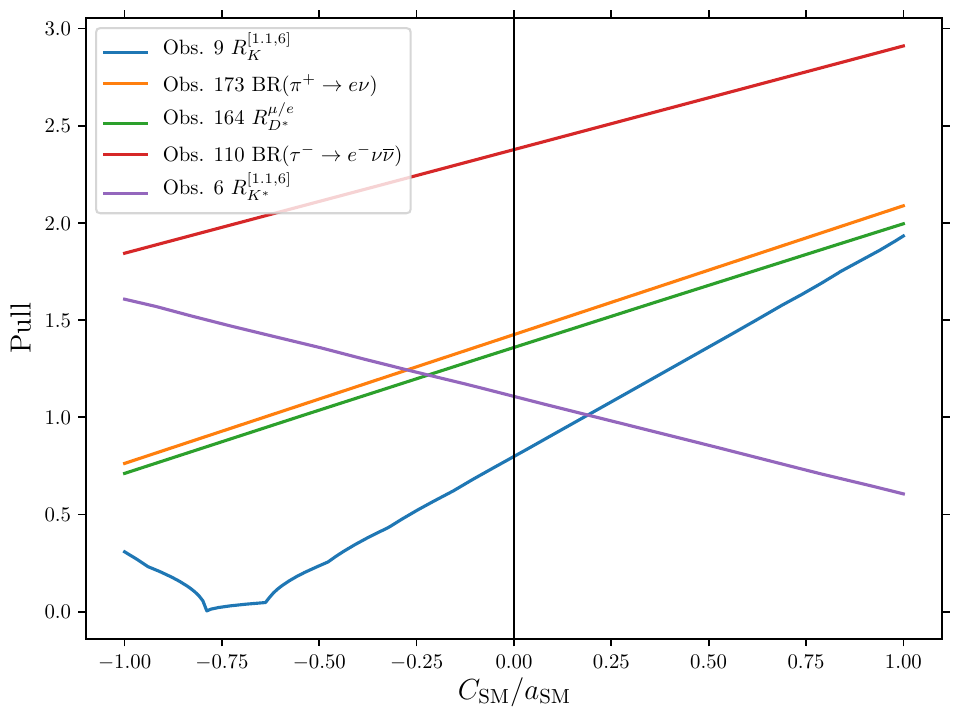}\\
(c)&(d)
\end{tabular}
\caption{Evolution of the pull of the observables in Table \ref{tab:deltas} along each axis of the ellipsoid (a)-(c) and the SM direction (d). }
\label{im:evoax}
\end{center}
\end{figure}

Fig.~\ref{im:evoax} represents the evolution of these observables along
the axes of the ellipsoid (see Eq.~(\ref{eq:Cellipsoid}) for
definitions of $C_1, C_2, C_3$). 
In the case of the first axis, $\delta
C_1/a_1 = -1$ corresponds to a suppression of NP in the $\tau$ sector,
which is preferred by the $\tau$ decays, while $\delta C_1/a_1 = 1$ is an
increase in $\tau$ effects with respect to the best fit, that
accommodates better the $\RDp$ anomalies, as was previously pointed 
out in~\cite{Capdevila:2017iqn}. In the second axis, the observables 
$A_\mathrm{FB}$ and $A_\tau$ favour a decrease in the flavour universal 
NP contribution, while $A_e$, $m_W$ and  $\Gamma_Z$  prefer lower
contributions, with the two latter observables attaining their
experimental values. In the case of axis 3,  $\delta
C_3/a_3 = 1$ favours NP effects in muons (it increases $\Clq^\mu \sim -
C_9$, and a deficit of muons needs a negative $C_9$) while $\delta
C_3/a_3 = -1$ favours NP effects in electrons: $R_K$ prefers a smaller
contribution to the muonic part while $R_{K^*}$ prefers a larger
contribution. This is consistent with Fig.~\ref{im:RK}, where the
prediction for $R_K$ is below its central experimental value and the
prediction of $R_{K^*}$ is above its experimental value. The other
LFUV observables also prefer smaller muonic NP effects.

The last columns of Table~\ref{tab:deltas} and Fig.~\ref{im:evoax}d 
show the observables that constrain the fit along
the direction connecting the SM and best fit point, that is in the
points with Wilson coefficients of the form $\Clq^i =
\Clq^i|_\mathrm{BF}\,(1 + \delta C_\mathrm{SM})$. We observe that this
direction is determined mostly by the LFUV observables $R_K^{[1.1,6]}$, 
$R_{K^*}^{[1.1,6]}$, $R_{D^*}^{\mu/e}$, the $\tau$ decay 
$\mathrm{BR}(\tau^-\to e^-\nu \overline{\nu})$ and $\mathrm{BR}(\pi^+\to
e^+\nu)$. These are the observables whose pulls change the most when
comparing the best fit and SM, and therefore the ones more relevant to 
constrain the fit. In particular, the fit is optimal for
$R_K^{[1.1,6]}$, a larger deviation would be needed for
$R_{K^*}^{[1.1,6]}$, while $R_{D^*}^{\mu/e}$, 
$\mathrm{BR}(\tau^-\to e^-\nu \overline{\nu})$ and $\mathrm{BR}(\pi^+\to
e^+\nu)$ would be better explained with a SM-like arrangement.

\section{Connection to leptoquark models}\label{sec:leptoq}

For completeness, we discuss in this section the phenomenological
    implications of our assumptions in the leptoquark models, concretely
    in the vector leptoquark model.
The vector leptoquark $U_1 = (\mathbf{\bar{3}},\mathbf{1})_{2/3}$
couples to left-handed and right-handed fermions as 
\begin{equation}
\mathcal{L}  = x^{ij}_L \bar{q}_i \gamma_\mu U_1^\mu \ell_j + x_R^{ij} \bar{d}_{Ri} \gamma_\mu U_1^\mu e_{Rj} + \mathrm{h.c.},
\end{equation}
where $d_R$ and $e_R$ are the $d$-type quark and charged lepton $SU(2)$ singlets, and $x_L$ and $x_R$ are the matrices of couplings of the $U_1$ leptoquark to left-handed and right-handed fermions respectively.

When matched with the SMEFT at the
scale $\Lambda$, an $U_1$ leptoquark with mass $M_U$ contributes to the following
Wilson coefficients~\cite{delAguila:2010mx}: 
\begin{align}
C_{\ell q(1)}^{ijkl} = C_{\ell q(3)}^{ijkl} &= \frac{- \Lambda^2}{2 M_U^2} x_L^{li} x_L^{kj*}\,, \nonumber \\
C_{ed}^{ijkl} = -\frac{1}{2} C_{ledq}^{ijkl} &= \frac{-\Lambda^2}{M_U^2} x_L^{li} x_R^{kj*}\,.
\end{align}

If we only allow couplings to the left-handed fermions, the leptoquark only affects $\Clq$, as we used in our assumptions. The coefficients used in Scenarios I through IX in terms of the leptoquarks couplings are
\begin{equation}
\Clq^e = -\frac{\Lambda^2}{2M_U^2}|x_L^{be}|^2 \qquad \Clq^\mu = -\frac{\Lambda^2}{2M_U^2}|x_L^{b\mu}|^2 \qquad \Clq^\tau = -\frac{\Lambda^2}{2M_U^2}|x_L^{b\tau}|^2\,,
\end{equation}
which obviously must be negative real numbers. 

According to the results of the fits in Table~\ref{tab:Fits}, the
scenarios that include NP contributions in the electronic or tau sectors
show preference for negative values of $\Clq^e$ and $\Clq^\tau$, and
thus can be described by a $U_1$ leptoquark. On the contrary, all the
fits to scenarios affecting the muon coupling show clear preference for 
positive values of the Wilson coefficient $\Clq^\mu$. In consequence,
with our assumptions, the leptoquark $U_1$ cannot describe the
anomalies in the muon sector and therefore, does not play an important 
role in describing the LFUV, as shown by the fact that the scenarios
with a greater pull from the SM, Scenarios IV, VII and IX, are not
compatible. These results confirm previous results  which have 
shown that the $U_1$ leptoquark models cannot describe the anomalies on
\RKp and can only address the deficit in this observable when it has
both couplings to $b \mu$ and $s \mu$ (see, for example~\cite{Hiller:2014yaa}). 

Other leptoquark models do not retain the $\Clqo = \Clqt$ condition
\cite{delAguila:2010mx,Bhattacharya:2016mcc}, and therefore produce
large contributions to the $B \to K^{(*)}\nu \bar{\nu}$ decays. That is
the case of the scalar $S_3 = (\mathbf{\bar{3}}, \mathbf{3})_{1/3}$,
that predicts $\Clqo = 3 \Clqt$, and the vector $U_3 =
(\mathbf{\bar{3}}, \mathbf{3})_{2/3}$, where $\Clqo = - 3 \Clqt$. The
scalar $S_1 = (\mathbf{\bar{3}}, \mathbf{1})_{1/3}$ is even less suited,
as it predicts $\Clqo = - \Clqt$, which would result in no NP
contributing to $b\to s \ell^+ \ell^-$ at all. New vector bosons $W'$
and $Z'$ would also be in conflict with the $B \to K^{(*)}\nu \bar{\nu}$
decays, as they predict $\Clqo = 0$ while \Clqt has a nonzero value.

\section{Conclusions}
\label{sec:conclu}

Several measurements of $B$ meson decays performed in the recent years
indicate a possible violation of Lepton Universality that may represent an 
indirect signal of New Physics. In this work, we provide an
analysis of the effects of the global fits to the Wilson
coefficients assuming a model independent effective Hamiltonian approach
and including a discussion of the consequences of our assumptions on the analysis in
leptoquark models. The global fit includes $b\to s \mu\mu$ observables (including the
Lepton Flavour Universality ratios $\RKp$, the angular observables $P_5'$ and the branching ratio of
$B_s \to \mu\mu$), as well as the $\RDp$, $b \to s \nu \bar{\nu}$ and electroweak precision
observables ($W$ and $Z$ decay widths and branching ratios to leptons). 

We consider different scenarios for the phenomenological analysis such
that New Physics is present in one, two or three of the Wilson
coefficients at a time (Table~\ref{tab:Fits}), with the choice of the effective operators
motivated by a $U(2)^3$ symmetry between light quarks. Our results are 
relevant for model-independent analysis, clarifying  which combinations
of the Wilson coefficients are constrained by the data. For all
scenarios, we compare the results of the global fit with respect to both the SM and the more
general and descriptive scenario: the best fit point of the three
independent Wilson coefficients scenario in which New Physics modifies 
each of the operators independently. 

We conclude that, when New Physics contributes to only one
lepton flavour operator at a time, the largest pull from the Standard
Model prediction, almost $3\ \sigma$ (Table~\ref{tab:Fits}), appears when the coupling to electrons
is added independently, corresponding to our Scenario I. 
In those scenarios in which New Physics is present in two of the Wilson 
coefficients simultaneously, the best fit corresponds to the case of 
Scenario IV, where the contributions to $\Clq^e$ and $\Clq^\mu$ are
favoured with a pull of $4.97\ \sigma$ with respect to the SM (Table~\ref{tab:Fits}). In this 
case, we confirm that the \RKp and \RDp constrain the linear combination 
$\Clq^e -\Clq^\mu$; while the LFU-conserving electroweak precision observables 
constrain $\Clq^e + \Clq^\mu$.

If we focus on the more general and descriptive scenario of three independent Wilson
coefficients, we found that the prediction of the $\RDp$ and $\RKp$ observables is
improved in the scenario in which the three $\Clq$ operators receive independent NP
contributions: Scenario VII. In this case, the pull from the Standard Model is
$4.97\ \sigma$ (Table~\ref{tab:Fits}) and the predictions for the \RKp observables are very similar to
the case of Scenario IV.
A better fit to $\RDp$ observables, and specially to $R_D^\ell$, is obtained in this scenario.
We have also analysed which observables constrain
the fit in each direction using the difference of their pulls: the
values of both $\Clq^e$ and $\Clq^\mu$ are constrained mostly by electroweak precision tests.
A clear separation between electroweak and LFU observables is
established, with electroweak precision observables focused around
a constant value of $\Clq^e + \Clq^\mu$, while the allowed region of the $\RKp$ observables is
focused around a constant value of $\Clq^e - \Clq^\mu$ requiring a large 
violation of Lepton Flavour Universality. From our analysis, we also
conclude that the more relevant observables in the global fit are the 
LFUV observables $R_K^{[1.1,6]}$, $R_{K^*}^{[1.1,6]}$, $R_{D^*}^{\mu/e}$  
and the branching ratio of the $\tau$ decay $\mathrm{BR}(\tau^-\to e^-\nu
\overline{\nu})$, given that these observables exhibit the larger change in their pulls along
the direction connecting the SM and best fit point, that is $\Clq^i =
\Clq^i|_\mathrm{BF}\,(1 + \delta C_\mathrm{SM})$.

Scenario IX (Table~\ref{tab:Fits}) represents a much more restricted
scenario with only one free Wilson coefficient; nevertheless, it
provides a good fit to experimental data, with a
pull of $5.55\ \sigma$ with respect to the SM, and it is compatible with
Scenario VII at $0.41\ \sigma$; therefore, it provides a similar description to
experimental data with less free parameters. 

Summarising, Scenario VII (three independent Wilson coefficients) is the
favoured one for explaining the tension between SM predictions and $B$-physics
anomalies, with Scenario IX providing a similar fit goodness with a smaller set
of free parameters.

Finally, we compare our setting to the $U_1$ leptoquark model. We conclude that, with our assumptions,
this model cannot describe the anomalies in the muon sector and, therefore, does not play an
important role in describing the LFUV. Other leptoquark models do not
contribute to the effective operators that we consider in this work. 

\section*{Acknowledgements}

The work of J.~A. and S.~P. is partially supported by Spanish grants 
MINECO/FEDER grant FPA2015-65745-P, PGC2018-095328-B-I00 
(FEDER/Agencia estatal de investigaci{\'o}n) and DGIID-DGA No. 2015-E24/2.
J.~A. is also supported by the 
Departamento de Innovaci\'on, Investigaci\'on y Universidad of Arag\'on
government, Grant No. DIIU-DGA. 
J.G. has been supported by MICIN under projects PID2019-105614GB-C22 and 
CEX2019-000918-M of ICCUB (\textit{Unit of Excellence Mar{\'\i}a de
  Maeztu 2020-2023}) and AGAUR (2017SGR754). 

\appendix
\section{Pulls of the observables in Scenario VII}
\label{app:pulls}

This table contains all observables that contribute to the global fit, as
well as their prediction in Scenario VII and their pull in both Scenario
VII (NP pull) and SM (SM pull). Predictions for dimensionful observables 
are expressed in the corresponding power of GeV (for example, $\Delta
M_s$ in GeV and $\sigma^0_\mathrm{had}$ in $\mathrm{GeV}^{-2}$). 
The notation $\langle\cdot\cdot\cdot\rangle$ means that the observable
is binned in the invariant mass-squared of the di-lepton system $q^2$, 
with the endpoints of the bin in $\mathrm{GeV}^2$ given in the
superscript. Accordingly, the notation $\frac{\langle
  \mathrm{BR}\rangle}{\mathrm{BR}}$
denotes a binned branching ratio normalised to the total branching ratio.
Observables are ordered according to their SM
pull, and color-coded according to the difference between the Scenario
VII and SM pulls: green observables have a better pull in Scenario VII, 
red observables have a better pull in the SM and white observables have 
a similar pull in both cases.\\
Notice that not all observables are affected by NP in our scenario. 
However, the inclusion of these observables does not alter the value of
the $\Delta \chi^2$, since their prediction and uncertainty are unchanged
from the SM, and the statistical significance of the fit remains unchanged.\\ 

{\scriptsize \input{pullsVII_21}}

\end{document}

%% file: RKplot_21.pgf
\begingroup%
\makeatletter%
\begin{pgfpicture}%
\pgfpathrectangle{\pgfpointorigin}{\pgfqpoint{6.400000in}{4.800000in}}%
\pgfusepath{use as bounding box, clip}%
\begin{pgfscope}%
\pgfsetbuttcap%
\pgfsetmiterjoin%
\definecolor{currentfill}{rgb}{1.000000,1.000000,1.000000}%
\pgfsetfillcolor{currentfill}%
\pgfsetlinewidth{0.000000pt}%
\definecolor{currentstroke}{rgb}{1.000000,1.000000,1.000000}%
\pgfsetstrokecolor{currentstroke}%
\pgfsetdash{}{0pt}%
\pgfpathmoveto{\pgfqpoint{0.000000in}{0.000000in}}%
\pgfpathlineto{\pgfqpoint{6.400000in}{0.000000in}}%
\pgfpathlineto{\pgfqpoint{6.400000in}{4.800000in}}%
\pgfpathlineto{\pgfqpoint{0.000000in}{4.800000in}}%
\pgfpathclose%
\pgfusepath{fill}%
\end{pgfscope}%
\begin{pgfscope}%
\pgfsetbuttcap%
\pgfsetmiterjoin%
\definecolor{currentfill}{rgb}{1.000000,1.000000,1.000000}%
\pgfsetfillcolor{currentfill}%
\pgfsetlinewidth{0.000000pt}%
\definecolor{currentstroke}{rgb}{0.000000,0.000000,0.000000}%
\pgfsetstrokecolor{currentstroke}%
\pgfsetstrokeopacity{0.000000}%
\pgfsetdash{}{0pt}%
\pgfpathmoveto{\pgfqpoint{0.800000in}{0.528000in}}%
\pgfpathlineto{\pgfqpoint{5.760000in}{0.528000in}}%
\pgfpathlineto{\pgfqpoint{5.760000in}{4.224000in}}%
\pgfpathlineto{\pgfqpoint{0.800000in}{4.224000in}}%
\pgfpathclose%
\pgfusepath{fill}%
\end{pgfscope}%
\begin{pgfscope}%
\pgfpathrectangle{\pgfqpoint{0.800000in}{0.528000in}}{\pgfqpoint{4.960000in}{3.696000in}}%
\pgfusepath{clip}%
\pgfsetbuttcap%
\pgfsetmiterjoin%
\definecolor{currentfill}{rgb}{1.000000,0.647059,0.000000}%
\pgfsetfillcolor{currentfill}%
\pgfsetfillopacity{0.700000}%
\pgfsetlinewidth{1.003750pt}%
\definecolor{currentstroke}{rgb}{1.000000,0.647059,0.000000}%
\pgfsetstrokecolor{currentstroke}%
\pgfsetstrokeopacity{0.700000}%
\pgfsetdash{}{0pt}%
\pgfpathmoveto{\pgfqpoint{0.800000in}{3.986968in}}%
\pgfpathlineto{\pgfqpoint{2.453333in}{3.986968in}}%
\pgfpathlineto{\pgfqpoint{2.453333in}{4.056000in}}%
\pgfpathlineto{\pgfqpoint{0.800000in}{4.056000in}}%
\pgfpathclose%
\pgfusepath{stroke,fill}%
\end{pgfscope}%
\begin{pgfscope}%
\pgfpathrectangle{\pgfqpoint{0.800000in}{0.528000in}}{\pgfqpoint{4.960000in}{3.696000in}}%
\pgfusepath{clip}%
\pgfsetbuttcap%
\pgfsetmiterjoin%
\definecolor{currentfill}{rgb}{0.000000,0.501961,0.000000}%
\pgfsetfillcolor{currentfill}%
\pgfsetfillopacity{0.700000}%
\pgfsetlinewidth{1.003750pt}%
\definecolor{currentstroke}{rgb}{0.000000,0.501961,0.000000}%
\pgfsetstrokecolor{currentstroke}%
\pgfsetstrokeopacity{0.700000}%
\pgfsetdash{}{0pt}%
\pgfpathmoveto{\pgfqpoint{0.800000in}{2.433842in}}%
\pgfpathlineto{\pgfqpoint{2.453333in}{2.433842in}}%
\pgfpathlineto{\pgfqpoint{2.453333in}{3.335030in}}%
\pgfpathlineto{\pgfqpoint{0.800000in}{3.335030in}}%
\pgfpathclose%
\pgfusepath{stroke,fill}%
\end{pgfscope}%
\begin{pgfscope}%
\pgfpathrectangle{\pgfqpoint{0.800000in}{0.528000in}}{\pgfqpoint{4.960000in}{3.696000in}}%
\pgfusepath{clip}%
\pgfsetbuttcap%
\pgfsetmiterjoin%
\definecolor{currentfill}{rgb}{0.000000,0.501961,0.000000}%
\pgfsetfillcolor{currentfill}%
\pgfsetfillopacity{0.700000}%
\pgfsetlinewidth{1.003750pt}%
\definecolor{currentstroke}{rgb}{0.000000,0.501961,0.000000}%
\pgfsetstrokecolor{currentstroke}%
\pgfsetstrokeopacity{0.700000}%
\pgfsetdash{}{0pt}%
\pgfpathmoveto{\pgfqpoint{2.453333in}{0.696000in}}%
\pgfpathlineto{\pgfqpoint{4.106667in}{0.696000in}}%
\pgfpathlineto{\pgfqpoint{4.106667in}{2.094754in}}%
\pgfpathlineto{\pgfqpoint{2.453333in}{2.094754in}}%
\pgfpathclose%
\pgfusepath{stroke,fill}%
\end{pgfscope}%
\begin{pgfscope}%
\pgfpathrectangle{\pgfqpoint{0.800000in}{0.528000in}}{\pgfqpoint{4.960000in}{3.696000in}}%
\pgfusepath{clip}%
\pgfsetbuttcap%
\pgfsetmiterjoin%
\definecolor{currentfill}{rgb}{1.000000,0.647059,0.000000}%
\pgfsetfillcolor{currentfill}%
\pgfsetfillopacity{0.700000}%
\pgfsetlinewidth{1.003750pt}%
\definecolor{currentstroke}{rgb}{1.000000,0.647059,0.000000}%
\pgfsetstrokecolor{currentstroke}%
\pgfsetstrokeopacity{0.700000}%
\pgfsetdash{}{0pt}%
\pgfpathmoveto{\pgfqpoint{2.453333in}{3.389456in}}%
\pgfpathlineto{\pgfqpoint{4.106667in}{3.389456in}}%
\pgfpathlineto{\pgfqpoint{4.106667in}{3.521372in}}%
\pgfpathlineto{\pgfqpoint{2.453333in}{3.521372in}}%
\pgfpathclose%
\pgfusepath{stroke,fill}%
\end{pgfscope}%
\begin{pgfscope}%
\pgfpathrectangle{\pgfqpoint{0.800000in}{0.528000in}}{\pgfqpoint{4.960000in}{3.696000in}}%
\pgfusepath{clip}%
\pgfsetbuttcap%
\pgfsetmiterjoin%
\definecolor{currentfill}{rgb}{0.000000,0.501961,0.000000}%
\pgfsetfillcolor{currentfill}%
\pgfsetfillopacity{0.700000}%
\pgfsetlinewidth{1.003750pt}%
\definecolor{currentstroke}{rgb}{0.000000,0.501961,0.000000}%
\pgfsetstrokecolor{currentstroke}%
\pgfsetstrokeopacity{0.700000}%
\pgfsetdash{}{0pt}%
\pgfpathmoveto{\pgfqpoint{4.106667in}{0.858280in}}%
\pgfpathlineto{\pgfqpoint{5.760000in}{0.858280in}}%
\pgfpathlineto{\pgfqpoint{5.760000in}{2.367050in}}%
\pgfpathlineto{\pgfqpoint{4.106667in}{2.367050in}}%
\pgfpathclose%
\pgfusepath{stroke,fill}%
\end{pgfscope}%
\begin{pgfscope}%
\pgfpathrectangle{\pgfqpoint{0.800000in}{0.528000in}}{\pgfqpoint{4.960000in}{3.696000in}}%
\pgfusepath{clip}%
\pgfsetbuttcap%
\pgfsetmiterjoin%
\definecolor{currentfill}{rgb}{1.000000,0.647059,0.000000}%
\pgfsetfillcolor{currentfill}%
\pgfsetfillopacity{0.700000}%
\pgfsetlinewidth{1.003750pt}%
\definecolor{currentstroke}{rgb}{1.000000,0.647059,0.000000}%
\pgfsetstrokecolor{currentstroke}%
\pgfsetstrokeopacity{0.700000}%
\pgfsetdash{}{0pt}%
\pgfpathmoveto{\pgfqpoint{4.106667in}{3.933929in}}%
\pgfpathlineto{\pgfqpoint{5.760000in}{3.933929in}}%
\pgfpathlineto{\pgfqpoint{5.760000in}{4.043614in}}%
\pgfpathlineto{\pgfqpoint{4.106667in}{4.043614in}}%
\pgfpathclose%
\pgfusepath{stroke,fill}%
\end{pgfscope}%
\begin{pgfscope}%
\pgfsetbuttcap%
\pgfsetroundjoin%
\definecolor{currentfill}{rgb}{0.000000,0.000000,0.000000}%
\pgfsetfillcolor{currentfill}%
\pgfsetlinewidth{0.803000pt}%
\definecolor{currentstroke}{rgb}{0.000000,0.000000,0.000000}%
\pgfsetstrokecolor{currentstroke}%
\pgfsetdash{}{0pt}%
\pgfsys@defobject{currentmarker}{\pgfqpoint{0.000000in}{-0.048611in}}{\pgfqpoint{0.000000in}{0.000000in}}{%
\pgfpathmoveto{\pgfqpoint{0.000000in}{0.000000in}}%
\pgfpathlineto{\pgfqpoint{0.000000in}{-0.048611in}}%
\pgfusepath{stroke,fill}%
}%
\begin{pgfscope}%
\pgfsys@transformshift{1.626667in}{0.528000in}%
\pgfsys@useobject{currentmarker}{}%
\end{pgfscope}%
\end{pgfscope}%
\begin{pgfscope}%
\definecolor{textcolor}{rgb}{0.000000,0.000000,0.000000}%
\pgfsetstrokecolor{textcolor}%
\pgfsetfillcolor{textcolor}%
\pgftext[x=1.626667in,y=0.430778in,,top]{\color{textcolor}\rmfamily\fontsize{16.000000}{19.200000}\selectfont \(\displaystyle R_K^{[1.1,6]}\)}%
\end{pgfscope}%
\begin{pgfscope}%
\pgfsetbuttcap%
\pgfsetroundjoin%
\definecolor{currentfill}{rgb}{0.000000,0.000000,0.000000}%
\pgfsetfillcolor{currentfill}%
\pgfsetlinewidth{0.803000pt}%
\definecolor{currentstroke}{rgb}{0.000000,0.000000,0.000000}%
\pgfsetstrokecolor{currentstroke}%
\pgfsetdash{}{0pt}%
\pgfsys@defobject{currentmarker}{\pgfqpoint{0.000000in}{-0.048611in}}{\pgfqpoint{0.000000in}{0.000000in}}{%
\pgfpathmoveto{\pgfqpoint{0.000000in}{0.000000in}}%
\pgfpathlineto{\pgfqpoint{0.000000in}{-0.048611in}}%
\pgfusepath{stroke,fill}%
}%
\begin{pgfscope}%
\pgfsys@transformshift{3.280000in}{0.528000in}%
\pgfsys@useobject{currentmarker}{}%
\end{pgfscope}%
\end{pgfscope}%
\begin{pgfscope}%
\definecolor{textcolor}{rgb}{0.000000,0.000000,0.000000}%
\pgfsetstrokecolor{textcolor}%
\pgfsetfillcolor{textcolor}%
\pgftext[x=3.280000in,y=0.430778in,,top]{\color{textcolor}\rmfamily\fontsize{16.000000}{19.200000}\selectfont \(\displaystyle R_{K^*}^{[0.045, 1.1]}\)}%
\end{pgfscope}%
\begin{pgfscope}%
\pgfsetbuttcap%
\pgfsetroundjoin%
\definecolor{currentfill}{rgb}{0.000000,0.000000,0.000000}%
\pgfsetfillcolor{currentfill}%
\pgfsetlinewidth{0.803000pt}%
\definecolor{currentstroke}{rgb}{0.000000,0.000000,0.000000}%
\pgfsetstrokecolor{currentstroke}%
\pgfsetdash{}{0pt}%
\pgfsys@defobject{currentmarker}{\pgfqpoint{0.000000in}{-0.048611in}}{\pgfqpoint{0.000000in}{0.000000in}}{%
\pgfpathmoveto{\pgfqpoint{0.000000in}{0.000000in}}%
\pgfpathlineto{\pgfqpoint{0.000000in}{-0.048611in}}%
\pgfusepath{stroke,fill}%
}%
\begin{pgfscope}%
\pgfsys@transformshift{4.933333in}{0.528000in}%
\pgfsys@useobject{currentmarker}{}%
\end{pgfscope}%
\end{pgfscope}%
\begin{pgfscope}%
\definecolor{textcolor}{rgb}{0.000000,0.000000,0.000000}%
\pgfsetstrokecolor{textcolor}%
\pgfsetfillcolor{textcolor}%
\pgftext[x=4.933333in,y=0.430778in,,top]{\color{textcolor}\rmfamily\fontsize{16.000000}{19.200000}\selectfont \(\displaystyle R_{K^*}^{[1.1, 6]}\)}%
\end{pgfscope}%
\begin{pgfscope}%
\pgfsetbuttcap%
\pgfsetroundjoin%
\definecolor{currentfill}{rgb}{0.000000,0.000000,0.000000}%
\pgfsetfillcolor{currentfill}%
\pgfsetlinewidth{0.803000pt}%
\definecolor{currentstroke}{rgb}{0.000000,0.000000,0.000000}%
\pgfsetstrokecolor{currentstroke}%
\pgfsetdash{}{0pt}%
\pgfsys@defobject{currentmarker}{\pgfqpoint{-0.048611in}{0.000000in}}{\pgfqpoint{0.000000in}{0.000000in}}{%
\pgfpathmoveto{\pgfqpoint{0.000000in}{0.000000in}}%
\pgfpathlineto{\pgfqpoint{-0.048611in}{0.000000in}}%
\pgfusepath{stroke,fill}%
}%
\begin{pgfscope}%
\pgfsys@transformshift{0.800000in}{0.999147in}%
\pgfsys@useobject{currentmarker}{}%
\end{pgfscope}%
\end{pgfscope}%
\begin{pgfscope}%
\definecolor{textcolor}{rgb}{0.000000,0.000000,0.000000}%
\pgfsetstrokecolor{textcolor}%
\pgfsetfillcolor{textcolor}%
\pgftext[x=0.417364in, y=0.915814in, left, base]{\color{textcolor}\rmfamily\fontsize{16.000000}{19.200000}\selectfont \(\displaystyle 0.6\)}%
\end{pgfscope}%
\begin{pgfscope}%
\pgfsetbuttcap%
\pgfsetroundjoin%
\definecolor{currentfill}{rgb}{0.000000,0.000000,0.000000}%
\pgfsetfillcolor{currentfill}%
\pgfsetlinewidth{0.803000pt}%
\definecolor{currentstroke}{rgb}{0.000000,0.000000,0.000000}%
\pgfsetstrokecolor{currentstroke}%
\pgfsetdash{}{0pt}%
\pgfsys@defobject{currentmarker}{\pgfqpoint{-0.048611in}{0.000000in}}{\pgfqpoint{0.000000in}{0.000000in}}{%
\pgfpathmoveto{\pgfqpoint{0.000000in}{0.000000in}}%
\pgfpathlineto{\pgfqpoint{-0.048611in}{0.000000in}}%
\pgfusepath{stroke,fill}%
}%
\begin{pgfscope}%
\pgfsys@transformshift{0.800000in}{1.753263in}%
\pgfsys@useobject{currentmarker}{}%
\end{pgfscope}%
\end{pgfscope}%
\begin{pgfscope}%
\definecolor{textcolor}{rgb}{0.000000,0.000000,0.000000}%
\pgfsetstrokecolor{textcolor}%
\pgfsetfillcolor{textcolor}%
\pgftext[x=0.417364in, y=1.669929in, left, base]{\color{textcolor}\rmfamily\fontsize{16.000000}{19.200000}\selectfont \(\displaystyle 0.7\)}%
\end{pgfscope}%
\begin{pgfscope}%
\pgfsetbuttcap%
\pgfsetroundjoin%
\definecolor{currentfill}{rgb}{0.000000,0.000000,0.000000}%
\pgfsetfillcolor{currentfill}%
\pgfsetlinewidth{0.803000pt}%
\definecolor{currentstroke}{rgb}{0.000000,0.000000,0.000000}%
\pgfsetstrokecolor{currentstroke}%
\pgfsetdash{}{0pt}%
\pgfsys@defobject{currentmarker}{\pgfqpoint{-0.048611in}{0.000000in}}{\pgfqpoint{0.000000in}{0.000000in}}{%
\pgfpathmoveto{\pgfqpoint{0.000000in}{0.000000in}}%
\pgfpathlineto{\pgfqpoint{-0.048611in}{0.000000in}}%
\pgfusepath{stroke,fill}%
}%
\begin{pgfscope}%
\pgfsys@transformshift{0.800000in}{2.507378in}%
\pgfsys@useobject{currentmarker}{}%
\end{pgfscope}%
\end{pgfscope}%
\begin{pgfscope}%
\definecolor{textcolor}{rgb}{0.000000,0.000000,0.000000}%
\pgfsetstrokecolor{textcolor}%
\pgfsetfillcolor{textcolor}%
\pgftext[x=0.417364in, y=2.424045in, left, base]{\color{textcolor}\rmfamily\fontsize{16.000000}{19.200000}\selectfont \(\displaystyle 0.8\)}%
\end{pgfscope}%
\begin{pgfscope}%
\pgfsetbuttcap%
\pgfsetroundjoin%
\definecolor{currentfill}{rgb}{0.000000,0.000000,0.000000}%
\pgfsetfillcolor{currentfill}%
\pgfsetlinewidth{0.803000pt}%
\definecolor{currentstroke}{rgb}{0.000000,0.000000,0.000000}%
\pgfsetstrokecolor{currentstroke}%
\pgfsetdash{}{0pt}%
\pgfsys@defobject{currentmarker}{\pgfqpoint{-0.048611in}{0.000000in}}{\pgfqpoint{0.000000in}{0.000000in}}{%
\pgfpathmoveto{\pgfqpoint{0.000000in}{0.000000in}}%
\pgfpathlineto{\pgfqpoint{-0.048611in}{0.000000in}}%
\pgfusepath{stroke,fill}%
}%
\begin{pgfscope}%
\pgfsys@transformshift{0.800000in}{3.261493in}%
\pgfsys@useobject{currentmarker}{}%
\end{pgfscope}%
\end{pgfscope}%
\begin{pgfscope}%
\definecolor{textcolor}{rgb}{0.000000,0.000000,0.000000}%
\pgfsetstrokecolor{textcolor}%
\pgfsetfillcolor{textcolor}%
\pgftext[x=0.417364in, y=3.178160in, left, base]{\color{textcolor}\rmfamily\fontsize{16.000000}{19.200000}\selectfont \(\displaystyle 0.9\)}%
\end{pgfscope}%
\begin{pgfscope}%
\pgfsetbuttcap%
\pgfsetroundjoin%
\definecolor{currentfill}{rgb}{0.000000,0.000000,0.000000}%
\pgfsetfillcolor{currentfill}%
\pgfsetlinewidth{0.803000pt}%
\definecolor{currentstroke}{rgb}{0.000000,0.000000,0.000000}%
\pgfsetstrokecolor{currentstroke}%
\pgfsetdash{}{0pt}%
\pgfsys@defobject{currentmarker}{\pgfqpoint{-0.048611in}{0.000000in}}{\pgfqpoint{0.000000in}{0.000000in}}{%
\pgfpathmoveto{\pgfqpoint{0.000000in}{0.000000in}}%
\pgfpathlineto{\pgfqpoint{-0.048611in}{0.000000in}}%
\pgfusepath{stroke,fill}%
}%
\begin{pgfscope}%
\pgfsys@transformshift{0.800000in}{4.015609in}%
\pgfsys@useobject{currentmarker}{}%
\end{pgfscope}%
\end{pgfscope}%
\begin{pgfscope}%
\definecolor{textcolor}{rgb}{0.000000,0.000000,0.000000}%
\pgfsetstrokecolor{textcolor}%
\pgfsetfillcolor{textcolor}%
\pgftext[x=0.417364in, y=3.932276in, left, base]{\color{textcolor}\rmfamily\fontsize{16.000000}{19.200000}\selectfont \(\displaystyle 1.0\)}%
\end{pgfscope}%
\begin{pgfscope}%
\pgfpathrectangle{\pgfqpoint{0.800000in}{0.528000in}}{\pgfqpoint{4.960000in}{3.696000in}}%
\pgfusepath{clip}%
\pgfsetbuttcap%
\pgfsetroundjoin%
\pgfsetlinewidth{1.505625pt}%
\definecolor{currentstroke}{rgb}{0.000000,0.000000,1.000000}%
\pgfsetstrokecolor{currentstroke}%
\pgfsetdash{}{0pt}%
\pgfpathmoveto{\pgfqpoint{1.213333in}{2.362084in}}%
\pgfpathlineto{\pgfqpoint{1.213333in}{2.619996in}}%
\pgfusepath{stroke}%
\end{pgfscope}%
\begin{pgfscope}%
\pgfpathrectangle{\pgfqpoint{0.800000in}{0.528000in}}{\pgfqpoint{4.960000in}{3.696000in}}%
\pgfusepath{clip}%
\pgfsetbuttcap%
\pgfsetroundjoin%
\pgfsetlinewidth{1.505625pt}%
\definecolor{currentstroke}{rgb}{0.000000,0.000000,1.000000}%
\pgfsetstrokecolor{currentstroke}%
\pgfsetdash{}{0pt}%
\pgfpathmoveto{\pgfqpoint{1.626667in}{2.366878in}}%
\pgfpathlineto{\pgfqpoint{1.626667in}{2.647809in}}%
\pgfusepath{stroke}%
\end{pgfscope}%
\begin{pgfscope}%
\pgfpathrectangle{\pgfqpoint{0.800000in}{0.528000in}}{\pgfqpoint{4.960000in}{3.696000in}}%
\pgfusepath{clip}%
\pgfsetbuttcap%
\pgfsetroundjoin%
\pgfsetlinewidth{1.505625pt}%
\definecolor{currentstroke}{rgb}{0.000000,0.000000,1.000000}%
\pgfsetstrokecolor{currentstroke}%
\pgfsetdash{}{0pt}%
\pgfpathmoveto{\pgfqpoint{2.040000in}{2.321566in}}%
\pgfpathlineto{\pgfqpoint{2.040000in}{2.638342in}}%
\pgfusepath{stroke}%
\end{pgfscope}%
\begin{pgfscope}%
\pgfpathrectangle{\pgfqpoint{0.800000in}{0.528000in}}{\pgfqpoint{4.960000in}{3.696000in}}%
\pgfusepath{clip}%
\pgfsetbuttcap%
\pgfsetroundjoin%
\pgfsetlinewidth{1.505625pt}%
\definecolor{currentstroke}{rgb}{0.000000,0.000000,1.000000}%
\pgfsetstrokecolor{currentstroke}%
\pgfsetdash{}{0pt}%
\pgfpathmoveto{\pgfqpoint{2.866667in}{2.967475in}}%
\pgfpathlineto{\pgfqpoint{2.866667in}{3.109717in}}%
\pgfusepath{stroke}%
\end{pgfscope}%
\begin{pgfscope}%
\pgfpathrectangle{\pgfqpoint{0.800000in}{0.528000in}}{\pgfqpoint{4.960000in}{3.696000in}}%
\pgfusepath{clip}%
\pgfsetbuttcap%
\pgfsetroundjoin%
\pgfsetlinewidth{1.505625pt}%
\definecolor{currentstroke}{rgb}{0.000000,0.000000,1.000000}%
\pgfsetstrokecolor{currentstroke}%
\pgfsetdash{}{0pt}%
\pgfpathmoveto{\pgfqpoint{3.280000in}{2.966898in}}%
\pgfpathlineto{\pgfqpoint{3.280000in}{3.119768in}}%
\pgfusepath{stroke}%
\end{pgfscope}%
\begin{pgfscope}%
\pgfpathrectangle{\pgfqpoint{0.800000in}{0.528000in}}{\pgfqpoint{4.960000in}{3.696000in}}%
\pgfusepath{clip}%
\pgfsetbuttcap%
\pgfsetroundjoin%
\pgfsetlinewidth{1.505625pt}%
\definecolor{currentstroke}{rgb}{0.000000,0.000000,1.000000}%
\pgfsetstrokecolor{currentstroke}%
\pgfsetdash{}{0pt}%
\pgfpathmoveto{\pgfqpoint{3.693333in}{2.958999in}}%
\pgfpathlineto{\pgfqpoint{3.693333in}{3.115872in}}%
\pgfusepath{stroke}%
\end{pgfscope}%
\begin{pgfscope}%
\pgfpathrectangle{\pgfqpoint{0.800000in}{0.528000in}}{\pgfqpoint{4.960000in}{3.696000in}}%
\pgfusepath{clip}%
\pgfsetbuttcap%
\pgfsetroundjoin%
\pgfsetlinewidth{1.505625pt}%
\definecolor{currentstroke}{rgb}{0.000000,0.000000,1.000000}%
\pgfsetstrokecolor{currentstroke}%
\pgfsetdash{}{0pt}%
\pgfpathmoveto{\pgfqpoint{4.520000in}{2.368161in}}%
\pgfpathlineto{\pgfqpoint{4.520000in}{2.643220in}}%
\pgfusepath{stroke}%
\end{pgfscope}%
\begin{pgfscope}%
\pgfpathrectangle{\pgfqpoint{0.800000in}{0.528000in}}{\pgfqpoint{4.960000in}{3.696000in}}%
\pgfusepath{clip}%
\pgfsetbuttcap%
\pgfsetroundjoin%
\pgfsetlinewidth{1.505625pt}%
\definecolor{currentstroke}{rgb}{0.000000,0.000000,1.000000}%
\pgfsetstrokecolor{currentstroke}%
\pgfsetdash{}{0pt}%
\pgfpathmoveto{\pgfqpoint{4.933333in}{2.379191in}}%
\pgfpathlineto{\pgfqpoint{4.933333in}{2.664008in}}%
\pgfusepath{stroke}%
\end{pgfscope}%
\begin{pgfscope}%
\pgfpathrectangle{\pgfqpoint{0.800000in}{0.528000in}}{\pgfqpoint{4.960000in}{3.696000in}}%
\pgfusepath{clip}%
\pgfsetbuttcap%
\pgfsetroundjoin%
\pgfsetlinewidth{1.505625pt}%
\definecolor{currentstroke}{rgb}{0.000000,0.000000,1.000000}%
\pgfsetstrokecolor{currentstroke}%
\pgfsetdash{}{0pt}%
\pgfpathmoveto{\pgfqpoint{5.346667in}{2.336603in}}%
\pgfpathlineto{\pgfqpoint{5.346667in}{2.656080in}}%
\pgfusepath{stroke}%
\end{pgfscope}%
\begin{pgfscope}%
\pgfpathrectangle{\pgfqpoint{0.800000in}{0.528000in}}{\pgfqpoint{4.960000in}{3.696000in}}%
\pgfusepath{clip}%
\pgfsetrectcap%
\pgfsetroundjoin%
\pgfsetlinewidth{1.505625pt}%
\definecolor{currentstroke}{rgb}{0.000000,0.000000,1.000000}%
\pgfsetstrokecolor{currentstroke}%
\pgfsetdash{}{0pt}%
\pgfpathmoveto{\pgfqpoint{1.213333in}{2.491040in}}%
\pgfusepath{stroke}%
\end{pgfscope}%
\begin{pgfscope}%
\pgfpathrectangle{\pgfqpoint{0.800000in}{0.528000in}}{\pgfqpoint{4.960000in}{3.696000in}}%
\pgfusepath{clip}%
\pgfsetbuttcap%
\pgfsetroundjoin%
\definecolor{currentfill}{rgb}{0.000000,0.000000,1.000000}%
\pgfsetfillcolor{currentfill}%
\pgfsetlinewidth{1.003750pt}%
\definecolor{currentstroke}{rgb}{0.000000,0.000000,1.000000}%
\pgfsetstrokecolor{currentstroke}%
\pgfsetdash{}{0pt}%
\pgfsys@defobject{currentmarker}{\pgfqpoint{-0.041667in}{-0.041667in}}{\pgfqpoint{0.041667in}{0.041667in}}{%
\pgfpathmoveto{\pgfqpoint{0.000000in}{-0.041667in}}%
\pgfpathcurveto{\pgfqpoint{0.011050in}{-0.041667in}}{\pgfqpoint{0.021649in}{-0.037276in}}{\pgfqpoint{0.029463in}{-0.029463in}}%
\pgfpathcurveto{\pgfqpoint{0.037276in}{-0.021649in}}{\pgfqpoint{0.041667in}{-0.011050in}}{\pgfqpoint{0.041667in}{0.000000in}}%
\pgfpathcurveto{\pgfqpoint{0.041667in}{0.011050in}}{\pgfqpoint{0.037276in}{0.021649in}}{\pgfqpoint{0.029463in}{0.029463in}}%
\pgfpathcurveto{\pgfqpoint{0.021649in}{0.037276in}}{\pgfqpoint{0.011050in}{0.041667in}}{\pgfqpoint{0.000000in}{0.041667in}}%
\pgfpathcurveto{\pgfqpoint{-0.011050in}{0.041667in}}{\pgfqpoint{-0.021649in}{0.037276in}}{\pgfqpoint{-0.029463in}{0.029463in}}%
\pgfpathcurveto{\pgfqpoint{-0.037276in}{0.021649in}}{\pgfqpoint{-0.041667in}{0.011050in}}{\pgfqpoint{-0.041667in}{0.000000in}}%
\pgfpathcurveto{\pgfqpoint{-0.041667in}{-0.011050in}}{\pgfqpoint{-0.037276in}{-0.021649in}}{\pgfqpoint{-0.029463in}{-0.029463in}}%
\pgfpathcurveto{\pgfqpoint{-0.021649in}{-0.037276in}}{\pgfqpoint{-0.011050in}{-0.041667in}}{\pgfqpoint{0.000000in}{-0.041667in}}%
\pgfpathclose%
\pgfusepath{stroke,fill}%
}%
\begin{pgfscope}%
\pgfsys@transformshift{1.213333in}{2.491040in}%
\pgfsys@useobject{currentmarker}{}%
\end{pgfscope}%
\end{pgfscope}%
\begin{pgfscope}%
\pgfpathrectangle{\pgfqpoint{0.800000in}{0.528000in}}{\pgfqpoint{4.960000in}{3.696000in}}%
\pgfusepath{clip}%
\pgfsetrectcap%
\pgfsetroundjoin%
\pgfsetlinewidth{1.505625pt}%
\definecolor{currentstroke}{rgb}{0.000000,0.000000,1.000000}%
\pgfsetstrokecolor{currentstroke}%
\pgfsetdash{}{0pt}%
\pgfpathmoveto{\pgfqpoint{1.626667in}{2.507344in}}%
\pgfusepath{stroke}%
\end{pgfscope}%
\begin{pgfscope}%
\pgfpathrectangle{\pgfqpoint{0.800000in}{0.528000in}}{\pgfqpoint{4.960000in}{3.696000in}}%
\pgfusepath{clip}%
\pgfsetbuttcap%
\pgfsetmiterjoin%
\definecolor{currentfill}{rgb}{0.000000,0.000000,1.000000}%
\pgfsetfillcolor{currentfill}%
\pgfsetlinewidth{1.003750pt}%
\definecolor{currentstroke}{rgb}{0.000000,0.000000,1.000000}%
\pgfsetstrokecolor{currentstroke}%
\pgfsetdash{}{0pt}%
\pgfsys@defobject{currentmarker}{\pgfqpoint{-0.041667in}{-0.041667in}}{\pgfqpoint{0.041667in}{0.041667in}}{%
\pgfpathmoveto{\pgfqpoint{0.000000in}{0.041667in}}%
\pgfpathlineto{\pgfqpoint{-0.041667in}{-0.041667in}}%
\pgfpathlineto{\pgfqpoint{0.041667in}{-0.041667in}}%
\pgfpathclose%
\pgfusepath{stroke,fill}%
}%
\begin{pgfscope}%
\pgfsys@transformshift{1.626667in}{2.507344in}%
\pgfsys@useobject{currentmarker}{}%
\end{pgfscope}%
\end{pgfscope}%
\begin{pgfscope}%
\pgfpathrectangle{\pgfqpoint{0.800000in}{0.528000in}}{\pgfqpoint{4.960000in}{3.696000in}}%
\pgfusepath{clip}%
\pgfsetrectcap%
\pgfsetroundjoin%
\pgfsetlinewidth{1.505625pt}%
\definecolor{currentstroke}{rgb}{0.000000,0.000000,1.000000}%
\pgfsetstrokecolor{currentstroke}%
\pgfsetdash{}{0pt}%
\pgfpathmoveto{\pgfqpoint{2.040000in}{2.479954in}}%
\pgfusepath{stroke}%
\end{pgfscope}%
\begin{pgfscope}%
\pgfpathrectangle{\pgfqpoint{0.800000in}{0.528000in}}{\pgfqpoint{4.960000in}{3.696000in}}%
\pgfusepath{clip}%
\pgfsetbuttcap%
\pgfsetmiterjoin%
\definecolor{currentfill}{rgb}{0.000000,0.000000,1.000000}%
\pgfsetfillcolor{currentfill}%
\pgfsetlinewidth{1.003750pt}%
\definecolor{currentstroke}{rgb}{0.000000,0.000000,1.000000}%
\pgfsetstrokecolor{currentstroke}%
\pgfsetdash{}{0pt}%
\pgfsys@defobject{currentmarker}{\pgfqpoint{-0.041667in}{-0.041667in}}{\pgfqpoint{0.041667in}{0.041667in}}{%
\pgfpathmoveto{\pgfqpoint{-0.041667in}{-0.041667in}}%
\pgfpathlineto{\pgfqpoint{0.041667in}{-0.041667in}}%
\pgfpathlineto{\pgfqpoint{0.041667in}{0.041667in}}%
\pgfpathlineto{\pgfqpoint{-0.041667in}{0.041667in}}%
\pgfpathclose%
\pgfusepath{stroke,fill}%
}%
\begin{pgfscope}%
\pgfsys@transformshift{2.040000in}{2.479954in}%
\pgfsys@useobject{currentmarker}{}%
\end{pgfscope}%
\end{pgfscope}%
\begin{pgfscope}%
\pgfpathrectangle{\pgfqpoint{0.800000in}{0.528000in}}{\pgfqpoint{4.960000in}{3.696000in}}%
\pgfusepath{clip}%
\pgfsetrectcap%
\pgfsetroundjoin%
\pgfsetlinewidth{1.505625pt}%
\definecolor{currentstroke}{rgb}{0.000000,0.000000,1.000000}%
\pgfsetstrokecolor{currentstroke}%
\pgfsetdash{}{0pt}%
\pgfpathmoveto{\pgfqpoint{2.866667in}{3.038596in}}%
\pgfusepath{stroke}%
\end{pgfscope}%
\begin{pgfscope}%
\pgfpathrectangle{\pgfqpoint{0.800000in}{0.528000in}}{\pgfqpoint{4.960000in}{3.696000in}}%
\pgfusepath{clip}%
\pgfsetbuttcap%
\pgfsetroundjoin%
\definecolor{currentfill}{rgb}{0.000000,0.000000,1.000000}%
\pgfsetfillcolor{currentfill}%
\pgfsetlinewidth{1.003750pt}%
\definecolor{currentstroke}{rgb}{0.000000,0.000000,1.000000}%
\pgfsetstrokecolor{currentstroke}%
\pgfsetdash{}{0pt}%
\pgfsys@defobject{currentmarker}{\pgfqpoint{-0.041667in}{-0.041667in}}{\pgfqpoint{0.041667in}{0.041667in}}{%
\pgfpathmoveto{\pgfqpoint{0.000000in}{-0.041667in}}%
\pgfpathcurveto{\pgfqpoint{0.011050in}{-0.041667in}}{\pgfqpoint{0.021649in}{-0.037276in}}{\pgfqpoint{0.029463in}{-0.029463in}}%
\pgfpathcurveto{\pgfqpoint{0.037276in}{-0.021649in}}{\pgfqpoint{0.041667in}{-0.011050in}}{\pgfqpoint{0.041667in}{0.000000in}}%
\pgfpathcurveto{\pgfqpoint{0.041667in}{0.011050in}}{\pgfqpoint{0.037276in}{0.021649in}}{\pgfqpoint{0.029463in}{0.029463in}}%
\pgfpathcurveto{\pgfqpoint{0.021649in}{0.037276in}}{\pgfqpoint{0.011050in}{0.041667in}}{\pgfqpoint{0.000000in}{0.041667in}}%
\pgfpathcurveto{\pgfqpoint{-0.011050in}{0.041667in}}{\pgfqpoint{-0.021649in}{0.037276in}}{\pgfqpoint{-0.029463in}{0.029463in}}%
\pgfpathcurveto{\pgfqpoint{-0.037276in}{0.021649in}}{\pgfqpoint{-0.041667in}{0.011050in}}{\pgfqpoint{-0.041667in}{0.000000in}}%
\pgfpathcurveto{\pgfqpoint{-0.041667in}{-0.011050in}}{\pgfqpoint{-0.037276in}{-0.021649in}}{\pgfqpoint{-0.029463in}{-0.029463in}}%
\pgfpathcurveto{\pgfqpoint{-0.021649in}{-0.037276in}}{\pgfqpoint{-0.011050in}{-0.041667in}}{\pgfqpoint{0.000000in}{-0.041667in}}%
\pgfpathclose%
\pgfusepath{stroke,fill}%
}%
\begin{pgfscope}%
\pgfsys@transformshift{2.866667in}{3.038596in}%
\pgfsys@useobject{currentmarker}{}%
\end{pgfscope}%
\end{pgfscope}%
\begin{pgfscope}%
\pgfpathrectangle{\pgfqpoint{0.800000in}{0.528000in}}{\pgfqpoint{4.960000in}{3.696000in}}%
\pgfusepath{clip}%
\pgfsetrectcap%
\pgfsetroundjoin%
\pgfsetlinewidth{1.505625pt}%
\definecolor{currentstroke}{rgb}{0.000000,0.000000,1.000000}%
\pgfsetstrokecolor{currentstroke}%
\pgfsetdash{}{0pt}%
\pgfpathmoveto{\pgfqpoint{3.280000in}{3.043333in}}%
\pgfusepath{stroke}%
\end{pgfscope}%
\begin{pgfscope}%
\pgfpathrectangle{\pgfqpoint{0.800000in}{0.528000in}}{\pgfqpoint{4.960000in}{3.696000in}}%
\pgfusepath{clip}%
\pgfsetbuttcap%
\pgfsetmiterjoin%
\definecolor{currentfill}{rgb}{0.000000,0.000000,1.000000}%
\pgfsetfillcolor{currentfill}%
\pgfsetlinewidth{1.003750pt}%
\definecolor{currentstroke}{rgb}{0.000000,0.000000,1.000000}%
\pgfsetstrokecolor{currentstroke}%
\pgfsetdash{}{0pt}%
\pgfsys@defobject{currentmarker}{\pgfqpoint{-0.041667in}{-0.041667in}}{\pgfqpoint{0.041667in}{0.041667in}}{%
\pgfpathmoveto{\pgfqpoint{0.000000in}{0.041667in}}%
\pgfpathlineto{\pgfqpoint{-0.041667in}{-0.041667in}}%
\pgfpathlineto{\pgfqpoint{0.041667in}{-0.041667in}}%
\pgfpathclose%
\pgfusepath{stroke,fill}%
}%
\begin{pgfscope}%
\pgfsys@transformshift{3.280000in}{3.043333in}%
\pgfsys@useobject{currentmarker}{}%
\end{pgfscope}%
\end{pgfscope}%
\begin{pgfscope}%
\pgfpathrectangle{\pgfqpoint{0.800000in}{0.528000in}}{\pgfqpoint{4.960000in}{3.696000in}}%
\pgfusepath{clip}%
\pgfsetrectcap%
\pgfsetroundjoin%
\pgfsetlinewidth{1.505625pt}%
\definecolor{currentstroke}{rgb}{0.000000,0.000000,1.000000}%
\pgfsetstrokecolor{currentstroke}%
\pgfsetdash{}{0pt}%
\pgfpathmoveto{\pgfqpoint{3.693333in}{3.037436in}}%
\pgfusepath{stroke}%
\end{pgfscope}%
\begin{pgfscope}%
\pgfpathrectangle{\pgfqpoint{0.800000in}{0.528000in}}{\pgfqpoint{4.960000in}{3.696000in}}%
\pgfusepath{clip}%
\pgfsetbuttcap%
\pgfsetmiterjoin%
\definecolor{currentfill}{rgb}{0.000000,0.000000,1.000000}%
\pgfsetfillcolor{currentfill}%
\pgfsetlinewidth{1.003750pt}%
\definecolor{currentstroke}{rgb}{0.000000,0.000000,1.000000}%
\pgfsetstrokecolor{currentstroke}%
\pgfsetdash{}{0pt}%
\pgfsys@defobject{currentmarker}{\pgfqpoint{-0.041667in}{-0.041667in}}{\pgfqpoint{0.041667in}{0.041667in}}{%
\pgfpathmoveto{\pgfqpoint{-0.041667in}{-0.041667in}}%
\pgfpathlineto{\pgfqpoint{0.041667in}{-0.041667in}}%
\pgfpathlineto{\pgfqpoint{0.041667in}{0.041667in}}%
\pgfpathlineto{\pgfqpoint{-0.041667in}{0.041667in}}%
\pgfpathclose%
\pgfusepath{stroke,fill}%
}%
\begin{pgfscope}%
\pgfsys@transformshift{3.693333in}{3.037436in}%
\pgfsys@useobject{currentmarker}{}%
\end{pgfscope}%
\end{pgfscope}%
\begin{pgfscope}%
\pgfpathrectangle{\pgfqpoint{0.800000in}{0.528000in}}{\pgfqpoint{4.960000in}{3.696000in}}%
\pgfusepath{clip}%
\pgfsetrectcap%
\pgfsetroundjoin%
\pgfsetlinewidth{1.505625pt}%
\definecolor{currentstroke}{rgb}{0.000000,0.000000,1.000000}%
\pgfsetstrokecolor{currentstroke}%
\pgfsetdash{}{0pt}%
\pgfpathmoveto{\pgfqpoint{4.520000in}{2.505691in}}%
\pgfusepath{stroke}%
\end{pgfscope}%
\begin{pgfscope}%
\pgfpathrectangle{\pgfqpoint{0.800000in}{0.528000in}}{\pgfqpoint{4.960000in}{3.696000in}}%
\pgfusepath{clip}%
\pgfsetbuttcap%
\pgfsetroundjoin%
\definecolor{currentfill}{rgb}{0.000000,0.000000,1.000000}%
\pgfsetfillcolor{currentfill}%
\pgfsetlinewidth{1.003750pt}%
\definecolor{currentstroke}{rgb}{0.000000,0.000000,1.000000}%
\pgfsetstrokecolor{currentstroke}%
\pgfsetdash{}{0pt}%
\pgfsys@defobject{currentmarker}{\pgfqpoint{-0.041667in}{-0.041667in}}{\pgfqpoint{0.041667in}{0.041667in}}{%
\pgfpathmoveto{\pgfqpoint{0.000000in}{-0.041667in}}%
\pgfpathcurveto{\pgfqpoint{0.011050in}{-0.041667in}}{\pgfqpoint{0.021649in}{-0.037276in}}{\pgfqpoint{0.029463in}{-0.029463in}}%
\pgfpathcurveto{\pgfqpoint{0.037276in}{-0.021649in}}{\pgfqpoint{0.041667in}{-0.011050in}}{\pgfqpoint{0.041667in}{0.000000in}}%
\pgfpathcurveto{\pgfqpoint{0.041667in}{0.011050in}}{\pgfqpoint{0.037276in}{0.021649in}}{\pgfqpoint{0.029463in}{0.029463in}}%
\pgfpathcurveto{\pgfqpoint{0.021649in}{0.037276in}}{\pgfqpoint{0.011050in}{0.041667in}}{\pgfqpoint{0.000000in}{0.041667in}}%
\pgfpathcurveto{\pgfqpoint{-0.011050in}{0.041667in}}{\pgfqpoint{-0.021649in}{0.037276in}}{\pgfqpoint{-0.029463in}{0.029463in}}%
\pgfpathcurveto{\pgfqpoint{-0.037276in}{0.021649in}}{\pgfqpoint{-0.041667in}{0.011050in}}{\pgfqpoint{-0.041667in}{0.000000in}}%
\pgfpathcurveto{\pgfqpoint{-0.041667in}{-0.011050in}}{\pgfqpoint{-0.037276in}{-0.021649in}}{\pgfqpoint{-0.029463in}{-0.029463in}}%
\pgfpathcurveto{\pgfqpoint{-0.021649in}{-0.037276in}}{\pgfqpoint{-0.011050in}{-0.041667in}}{\pgfqpoint{0.000000in}{-0.041667in}}%
\pgfpathclose%
\pgfusepath{stroke,fill}%
}%
\begin{pgfscope}%
\pgfsys@transformshift{4.520000in}{2.505691in}%
\pgfsys@useobject{currentmarker}{}%
\end{pgfscope}%
\end{pgfscope}%
\begin{pgfscope}%
\pgfpathrectangle{\pgfqpoint{0.800000in}{0.528000in}}{\pgfqpoint{4.960000in}{3.696000in}}%
\pgfusepath{clip}%
\pgfsetrectcap%
\pgfsetroundjoin%
\pgfsetlinewidth{1.505625pt}%
\definecolor{currentstroke}{rgb}{0.000000,0.000000,1.000000}%
\pgfsetstrokecolor{currentstroke}%
\pgfsetdash{}{0pt}%
\pgfpathmoveto{\pgfqpoint{4.933333in}{2.521600in}}%
\pgfusepath{stroke}%
\end{pgfscope}%
\begin{pgfscope}%
\pgfpathrectangle{\pgfqpoint{0.800000in}{0.528000in}}{\pgfqpoint{4.960000in}{3.696000in}}%
\pgfusepath{clip}%
\pgfsetbuttcap%
\pgfsetmiterjoin%
\definecolor{currentfill}{rgb}{0.000000,0.000000,1.000000}%
\pgfsetfillcolor{currentfill}%
\pgfsetlinewidth{1.003750pt}%
\definecolor{currentstroke}{rgb}{0.000000,0.000000,1.000000}%
\pgfsetstrokecolor{currentstroke}%
\pgfsetdash{}{0pt}%
\pgfsys@defobject{currentmarker}{\pgfqpoint{-0.041667in}{-0.041667in}}{\pgfqpoint{0.041667in}{0.041667in}}{%
\pgfpathmoveto{\pgfqpoint{0.000000in}{0.041667in}}%
\pgfpathlineto{\pgfqpoint{-0.041667in}{-0.041667in}}%
\pgfpathlineto{\pgfqpoint{0.041667in}{-0.041667in}}%
\pgfpathclose%
\pgfusepath{stroke,fill}%
}%
\begin{pgfscope}%
\pgfsys@transformshift{4.933333in}{2.521600in}%
\pgfsys@useobject{currentmarker}{}%
\end{pgfscope}%
\end{pgfscope}%
\begin{pgfscope}%
\pgfpathrectangle{\pgfqpoint{0.800000in}{0.528000in}}{\pgfqpoint{4.960000in}{3.696000in}}%
\pgfusepath{clip}%
\pgfsetrectcap%
\pgfsetroundjoin%
\pgfsetlinewidth{1.505625pt}%
\definecolor{currentstroke}{rgb}{0.000000,0.000000,1.000000}%
\pgfsetstrokecolor{currentstroke}%
\pgfsetdash{}{0pt}%
\pgfpathmoveto{\pgfqpoint{5.346667in}{2.496341in}}%
\pgfusepath{stroke}%
\end{pgfscope}%
\begin{pgfscope}%
\pgfpathrectangle{\pgfqpoint{0.800000in}{0.528000in}}{\pgfqpoint{4.960000in}{3.696000in}}%
\pgfusepath{clip}%
\pgfsetbuttcap%
\pgfsetmiterjoin%
\definecolor{currentfill}{rgb}{0.000000,0.000000,1.000000}%
\pgfsetfillcolor{currentfill}%
\pgfsetlinewidth{1.003750pt}%
\definecolor{currentstroke}{rgb}{0.000000,0.000000,1.000000}%
\pgfsetstrokecolor{currentstroke}%
\pgfsetdash{}{0pt}%
\pgfsys@defobject{currentmarker}{\pgfqpoint{-0.041667in}{-0.041667in}}{\pgfqpoint{0.041667in}{0.041667in}}{%
\pgfpathmoveto{\pgfqpoint{-0.041667in}{-0.041667in}}%
\pgfpathlineto{\pgfqpoint{0.041667in}{-0.041667in}}%
\pgfpathlineto{\pgfqpoint{0.041667in}{0.041667in}}%
\pgfpathlineto{\pgfqpoint{-0.041667in}{0.041667in}}%
\pgfpathclose%
\pgfusepath{stroke,fill}%
}%
\begin{pgfscope}%
\pgfsys@transformshift{5.346667in}{2.496341in}%
\pgfsys@useobject{currentmarker}{}%
\end{pgfscope}%
\end{pgfscope}%
\begin{pgfscope}%
\pgfpathrectangle{\pgfqpoint{0.800000in}{0.528000in}}{\pgfqpoint{4.960000in}{3.696000in}}%
\pgfusepath{clip}%
\pgfsetrectcap%
\pgfsetroundjoin%
\pgfsetlinewidth{1.505625pt}%
\definecolor{currentstroke}{rgb}{0.000000,0.000000,1.000000}%
\pgfsetstrokecolor{currentstroke}%
\pgfsetdash{}{0pt}%
\pgfpathmoveto{\pgfqpoint{1.213333in}{2.491040in}}%
\pgfusepath{stroke}%
\end{pgfscope}%
\begin{pgfscope}%
\pgfpathrectangle{\pgfqpoint{0.800000in}{0.528000in}}{\pgfqpoint{4.960000in}{3.696000in}}%
\pgfusepath{clip}%
\pgfsetrectcap%
\pgfsetroundjoin%
\pgfsetlinewidth{1.505625pt}%
\definecolor{currentstroke}{rgb}{0.000000,0.000000,1.000000}%
\pgfsetstrokecolor{currentstroke}%
\pgfsetdash{}{0pt}%
\pgfpathmoveto{\pgfqpoint{1.626667in}{2.507344in}}%
\pgfusepath{stroke}%
\end{pgfscope}%
\begin{pgfscope}%
\pgfpathrectangle{\pgfqpoint{0.800000in}{0.528000in}}{\pgfqpoint{4.960000in}{3.696000in}}%
\pgfusepath{clip}%
\pgfsetrectcap%
\pgfsetroundjoin%
\pgfsetlinewidth{1.505625pt}%
\definecolor{currentstroke}{rgb}{0.000000,0.000000,1.000000}%
\pgfsetstrokecolor{currentstroke}%
\pgfsetdash{}{0pt}%
\pgfpathmoveto{\pgfqpoint{2.040000in}{2.479954in}}%
\pgfusepath{stroke}%
\end{pgfscope}%
\begin{pgfscope}%
\pgfpathrectangle{\pgfqpoint{0.800000in}{0.528000in}}{\pgfqpoint{4.960000in}{3.696000in}}%
\pgfusepath{clip}%
\pgfsetrectcap%
\pgfsetroundjoin%
\pgfsetlinewidth{1.505625pt}%
\definecolor{currentstroke}{rgb}{0.000000,0.000000,1.000000}%
\pgfsetstrokecolor{currentstroke}%
\pgfsetdash{}{0pt}%
\pgfpathmoveto{\pgfqpoint{2.866667in}{3.038596in}}%
\pgfusepath{stroke}%
\end{pgfscope}%
\begin{pgfscope}%
\pgfpathrectangle{\pgfqpoint{0.800000in}{0.528000in}}{\pgfqpoint{4.960000in}{3.696000in}}%
\pgfusepath{clip}%
\pgfsetrectcap%
\pgfsetroundjoin%
\pgfsetlinewidth{1.505625pt}%
\definecolor{currentstroke}{rgb}{0.000000,0.000000,1.000000}%
\pgfsetstrokecolor{currentstroke}%
\pgfsetdash{}{0pt}%
\pgfpathmoveto{\pgfqpoint{3.280000in}{3.043333in}}%
\pgfusepath{stroke}%
\end{pgfscope}%
\begin{pgfscope}%
\pgfpathrectangle{\pgfqpoint{0.800000in}{0.528000in}}{\pgfqpoint{4.960000in}{3.696000in}}%
\pgfusepath{clip}%
\pgfsetrectcap%
\pgfsetroundjoin%
\pgfsetlinewidth{1.505625pt}%
\definecolor{currentstroke}{rgb}{0.000000,0.000000,1.000000}%
\pgfsetstrokecolor{currentstroke}%
\pgfsetdash{}{0pt}%
\pgfpathmoveto{\pgfqpoint{3.693333in}{3.037436in}}%
\pgfusepath{stroke}%
\end{pgfscope}%
\begin{pgfscope}%
\pgfpathrectangle{\pgfqpoint{0.800000in}{0.528000in}}{\pgfqpoint{4.960000in}{3.696000in}}%
\pgfusepath{clip}%
\pgfsetrectcap%
\pgfsetroundjoin%
\pgfsetlinewidth{1.505625pt}%
\definecolor{currentstroke}{rgb}{0.000000,0.000000,1.000000}%
\pgfsetstrokecolor{currentstroke}%
\pgfsetdash{}{0pt}%
\pgfpathmoveto{\pgfqpoint{4.520000in}{2.505691in}}%
\pgfusepath{stroke}%
\end{pgfscope}%
\begin{pgfscope}%
\pgfpathrectangle{\pgfqpoint{0.800000in}{0.528000in}}{\pgfqpoint{4.960000in}{3.696000in}}%
\pgfusepath{clip}%
\pgfsetrectcap%
\pgfsetroundjoin%
\pgfsetlinewidth{1.505625pt}%
\definecolor{currentstroke}{rgb}{0.000000,0.000000,1.000000}%
\pgfsetstrokecolor{currentstroke}%
\pgfsetdash{}{0pt}%
\pgfpathmoveto{\pgfqpoint{4.933333in}{2.521600in}}%
\pgfusepath{stroke}%
\end{pgfscope}%
\begin{pgfscope}%
\pgfpathrectangle{\pgfqpoint{0.800000in}{0.528000in}}{\pgfqpoint{4.960000in}{3.696000in}}%
\pgfusepath{clip}%
\pgfsetrectcap%
\pgfsetroundjoin%
\pgfsetlinewidth{1.505625pt}%
\definecolor{currentstroke}{rgb}{0.000000,0.000000,1.000000}%
\pgfsetstrokecolor{currentstroke}%
\pgfsetdash{}{0pt}%
\pgfpathmoveto{\pgfqpoint{5.346667in}{2.496341in}}%
\pgfusepath{stroke}%
\end{pgfscope}%
\begin{pgfscope}%
\pgfsetrectcap%
\pgfsetmiterjoin%
\pgfsetlinewidth{0.803000pt}%
\definecolor{currentstroke}{rgb}{0.000000,0.000000,0.000000}%
\pgfsetstrokecolor{currentstroke}%
\pgfsetdash{}{0pt}%
\pgfpathmoveto{\pgfqpoint{0.800000in}{0.528000in}}%
\pgfpathlineto{\pgfqpoint{0.800000in}{4.224000in}}%
\pgfusepath{stroke}%
\end{pgfscope}%
\begin{pgfscope}%
\pgfsetrectcap%
\pgfsetmiterjoin%
\pgfsetlinewidth{0.803000pt}%
\definecolor{currentstroke}{rgb}{0.000000,0.000000,0.000000}%
\pgfsetstrokecolor{currentstroke}%
\pgfsetdash{}{0pt}%
\pgfpathmoveto{\pgfqpoint{5.760000in}{0.528000in}}%
\pgfpathlineto{\pgfqpoint{5.760000in}{4.224000in}}%
\pgfusepath{stroke}%
\end{pgfscope}%
\begin{pgfscope}%
\pgfsetrectcap%
\pgfsetmiterjoin%
\pgfsetlinewidth{0.803000pt}%
\definecolor{currentstroke}{rgb}{0.000000,0.000000,0.000000}%
\pgfsetstrokecolor{currentstroke}%
\pgfsetdash{}{0pt}%
\pgfpathmoveto{\pgfqpoint{0.800000in}{0.528000in}}%
\pgfpathlineto{\pgfqpoint{5.760000in}{0.528000in}}%
\pgfusepath{stroke}%
\end{pgfscope}%
\begin{pgfscope}%
\pgfsetrectcap%
\pgfsetmiterjoin%
\pgfsetlinewidth{0.803000pt}%
\definecolor{currentstroke}{rgb}{0.000000,0.000000,0.000000}%
\pgfsetstrokecolor{currentstroke}%
\pgfsetdash{}{0pt}%
\pgfpathmoveto{\pgfqpoint{0.800000in}{4.224000in}}%
\pgfpathlineto{\pgfqpoint{5.760000in}{4.224000in}}%
\pgfusepath{stroke}%
\end{pgfscope}%
\begin{pgfscope}%
\pgfsetbuttcap%
\pgfsetmiterjoin%
\definecolor{currentfill}{rgb}{1.000000,1.000000,1.000000}%
\pgfsetfillcolor{currentfill}%
\pgfsetfillopacity{0.800000}%
\pgfsetlinewidth{1.003750pt}%
\definecolor{currentstroke}{rgb}{0.800000,0.800000,0.800000}%
\pgfsetstrokecolor{currentstroke}%
\pgfsetstrokeopacity{0.800000}%
\pgfsetdash{}{0pt}%
\pgfpathmoveto{\pgfqpoint{0.936111in}{0.625222in}}%
\pgfpathlineto{\pgfqpoint{2.708805in}{0.625222in}}%
\pgfpathquadraticcurveto{\pgfqpoint{2.747694in}{0.625222in}}{\pgfqpoint{2.747694in}{0.664111in}}%
\pgfpathlineto{\pgfqpoint{2.747694in}{2.019665in}}%
\pgfpathquadraticcurveto{\pgfqpoint{2.747694in}{2.058553in}}{\pgfqpoint{2.708805in}{2.058553in}}%
\pgfpathlineto{\pgfqpoint{0.936111in}{2.058553in}}%
\pgfpathquadraticcurveto{\pgfqpoint{0.897222in}{2.058553in}}{\pgfqpoint{0.897222in}{2.019665in}}%
\pgfpathlineto{\pgfqpoint{0.897222in}{0.664111in}}%
\pgfpathquadraticcurveto{\pgfqpoint{0.897222in}{0.625222in}}{\pgfqpoint{0.936111in}{0.625222in}}%
\pgfpathclose%
\pgfusepath{stroke,fill}%
\end{pgfscope}%
\begin{pgfscope}%
\pgfsetrectcap%
\pgfsetroundjoin%
\pgfsetlinewidth{1.505625pt}%
\definecolor{currentstroke}{rgb}{0.000000,0.000000,1.000000}%
\pgfsetstrokecolor{currentstroke}%
\pgfsetdash{}{0pt}%
\pgfpathmoveto{\pgfqpoint{0.975000in}{1.909943in}}%
\pgfpathlineto{\pgfqpoint{1.363889in}{1.909943in}}%
\pgfusepath{stroke}%
\end{pgfscope}%
\begin{pgfscope}%
\pgfsetbuttcap%
\pgfsetroundjoin%
\definecolor{currentfill}{rgb}{0.000000,0.000000,1.000000}%
\pgfsetfillcolor{currentfill}%
\pgfsetlinewidth{1.003750pt}%
\definecolor{currentstroke}{rgb}{0.000000,0.000000,1.000000}%
\pgfsetstrokecolor{currentstroke}%
\pgfsetdash{}{0pt}%
\pgfsys@defobject{currentmarker}{\pgfqpoint{-0.041667in}{-0.041667in}}{\pgfqpoint{0.041667in}{0.041667in}}{%
\pgfpathmoveto{\pgfqpoint{0.000000in}{-0.041667in}}%
\pgfpathcurveto{\pgfqpoint{0.011050in}{-0.041667in}}{\pgfqpoint{0.021649in}{-0.037276in}}{\pgfqpoint{0.029463in}{-0.029463in}}%
\pgfpathcurveto{\pgfqpoint{0.037276in}{-0.021649in}}{\pgfqpoint{0.041667in}{-0.011050in}}{\pgfqpoint{0.041667in}{0.000000in}}%
\pgfpathcurveto{\pgfqpoint{0.041667in}{0.011050in}}{\pgfqpoint{0.037276in}{0.021649in}}{\pgfqpoint{0.029463in}{0.029463in}}%
\pgfpathcurveto{\pgfqpoint{0.021649in}{0.037276in}}{\pgfqpoint{0.011050in}{0.041667in}}{\pgfqpoint{0.000000in}{0.041667in}}%
\pgfpathcurveto{\pgfqpoint{-0.011050in}{0.041667in}}{\pgfqpoint{-0.021649in}{0.037276in}}{\pgfqpoint{-0.029463in}{0.029463in}}%
\pgfpathcurveto{\pgfqpoint{-0.037276in}{0.021649in}}{\pgfqpoint{-0.041667in}{0.011050in}}{\pgfqpoint{-0.041667in}{0.000000in}}%
\pgfpathcurveto{\pgfqpoint{-0.041667in}{-0.011050in}}{\pgfqpoint{-0.037276in}{-0.021649in}}{\pgfqpoint{-0.029463in}{-0.029463in}}%
\pgfpathcurveto{\pgfqpoint{-0.021649in}{-0.037276in}}{\pgfqpoint{-0.011050in}{-0.041667in}}{\pgfqpoint{0.000000in}{-0.041667in}}%
\pgfpathclose%
\pgfusepath{stroke,fill}%
}%
\begin{pgfscope}%
\pgfsys@transformshift{1.169444in}{1.909943in}%
\pgfsys@useobject{currentmarker}{}%
\end{pgfscope}%
\end{pgfscope}%
\begin{pgfscope}%
\definecolor{textcolor}{rgb}{0.000000,0.000000,0.000000}%
\pgfsetstrokecolor{textcolor}%
\pgfsetfillcolor{textcolor}%
\pgftext[x=1.519444in,y=1.841887in,left,base]{\color{textcolor}\rmfamily\fontsize{14.000000}{16.800000}\selectfont Scenario IV}%
\end{pgfscope}%
\begin{pgfscope}%
\pgfsetrectcap%
\pgfsetroundjoin%
\pgfsetlinewidth{1.505625pt}%
\definecolor{currentstroke}{rgb}{0.000000,0.000000,1.000000}%
\pgfsetstrokecolor{currentstroke}%
\pgfsetdash{}{0pt}%
\pgfpathmoveto{\pgfqpoint{0.975000in}{1.634943in}}%
\pgfpathlineto{\pgfqpoint{1.363889in}{1.634943in}}%
\pgfusepath{stroke}%
\end{pgfscope}%
\begin{pgfscope}%
\pgfsetbuttcap%
\pgfsetmiterjoin%
\definecolor{currentfill}{rgb}{0.000000,0.000000,1.000000}%
\pgfsetfillcolor{currentfill}%
\pgfsetlinewidth{1.003750pt}%
\definecolor{currentstroke}{rgb}{0.000000,0.000000,1.000000}%
\pgfsetstrokecolor{currentstroke}%
\pgfsetdash{}{0pt}%
\pgfsys@defobject{currentmarker}{\pgfqpoint{-0.041667in}{-0.041667in}}{\pgfqpoint{0.041667in}{0.041667in}}{%
\pgfpathmoveto{\pgfqpoint{0.000000in}{0.041667in}}%
\pgfpathlineto{\pgfqpoint{-0.041667in}{-0.041667in}}%
\pgfpathlineto{\pgfqpoint{0.041667in}{-0.041667in}}%
\pgfpathclose%
\pgfusepath{stroke,fill}%
}%
\begin{pgfscope}%
\pgfsys@transformshift{1.169444in}{1.634943in}%
\pgfsys@useobject{currentmarker}{}%
\end{pgfscope}%
\end{pgfscope}%
\begin{pgfscope}%
\definecolor{textcolor}{rgb}{0.000000,0.000000,0.000000}%
\pgfsetstrokecolor{textcolor}%
\pgfsetfillcolor{textcolor}%
\pgftext[x=1.519444in,y=1.566887in,left,base]{\color{textcolor}\rmfamily\fontsize{14.000000}{16.800000}\selectfont Scenario VII}%
\end{pgfscope}%
\begin{pgfscope}%
\pgfsetrectcap%
\pgfsetroundjoin%
\pgfsetlinewidth{1.505625pt}%
\definecolor{currentstroke}{rgb}{0.000000,0.000000,1.000000}%
\pgfsetstrokecolor{currentstroke}%
\pgfsetdash{}{0pt}%
\pgfpathmoveto{\pgfqpoint{0.975000in}{1.359943in}}%
\pgfpathlineto{\pgfqpoint{1.363889in}{1.359943in}}%
\pgfusepath{stroke}%
\end{pgfscope}%
\begin{pgfscope}%
\pgfsetbuttcap%
\pgfsetmiterjoin%
\definecolor{currentfill}{rgb}{0.000000,0.000000,1.000000}%
\pgfsetfillcolor{currentfill}%
\pgfsetlinewidth{1.003750pt}%
\definecolor{currentstroke}{rgb}{0.000000,0.000000,1.000000}%
\pgfsetstrokecolor{currentstroke}%
\pgfsetdash{}{0pt}%
\pgfsys@defobject{currentmarker}{\pgfqpoint{-0.041667in}{-0.041667in}}{\pgfqpoint{0.041667in}{0.041667in}}{%
\pgfpathmoveto{\pgfqpoint{-0.041667in}{-0.041667in}}%
\pgfpathlineto{\pgfqpoint{0.041667in}{-0.041667in}}%
\pgfpathlineto{\pgfqpoint{0.041667in}{0.041667in}}%
\pgfpathlineto{\pgfqpoint{-0.041667in}{0.041667in}}%
\pgfpathclose%
\pgfusepath{stroke,fill}%
}%
\begin{pgfscope}%
\pgfsys@transformshift{1.169444in}{1.359943in}%
\pgfsys@useobject{currentmarker}{}%
\end{pgfscope}%
\end{pgfscope}%
\begin{pgfscope}%
\definecolor{textcolor}{rgb}{0.000000,0.000000,0.000000}%
\pgfsetstrokecolor{textcolor}%
\pgfsetfillcolor{textcolor}%
\pgftext[x=1.519444in,y=1.291888in,left,base]{\color{textcolor}\rmfamily\fontsize{14.000000}{16.800000}\selectfont Scenario IX}%
\end{pgfscope}%
\begin{pgfscope}%
\pgfsetbuttcap%
\pgfsetmiterjoin%
\definecolor{currentfill}{rgb}{1.000000,0.647059,0.000000}%
\pgfsetfillcolor{currentfill}%
\pgfsetfillopacity{0.700000}%
\pgfsetlinewidth{1.003750pt}%
\definecolor{currentstroke}{rgb}{1.000000,0.647059,0.000000}%
\pgfsetstrokecolor{currentstroke}%
\pgfsetstrokeopacity{0.700000}%
\pgfsetdash{}{0pt}%
\pgfpathmoveto{\pgfqpoint{0.975000in}{1.016888in}}%
\pgfpathlineto{\pgfqpoint{1.363889in}{1.016888in}}%
\pgfpathlineto{\pgfqpoint{1.363889in}{1.152999in}}%
\pgfpathlineto{\pgfqpoint{0.975000in}{1.152999in}}%
\pgfpathclose%
\pgfusepath{stroke,fill}%
\end{pgfscope}%
\begin{pgfscope}%
\definecolor{textcolor}{rgb}{0.000000,0.000000,0.000000}%
\pgfsetstrokecolor{textcolor}%
\pgfsetfillcolor{textcolor}%
\pgftext[x=1.519444in,y=1.016888in,left,base]{\color{textcolor}\rmfamily\fontsize{14.000000}{16.800000}\selectfont SM}%
\end{pgfscope}%
\begin{pgfscope}%
\pgfsetbuttcap%
\pgfsetmiterjoin%
\definecolor{currentfill}{rgb}{0.000000,0.501961,0.000000}%
\pgfsetfillcolor{currentfill}%
\pgfsetfillopacity{0.700000}%
\pgfsetlinewidth{1.003750pt}%
\definecolor{currentstroke}{rgb}{0.000000,0.501961,0.000000}%
\pgfsetstrokecolor{currentstroke}%
\pgfsetstrokeopacity{0.700000}%
\pgfsetdash{}{0pt}%
\pgfpathmoveto{\pgfqpoint{0.975000in}{0.741889in}}%
\pgfpathlineto{\pgfqpoint{1.363889in}{0.741889in}}%
\pgfpathlineto{\pgfqpoint{1.363889in}{0.878000in}}%
\pgfpathlineto{\pgfqpoint{0.975000in}{0.878000in}}%
\pgfpathclose%
\pgfusepath{stroke,fill}%
\end{pgfscope}%
\begin{pgfscope}%
\definecolor{textcolor}{rgb}{0.000000,0.000000,0.000000}%
\pgfsetstrokecolor{textcolor}%
\pgfsetfillcolor{textcolor}%
\pgftext[x=1.519444in,y=0.741889in,left,base]{\color{textcolor}\rmfamily\fontsize{14.000000}{16.800000}\selectfont Experimental}%
\end{pgfscope}%
\end{pgfpicture}%
\makeatother%
\endgroup%

%% file: RDplot_21.pgf
\begingroup%
\makeatletter%
\begin{pgfpicture}%
\pgfpathrectangle{\pgfpointorigin}{\pgfqpoint{6.400000in}{4.800000in}}%
\pgfusepath{use as bounding box, clip}%
\begin{pgfscope}%
\pgfsetbuttcap%
\pgfsetmiterjoin%
\definecolor{currentfill}{rgb}{1.000000,1.000000,1.000000}%
\pgfsetfillcolor{currentfill}%
\pgfsetlinewidth{0.000000pt}%
\definecolor{currentstroke}{rgb}{1.000000,1.000000,1.000000}%
\pgfsetstrokecolor{currentstroke}%
\pgfsetdash{}{0pt}%
\pgfpathmoveto{\pgfqpoint{0.000000in}{0.000000in}}%
\pgfpathlineto{\pgfqpoint{6.400000in}{0.000000in}}%
\pgfpathlineto{\pgfqpoint{6.400000in}{4.800000in}}%
\pgfpathlineto{\pgfqpoint{0.000000in}{4.800000in}}%
\pgfpathclose%
\pgfusepath{fill}%
\end{pgfscope}%
\begin{pgfscope}%
\pgfsetbuttcap%
\pgfsetmiterjoin%
\definecolor{currentfill}{rgb}{1.000000,1.000000,1.000000}%
\pgfsetfillcolor{currentfill}%
\pgfsetlinewidth{0.000000pt}%
\definecolor{currentstroke}{rgb}{0.000000,0.000000,0.000000}%
\pgfsetstrokecolor{currentstroke}%
\pgfsetstrokeopacity{0.000000}%
\pgfsetdash{}{0pt}%
\pgfpathmoveto{\pgfqpoint{0.800000in}{0.528000in}}%
\pgfpathlineto{\pgfqpoint{5.760000in}{0.528000in}}%
\pgfpathlineto{\pgfqpoint{5.760000in}{4.224000in}}%
\pgfpathlineto{\pgfqpoint{0.800000in}{4.224000in}}%
\pgfpathclose%
\pgfusepath{fill}%
\end{pgfscope}%
\begin{pgfscope}%
\pgfpathrectangle{\pgfqpoint{0.800000in}{0.528000in}}{\pgfqpoint{4.960000in}{3.696000in}}%
\pgfusepath{clip}%
\pgfsetbuttcap%
\pgfsetmiterjoin%
\definecolor{currentfill}{rgb}{1.000000,0.647059,0.000000}%
\pgfsetfillcolor{currentfill}%
\pgfsetfillopacity{0.700000}%
\pgfsetlinewidth{1.003750pt}%
\definecolor{currentstroke}{rgb}{1.000000,0.647059,0.000000}%
\pgfsetstrokecolor{currentstroke}%
\pgfsetstrokeopacity{0.700000}%
\pgfsetdash{}{0pt}%
\pgfpathmoveto{\pgfqpoint{0.800000in}{1.943873in}}%
\pgfpathlineto{\pgfqpoint{2.453333in}{1.943873in}}%
\pgfpathlineto{\pgfqpoint{2.453333in}{2.222221in}}%
\pgfpathlineto{\pgfqpoint{0.800000in}{2.222221in}}%
\pgfpathclose%
\pgfusepath{stroke,fill}%
\end{pgfscope}%
\begin{pgfscope}%
\pgfpathrectangle{\pgfqpoint{0.800000in}{0.528000in}}{\pgfqpoint{4.960000in}{3.696000in}}%
\pgfusepath{clip}%
\pgfsetbuttcap%
\pgfsetmiterjoin%
\definecolor{currentfill}{rgb}{0.000000,0.501961,0.000000}%
\pgfsetfillcolor{currentfill}%
\pgfsetfillopacity{0.700000}%
\pgfsetlinewidth{1.003750pt}%
\definecolor{currentstroke}{rgb}{0.000000,0.501961,0.000000}%
\pgfsetstrokecolor{currentstroke}%
\pgfsetstrokeopacity{0.700000}%
\pgfsetdash{}{0pt}%
\pgfpathmoveto{\pgfqpoint{0.800000in}{2.402200in}}%
\pgfpathlineto{\pgfqpoint{2.453333in}{2.402200in}}%
\pgfpathlineto{\pgfqpoint{2.453333in}{4.056000in}}%
\pgfpathlineto{\pgfqpoint{0.800000in}{4.056000in}}%
\pgfpathclose%
\pgfusepath{stroke,fill}%
\end{pgfscope}%
\begin{pgfscope}%
\pgfpathrectangle{\pgfqpoint{0.800000in}{0.528000in}}{\pgfqpoint{4.960000in}{3.696000in}}%
\pgfusepath{clip}%
\pgfsetbuttcap%
\pgfsetmiterjoin%
\definecolor{currentfill}{rgb}{0.000000,0.501961,0.000000}%
\pgfsetfillcolor{currentfill}%
\pgfsetfillopacity{0.700000}%
\pgfsetlinewidth{1.003750pt}%
\definecolor{currentstroke}{rgb}{0.000000,0.501961,0.000000}%
\pgfsetstrokecolor{currentstroke}%
\pgfsetstrokeopacity{0.700000}%
\pgfsetdash{}{0pt}%
\pgfpathmoveto{\pgfqpoint{2.453333in}{1.491806in}}%
\pgfpathlineto{\pgfqpoint{4.106667in}{1.491806in}}%
\pgfpathlineto{\pgfqpoint{4.106667in}{2.312161in}}%
\pgfpathlineto{\pgfqpoint{2.453333in}{2.312161in}}%
\pgfpathclose%
\pgfusepath{stroke,fill}%
\end{pgfscope}%
\begin{pgfscope}%
\pgfpathrectangle{\pgfqpoint{0.800000in}{0.528000in}}{\pgfqpoint{4.960000in}{3.696000in}}%
\pgfusepath{clip}%
\pgfsetbuttcap%
\pgfsetmiterjoin%
\definecolor{currentfill}{rgb}{1.000000,0.647059,0.000000}%
\pgfsetfillcolor{currentfill}%
\pgfsetfillopacity{0.700000}%
\pgfsetlinewidth{1.003750pt}%
\definecolor{currentstroke}{rgb}{1.000000,0.647059,0.000000}%
\pgfsetstrokecolor{currentstroke}%
\pgfsetstrokeopacity{0.700000}%
\pgfsetdash{}{0pt}%
\pgfpathmoveto{\pgfqpoint{2.453333in}{0.715580in}}%
\pgfpathlineto{\pgfqpoint{4.106667in}{0.715580in}}%
\pgfpathlineto{\pgfqpoint{4.106667in}{0.904491in}}%
\pgfpathlineto{\pgfqpoint{2.453333in}{0.904491in}}%
\pgfpathclose%
\pgfusepath{stroke,fill}%
\end{pgfscope}%
\begin{pgfscope}%
\pgfpathrectangle{\pgfqpoint{0.800000in}{0.528000in}}{\pgfqpoint{4.960000in}{3.696000in}}%
\pgfusepath{clip}%
\pgfsetbuttcap%
\pgfsetmiterjoin%
\definecolor{currentfill}{rgb}{0.000000,0.501961,0.000000}%
\pgfsetfillcolor{currentfill}%
\pgfsetfillopacity{0.700000}%
\pgfsetlinewidth{1.003750pt}%
\definecolor{currentstroke}{rgb}{0.000000,0.501961,0.000000}%
\pgfsetstrokecolor{currentstroke}%
\pgfsetstrokeopacity{0.700000}%
\pgfsetdash{}{0pt}%
\pgfpathmoveto{\pgfqpoint{4.106667in}{1.576426in}}%
\pgfpathlineto{\pgfqpoint{5.760000in}{1.576426in}}%
\pgfpathlineto{\pgfqpoint{5.760000in}{2.966794in}}%
\pgfpathlineto{\pgfqpoint{4.106667in}{2.966794in}}%
\pgfpathclose%
\pgfusepath{stroke,fill}%
\end{pgfscope}%
\begin{pgfscope}%
\pgfpathrectangle{\pgfqpoint{0.800000in}{0.528000in}}{\pgfqpoint{4.960000in}{3.696000in}}%
\pgfusepath{clip}%
\pgfsetbuttcap%
\pgfsetmiterjoin%
\definecolor{currentfill}{rgb}{1.000000,0.647059,0.000000}%
\pgfsetfillcolor{currentfill}%
\pgfsetfillopacity{0.700000}%
\pgfsetlinewidth{1.003750pt}%
\definecolor{currentstroke}{rgb}{1.000000,0.647059,0.000000}%
\pgfsetstrokecolor{currentstroke}%
\pgfsetstrokeopacity{0.700000}%
\pgfsetdash{}{0pt}%
\pgfpathmoveto{\pgfqpoint{4.106667in}{0.744157in}}%
\pgfpathlineto{\pgfqpoint{5.760000in}{0.744157in}}%
\pgfpathlineto{\pgfqpoint{5.760000in}{0.897439in}}%
\pgfpathlineto{\pgfqpoint{4.106667in}{0.897439in}}%
\pgfpathclose%
\pgfusepath{stroke,fill}%
\end{pgfscope}%
\begin{pgfscope}%
\pgfsetbuttcap%
\pgfsetroundjoin%
\definecolor{currentfill}{rgb}{0.000000,0.000000,0.000000}%
\pgfsetfillcolor{currentfill}%
\pgfsetlinewidth{0.803000pt}%
\definecolor{currentstroke}{rgb}{0.000000,0.000000,0.000000}%
\pgfsetstrokecolor{currentstroke}%
\pgfsetdash{}{0pt}%
\pgfsys@defobject{currentmarker}{\pgfqpoint{0.000000in}{-0.048611in}}{\pgfqpoint{0.000000in}{0.000000in}}{%
\pgfpathmoveto{\pgfqpoint{0.000000in}{0.000000in}}%
\pgfpathlineto{\pgfqpoint{0.000000in}{-0.048611in}}%
\pgfusepath{stroke,fill}%
}%
\begin{pgfscope}%
\pgfsys@transformshift{1.626667in}{0.528000in}%
\pgfsys@useobject{currentmarker}{}%
\end{pgfscope}%
\end{pgfscope}%
\begin{pgfscope}%
\definecolor{textcolor}{rgb}{0.000000,0.000000,0.000000}%
\pgfsetstrokecolor{textcolor}%
\pgfsetfillcolor{textcolor}%
\pgftext[x=1.626667in,y=0.430778in,,top]{\color{textcolor}\rmfamily\fontsize{16.000000}{19.200000}\selectfont \(\displaystyle R_D^\ell\)}%
\end{pgfscope}%
\begin{pgfscope}%
\pgfsetbuttcap%
\pgfsetroundjoin%
\definecolor{currentfill}{rgb}{0.000000,0.000000,0.000000}%
\pgfsetfillcolor{currentfill}%
\pgfsetlinewidth{0.803000pt}%
\definecolor{currentstroke}{rgb}{0.000000,0.000000,0.000000}%
\pgfsetstrokecolor{currentstroke}%
\pgfsetdash{}{0pt}%
\pgfsys@defobject{currentmarker}{\pgfqpoint{0.000000in}{-0.048611in}}{\pgfqpoint{0.000000in}{0.000000in}}{%
\pgfpathmoveto{\pgfqpoint{0.000000in}{0.000000in}}%
\pgfpathlineto{\pgfqpoint{0.000000in}{-0.048611in}}%
\pgfusepath{stroke,fill}%
}%
\begin{pgfscope}%
\pgfsys@transformshift{3.280000in}{0.528000in}%
\pgfsys@useobject{currentmarker}{}%
\end{pgfscope}%
\end{pgfscope}%
\begin{pgfscope}%
\definecolor{textcolor}{rgb}{0.000000,0.000000,0.000000}%
\pgfsetstrokecolor{textcolor}%
\pgfsetfillcolor{textcolor}%
\pgftext[x=3.280000in,y=0.430778in,,top]{\color{textcolor}\rmfamily\fontsize{16.000000}{19.200000}\selectfont \(\displaystyle R_{D^*}^\ell\)}%
\end{pgfscope}%
\begin{pgfscope}%
\pgfsetbuttcap%
\pgfsetroundjoin%
\definecolor{currentfill}{rgb}{0.000000,0.000000,0.000000}%
\pgfsetfillcolor{currentfill}%
\pgfsetlinewidth{0.803000pt}%
\definecolor{currentstroke}{rgb}{0.000000,0.000000,0.000000}%
\pgfsetstrokecolor{currentstroke}%
\pgfsetdash{}{0pt}%
\pgfsys@defobject{currentmarker}{\pgfqpoint{0.000000in}{-0.048611in}}{\pgfqpoint{0.000000in}{0.000000in}}{%
\pgfpathmoveto{\pgfqpoint{0.000000in}{0.000000in}}%
\pgfpathlineto{\pgfqpoint{0.000000in}{-0.048611in}}%
\pgfusepath{stroke,fill}%
}%
\begin{pgfscope}%
\pgfsys@transformshift{4.933333in}{0.528000in}%
\pgfsys@useobject{currentmarker}{}%
\end{pgfscope}%
\end{pgfscope}%
\begin{pgfscope}%
\definecolor{textcolor}{rgb}{0.000000,0.000000,0.000000}%
\pgfsetstrokecolor{textcolor}%
\pgfsetfillcolor{textcolor}%
\pgftext[x=4.933333in,y=0.430778in,,top]{\color{textcolor}\rmfamily\fontsize{16.000000}{19.200000}\selectfont \(\displaystyle R_{D^*}^\mu\)}%
\end{pgfscope}%
\begin{pgfscope}%
\pgfsetbuttcap%
\pgfsetroundjoin%
\definecolor{currentfill}{rgb}{0.000000,0.000000,0.000000}%
\pgfsetfillcolor{currentfill}%
\pgfsetlinewidth{0.803000pt}%
\definecolor{currentstroke}{rgb}{0.000000,0.000000,0.000000}%
\pgfsetstrokecolor{currentstroke}%
\pgfsetdash{}{0pt}%
\pgfsys@defobject{currentmarker}{\pgfqpoint{-0.048611in}{0.000000in}}{\pgfqpoint{0.000000in}{0.000000in}}{%
\pgfpathmoveto{\pgfqpoint{0.000000in}{0.000000in}}%
\pgfpathlineto{\pgfqpoint{-0.048611in}{0.000000in}}%
\pgfusepath{stroke,fill}%
}%
\begin{pgfscope}%
\pgfsys@transformshift{0.800000in}{0.947417in}%
\pgfsys@useobject{currentmarker}{}%
\end{pgfscope}%
\end{pgfscope}%
\begin{pgfscope}%
\definecolor{textcolor}{rgb}{0.000000,0.000000,0.000000}%
\pgfsetstrokecolor{textcolor}%
\pgfsetfillcolor{textcolor}%
\pgftext[x=0.307296in, y=0.864084in, left, base]{\color{textcolor}\rmfamily\fontsize{16.000000}{19.200000}\selectfont \(\displaystyle 0.26\)}%
\end{pgfscope}%
\begin{pgfscope}%
\pgfsetbuttcap%
\pgfsetroundjoin%
\definecolor{currentfill}{rgb}{0.000000,0.000000,0.000000}%
\pgfsetfillcolor{currentfill}%
\pgfsetlinewidth{0.803000pt}%
\definecolor{currentstroke}{rgb}{0.000000,0.000000,0.000000}%
\pgfsetstrokecolor{currentstroke}%
\pgfsetdash{}{0pt}%
\pgfsys@defobject{currentmarker}{\pgfqpoint{-0.048611in}{0.000000in}}{\pgfqpoint{0.000000in}{0.000000in}}{%
\pgfpathmoveto{\pgfqpoint{0.000000in}{0.000000in}}%
\pgfpathlineto{\pgfqpoint{-0.048611in}{0.000000in}}%
\pgfusepath{stroke,fill}%
}%
\begin{pgfscope}%
\pgfsys@transformshift{0.800000in}{1.475585in}%
\pgfsys@useobject{currentmarker}{}%
\end{pgfscope}%
\end{pgfscope}%
\begin{pgfscope}%
\definecolor{textcolor}{rgb}{0.000000,0.000000,0.000000}%
\pgfsetstrokecolor{textcolor}%
\pgfsetfillcolor{textcolor}%
\pgftext[x=0.307296in, y=1.392252in, left, base]{\color{textcolor}\rmfamily\fontsize{16.000000}{19.200000}\selectfont \(\displaystyle 0.28\)}%
\end{pgfscope}%
\begin{pgfscope}%
\pgfsetbuttcap%
\pgfsetroundjoin%
\definecolor{currentfill}{rgb}{0.000000,0.000000,0.000000}%
\pgfsetfillcolor{currentfill}%
\pgfsetlinewidth{0.803000pt}%
\definecolor{currentstroke}{rgb}{0.000000,0.000000,0.000000}%
\pgfsetstrokecolor{currentstroke}%
\pgfsetdash{}{0pt}%
\pgfsys@defobject{currentmarker}{\pgfqpoint{-0.048611in}{0.000000in}}{\pgfqpoint{0.000000in}{0.000000in}}{%
\pgfpathmoveto{\pgfqpoint{0.000000in}{0.000000in}}%
\pgfpathlineto{\pgfqpoint{-0.048611in}{0.000000in}}%
\pgfusepath{stroke,fill}%
}%
\begin{pgfscope}%
\pgfsys@transformshift{0.800000in}{2.003753in}%
\pgfsys@useobject{currentmarker}{}%
\end{pgfscope}%
\end{pgfscope}%
\begin{pgfscope}%
\definecolor{textcolor}{rgb}{0.000000,0.000000,0.000000}%
\pgfsetstrokecolor{textcolor}%
\pgfsetfillcolor{textcolor}%
\pgftext[x=0.307296in, y=1.920420in, left, base]{\color{textcolor}\rmfamily\fontsize{16.000000}{19.200000}\selectfont \(\displaystyle 0.30\)}%
\end{pgfscope}%
\begin{pgfscope}%
\pgfsetbuttcap%
\pgfsetroundjoin%
\definecolor{currentfill}{rgb}{0.000000,0.000000,0.000000}%
\pgfsetfillcolor{currentfill}%
\pgfsetlinewidth{0.803000pt}%
\definecolor{currentstroke}{rgb}{0.000000,0.000000,0.000000}%
\pgfsetstrokecolor{currentstroke}%
\pgfsetdash{}{0pt}%
\pgfsys@defobject{currentmarker}{\pgfqpoint{-0.048611in}{0.000000in}}{\pgfqpoint{0.000000in}{0.000000in}}{%
\pgfpathmoveto{\pgfqpoint{0.000000in}{0.000000in}}%
\pgfpathlineto{\pgfqpoint{-0.048611in}{0.000000in}}%
\pgfusepath{stroke,fill}%
}%
\begin{pgfscope}%
\pgfsys@transformshift{0.800000in}{2.531921in}%
\pgfsys@useobject{currentmarker}{}%
\end{pgfscope}%
\end{pgfscope}%
\begin{pgfscope}%
\definecolor{textcolor}{rgb}{0.000000,0.000000,0.000000}%
\pgfsetstrokecolor{textcolor}%
\pgfsetfillcolor{textcolor}%
\pgftext[x=0.307296in, y=2.448588in, left, base]{\color{textcolor}\rmfamily\fontsize{16.000000}{19.200000}\selectfont \(\displaystyle 0.32\)}%
\end{pgfscope}%
\begin{pgfscope}%
\pgfsetbuttcap%
\pgfsetroundjoin%
\definecolor{currentfill}{rgb}{0.000000,0.000000,0.000000}%
\pgfsetfillcolor{currentfill}%
\pgfsetlinewidth{0.803000pt}%
\definecolor{currentstroke}{rgb}{0.000000,0.000000,0.000000}%
\pgfsetstrokecolor{currentstroke}%
\pgfsetdash{}{0pt}%
\pgfsys@defobject{currentmarker}{\pgfqpoint{-0.048611in}{0.000000in}}{\pgfqpoint{0.000000in}{0.000000in}}{%
\pgfpathmoveto{\pgfqpoint{0.000000in}{0.000000in}}%
\pgfpathlineto{\pgfqpoint{-0.048611in}{0.000000in}}%
\pgfusepath{stroke,fill}%
}%
\begin{pgfscope}%
\pgfsys@transformshift{0.800000in}{3.060089in}%
\pgfsys@useobject{currentmarker}{}%
\end{pgfscope}%
\end{pgfscope}%
\begin{pgfscope}%
\definecolor{textcolor}{rgb}{0.000000,0.000000,0.000000}%
\pgfsetstrokecolor{textcolor}%
\pgfsetfillcolor{textcolor}%
\pgftext[x=0.307296in, y=2.976756in, left, base]{\color{textcolor}\rmfamily\fontsize{16.000000}{19.200000}\selectfont \(\displaystyle 0.34\)}%
\end{pgfscope}%
\begin{pgfscope}%
\pgfsetbuttcap%
\pgfsetroundjoin%
\definecolor{currentfill}{rgb}{0.000000,0.000000,0.000000}%
\pgfsetfillcolor{currentfill}%
\pgfsetlinewidth{0.803000pt}%
\definecolor{currentstroke}{rgb}{0.000000,0.000000,0.000000}%
\pgfsetstrokecolor{currentstroke}%
\pgfsetdash{}{0pt}%
\pgfsys@defobject{currentmarker}{\pgfqpoint{-0.048611in}{0.000000in}}{\pgfqpoint{0.000000in}{0.000000in}}{%
\pgfpathmoveto{\pgfqpoint{0.000000in}{0.000000in}}%
\pgfpathlineto{\pgfqpoint{-0.048611in}{0.000000in}}%
\pgfusepath{stroke,fill}%
}%
\begin{pgfscope}%
\pgfsys@transformshift{0.800000in}{3.588257in}%
\pgfsys@useobject{currentmarker}{}%
\end{pgfscope}%
\end{pgfscope}%
\begin{pgfscope}%
\definecolor{textcolor}{rgb}{0.000000,0.000000,0.000000}%
\pgfsetstrokecolor{textcolor}%
\pgfsetfillcolor{textcolor}%
\pgftext[x=0.307296in, y=3.504924in, left, base]{\color{textcolor}\rmfamily\fontsize{16.000000}{19.200000}\selectfont \(\displaystyle 0.36\)}%
\end{pgfscope}%
\begin{pgfscope}%
\pgfsetbuttcap%
\pgfsetroundjoin%
\definecolor{currentfill}{rgb}{0.000000,0.000000,0.000000}%
\pgfsetfillcolor{currentfill}%
\pgfsetlinewidth{0.803000pt}%
\definecolor{currentstroke}{rgb}{0.000000,0.000000,0.000000}%
\pgfsetstrokecolor{currentstroke}%
\pgfsetdash{}{0pt}%
\pgfsys@defobject{currentmarker}{\pgfqpoint{-0.048611in}{0.000000in}}{\pgfqpoint{0.000000in}{0.000000in}}{%
\pgfpathmoveto{\pgfqpoint{0.000000in}{0.000000in}}%
\pgfpathlineto{\pgfqpoint{-0.048611in}{0.000000in}}%
\pgfusepath{stroke,fill}%
}%
\begin{pgfscope}%
\pgfsys@transformshift{0.800000in}{4.116426in}%
\pgfsys@useobject{currentmarker}{}%
\end{pgfscope}%
\end{pgfscope}%
\begin{pgfscope}%
\definecolor{textcolor}{rgb}{0.000000,0.000000,0.000000}%
\pgfsetstrokecolor{textcolor}%
\pgfsetfillcolor{textcolor}%
\pgftext[x=0.307296in, y=4.033092in, left, base]{\color{textcolor}\rmfamily\fontsize{16.000000}{19.200000}\selectfont \(\displaystyle 0.38\)}%
\end{pgfscope}%
\begin{pgfscope}%
\pgfpathrectangle{\pgfqpoint{0.800000in}{0.528000in}}{\pgfqpoint{4.960000in}{3.696000in}}%
\pgfusepath{clip}%
\pgfsetbuttcap%
\pgfsetroundjoin%
\pgfsetlinewidth{1.505625pt}%
\definecolor{currentstroke}{rgb}{0.000000,0.000000,1.000000}%
\pgfsetstrokecolor{currentstroke}%
\pgfsetdash{}{0pt}%
\pgfpathmoveto{\pgfqpoint{1.213333in}{1.930766in}}%
\pgfpathlineto{\pgfqpoint{1.213333in}{2.203834in}}%
\pgfusepath{stroke}%
\end{pgfscope}%
\begin{pgfscope}%
\pgfpathrectangle{\pgfqpoint{0.800000in}{0.528000in}}{\pgfqpoint{4.960000in}{3.696000in}}%
\pgfusepath{clip}%
\pgfsetbuttcap%
\pgfsetroundjoin%
\pgfsetlinewidth{1.505625pt}%
\definecolor{currentstroke}{rgb}{0.000000,0.000000,1.000000}%
\pgfsetstrokecolor{currentstroke}%
\pgfsetdash{}{0pt}%
\pgfpathmoveto{\pgfqpoint{1.626667in}{2.192301in}}%
\pgfpathlineto{\pgfqpoint{1.626667in}{2.571729in}}%
\pgfusepath{stroke}%
\end{pgfscope}%
\begin{pgfscope}%
\pgfpathrectangle{\pgfqpoint{0.800000in}{0.528000in}}{\pgfqpoint{4.960000in}{3.696000in}}%
\pgfusepath{clip}%
\pgfsetbuttcap%
\pgfsetroundjoin%
\pgfsetlinewidth{1.505625pt}%
\definecolor{currentstroke}{rgb}{0.000000,0.000000,1.000000}%
\pgfsetstrokecolor{currentstroke}%
\pgfsetdash{}{0pt}%
\pgfpathmoveto{\pgfqpoint{2.040000in}{2.155559in}}%
\pgfpathlineto{\pgfqpoint{2.040000in}{2.443744in}}%
\pgfusepath{stroke}%
\end{pgfscope}%
\begin{pgfscope}%
\pgfpathrectangle{\pgfqpoint{0.800000in}{0.528000in}}{\pgfqpoint{4.960000in}{3.696000in}}%
\pgfusepath{clip}%
\pgfsetbuttcap%
\pgfsetroundjoin%
\pgfsetlinewidth{1.505625pt}%
\definecolor{currentstroke}{rgb}{0.000000,0.000000,1.000000}%
\pgfsetstrokecolor{currentstroke}%
\pgfsetdash{}{0pt}%
\pgfpathmoveto{\pgfqpoint{2.866667in}{0.696000in}}%
\pgfpathlineto{\pgfqpoint{2.866667in}{0.897047in}}%
\pgfusepath{stroke}%
\end{pgfscope}%
\begin{pgfscope}%
\pgfpathrectangle{\pgfqpoint{0.800000in}{0.528000in}}{\pgfqpoint{4.960000in}{3.696000in}}%
\pgfusepath{clip}%
\pgfsetbuttcap%
\pgfsetroundjoin%
\pgfsetlinewidth{1.505625pt}%
\definecolor{currentstroke}{rgb}{0.000000,0.000000,1.000000}%
\pgfsetstrokecolor{currentstroke}%
\pgfsetdash{}{0pt}%
\pgfpathmoveto{\pgfqpoint{3.280000in}{0.954698in}}%
\pgfpathlineto{\pgfqpoint{3.280000in}{1.167631in}}%
\pgfusepath{stroke}%
\end{pgfscope}%
\begin{pgfscope}%
\pgfpathrectangle{\pgfqpoint{0.800000in}{0.528000in}}{\pgfqpoint{4.960000in}{3.696000in}}%
\pgfusepath{clip}%
\pgfsetbuttcap%
\pgfsetroundjoin%
\pgfsetlinewidth{1.505625pt}%
\definecolor{currentstroke}{rgb}{0.000000,0.000000,1.000000}%
\pgfsetstrokecolor{currentstroke}%
\pgfsetdash{}{0pt}%
\pgfpathmoveto{\pgfqpoint{3.693333in}{0.892479in}}%
\pgfpathlineto{\pgfqpoint{3.693333in}{1.091324in}}%
\pgfusepath{stroke}%
\end{pgfscope}%
\begin{pgfscope}%
\pgfpathrectangle{\pgfqpoint{0.800000in}{0.528000in}}{\pgfqpoint{4.960000in}{3.696000in}}%
\pgfusepath{clip}%
\pgfsetbuttcap%
\pgfsetroundjoin%
\pgfsetlinewidth{1.505625pt}%
\definecolor{currentstroke}{rgb}{0.000000,0.000000,1.000000}%
\pgfsetstrokecolor{currentstroke}%
\pgfsetdash{}{0pt}%
\pgfpathmoveto{\pgfqpoint{4.520000in}{0.881098in}}%
\pgfpathlineto{\pgfqpoint{4.520000in}{1.106367in}}%
\pgfusepath{stroke}%
\end{pgfscope}%
\begin{pgfscope}%
\pgfpathrectangle{\pgfqpoint{0.800000in}{0.528000in}}{\pgfqpoint{4.960000in}{3.696000in}}%
\pgfusepath{clip}%
\pgfsetbuttcap%
\pgfsetroundjoin%
\pgfsetlinewidth{1.505625pt}%
\definecolor{currentstroke}{rgb}{0.000000,0.000000,1.000000}%
\pgfsetstrokecolor{currentstroke}%
\pgfsetdash{}{0pt}%
\pgfpathmoveto{\pgfqpoint{4.933333in}{1.159220in}}%
\pgfpathlineto{\pgfqpoint{4.933333in}{1.368364in}}%
\pgfusepath{stroke}%
\end{pgfscope}%
\begin{pgfscope}%
\pgfpathrectangle{\pgfqpoint{0.800000in}{0.528000in}}{\pgfqpoint{4.960000in}{3.696000in}}%
\pgfusepath{clip}%
\pgfsetbuttcap%
\pgfsetroundjoin%
\pgfsetlinewidth{1.505625pt}%
\definecolor{currentstroke}{rgb}{0.000000,0.000000,1.000000}%
\pgfsetstrokecolor{currentstroke}%
\pgfsetdash{}{0pt}%
\pgfpathmoveto{\pgfqpoint{5.346667in}{1.099250in}}%
\pgfpathlineto{\pgfqpoint{5.346667in}{1.292480in}}%
\pgfusepath{stroke}%
\end{pgfscope}%
\begin{pgfscope}%
\pgfpathrectangle{\pgfqpoint{0.800000in}{0.528000in}}{\pgfqpoint{4.960000in}{3.696000in}}%
\pgfusepath{clip}%
\pgfsetrectcap%
\pgfsetroundjoin%
\pgfsetlinewidth{1.505625pt}%
\definecolor{currentstroke}{rgb}{0.000000,0.000000,1.000000}%
\pgfsetstrokecolor{currentstroke}%
\pgfsetdash{}{0pt}%
\pgfpathmoveto{\pgfqpoint{1.213333in}{2.067300in}}%
\pgfusepath{stroke}%
\end{pgfscope}%
\begin{pgfscope}%
\pgfpathrectangle{\pgfqpoint{0.800000in}{0.528000in}}{\pgfqpoint{4.960000in}{3.696000in}}%
\pgfusepath{clip}%
\pgfsetbuttcap%
\pgfsetroundjoin%
\definecolor{currentfill}{rgb}{0.000000,0.000000,1.000000}%
\pgfsetfillcolor{currentfill}%
\pgfsetlinewidth{1.003750pt}%
\definecolor{currentstroke}{rgb}{0.000000,0.000000,1.000000}%
\pgfsetstrokecolor{currentstroke}%
\pgfsetdash{}{0pt}%
\pgfsys@defobject{currentmarker}{\pgfqpoint{-0.041667in}{-0.041667in}}{\pgfqpoint{0.041667in}{0.041667in}}{%
\pgfpathmoveto{\pgfqpoint{0.000000in}{-0.041667in}}%
\pgfpathcurveto{\pgfqpoint{0.011050in}{-0.041667in}}{\pgfqpoint{0.021649in}{-0.037276in}}{\pgfqpoint{0.029463in}{-0.029463in}}%
\pgfpathcurveto{\pgfqpoint{0.037276in}{-0.021649in}}{\pgfqpoint{0.041667in}{-0.011050in}}{\pgfqpoint{0.041667in}{0.000000in}}%
\pgfpathcurveto{\pgfqpoint{0.041667in}{0.011050in}}{\pgfqpoint{0.037276in}{0.021649in}}{\pgfqpoint{0.029463in}{0.029463in}}%
\pgfpathcurveto{\pgfqpoint{0.021649in}{0.037276in}}{\pgfqpoint{0.011050in}{0.041667in}}{\pgfqpoint{0.000000in}{0.041667in}}%
\pgfpathcurveto{\pgfqpoint{-0.011050in}{0.041667in}}{\pgfqpoint{-0.021649in}{0.037276in}}{\pgfqpoint{-0.029463in}{0.029463in}}%
\pgfpathcurveto{\pgfqpoint{-0.037276in}{0.021649in}}{\pgfqpoint{-0.041667in}{0.011050in}}{\pgfqpoint{-0.041667in}{0.000000in}}%
\pgfpathcurveto{\pgfqpoint{-0.041667in}{-0.011050in}}{\pgfqpoint{-0.037276in}{-0.021649in}}{\pgfqpoint{-0.029463in}{-0.029463in}}%
\pgfpathcurveto{\pgfqpoint{-0.021649in}{-0.037276in}}{\pgfqpoint{-0.011050in}{-0.041667in}}{\pgfqpoint{0.000000in}{-0.041667in}}%
\pgfpathclose%
\pgfusepath{stroke,fill}%
}%
\begin{pgfscope}%
\pgfsys@transformshift{1.213333in}{2.067300in}%
\pgfsys@useobject{currentmarker}{}%
\end{pgfscope}%
\end{pgfscope}%
\begin{pgfscope}%
\pgfpathrectangle{\pgfqpoint{0.800000in}{0.528000in}}{\pgfqpoint{4.960000in}{3.696000in}}%
\pgfusepath{clip}%
\pgfsetrectcap%
\pgfsetroundjoin%
\pgfsetlinewidth{1.505625pt}%
\definecolor{currentstroke}{rgb}{0.000000,0.000000,1.000000}%
\pgfsetstrokecolor{currentstroke}%
\pgfsetdash{}{0pt}%
\pgfpathmoveto{\pgfqpoint{1.626667in}{2.382015in}}%
\pgfusepath{stroke}%
\end{pgfscope}%
\begin{pgfscope}%
\pgfpathrectangle{\pgfqpoint{0.800000in}{0.528000in}}{\pgfqpoint{4.960000in}{3.696000in}}%
\pgfusepath{clip}%
\pgfsetbuttcap%
\pgfsetmiterjoin%
\definecolor{currentfill}{rgb}{0.000000,0.000000,1.000000}%
\pgfsetfillcolor{currentfill}%
\pgfsetlinewidth{1.003750pt}%
\definecolor{currentstroke}{rgb}{0.000000,0.000000,1.000000}%
\pgfsetstrokecolor{currentstroke}%
\pgfsetdash{}{0pt}%
\pgfsys@defobject{currentmarker}{\pgfqpoint{-0.041667in}{-0.041667in}}{\pgfqpoint{0.041667in}{0.041667in}}{%
\pgfpathmoveto{\pgfqpoint{0.000000in}{0.041667in}}%
\pgfpathlineto{\pgfqpoint{-0.041667in}{-0.041667in}}%
\pgfpathlineto{\pgfqpoint{0.041667in}{-0.041667in}}%
\pgfpathclose%
\pgfusepath{stroke,fill}%
}%
\begin{pgfscope}%
\pgfsys@transformshift{1.626667in}{2.382015in}%
\pgfsys@useobject{currentmarker}{}%
\end{pgfscope}%
\end{pgfscope}%
\begin{pgfscope}%
\pgfpathrectangle{\pgfqpoint{0.800000in}{0.528000in}}{\pgfqpoint{4.960000in}{3.696000in}}%
\pgfusepath{clip}%
\pgfsetrectcap%
\pgfsetroundjoin%
\pgfsetlinewidth{1.505625pt}%
\definecolor{currentstroke}{rgb}{0.000000,0.000000,1.000000}%
\pgfsetstrokecolor{currentstroke}%
\pgfsetdash{}{0pt}%
\pgfpathmoveto{\pgfqpoint{2.040000in}{2.299652in}}%
\pgfusepath{stroke}%
\end{pgfscope}%
\begin{pgfscope}%
\pgfpathrectangle{\pgfqpoint{0.800000in}{0.528000in}}{\pgfqpoint{4.960000in}{3.696000in}}%
\pgfusepath{clip}%
\pgfsetbuttcap%
\pgfsetmiterjoin%
\definecolor{currentfill}{rgb}{0.000000,0.000000,1.000000}%
\pgfsetfillcolor{currentfill}%
\pgfsetlinewidth{1.003750pt}%
\definecolor{currentstroke}{rgb}{0.000000,0.000000,1.000000}%
\pgfsetstrokecolor{currentstroke}%
\pgfsetdash{}{0pt}%
\pgfsys@defobject{currentmarker}{\pgfqpoint{-0.041667in}{-0.041667in}}{\pgfqpoint{0.041667in}{0.041667in}}{%
\pgfpathmoveto{\pgfqpoint{-0.041667in}{-0.041667in}}%
\pgfpathlineto{\pgfqpoint{0.041667in}{-0.041667in}}%
\pgfpathlineto{\pgfqpoint{0.041667in}{0.041667in}}%
\pgfpathlineto{\pgfqpoint{-0.041667in}{0.041667in}}%
\pgfpathclose%
\pgfusepath{stroke,fill}%
}%
\begin{pgfscope}%
\pgfsys@transformshift{2.040000in}{2.299652in}%
\pgfsys@useobject{currentmarker}{}%
\end{pgfscope}%
\end{pgfscope}%
\begin{pgfscope}%
\pgfpathrectangle{\pgfqpoint{0.800000in}{0.528000in}}{\pgfqpoint{4.960000in}{3.696000in}}%
\pgfusepath{clip}%
\pgfsetrectcap%
\pgfsetroundjoin%
\pgfsetlinewidth{1.505625pt}%
\definecolor{currentstroke}{rgb}{0.000000,0.000000,1.000000}%
\pgfsetstrokecolor{currentstroke}%
\pgfsetdash{}{0pt}%
\pgfpathmoveto{\pgfqpoint{2.866667in}{0.796524in}}%
\pgfusepath{stroke}%
\end{pgfscope}%
\begin{pgfscope}%
\pgfpathrectangle{\pgfqpoint{0.800000in}{0.528000in}}{\pgfqpoint{4.960000in}{3.696000in}}%
\pgfusepath{clip}%
\pgfsetbuttcap%
\pgfsetroundjoin%
\definecolor{currentfill}{rgb}{0.000000,0.000000,1.000000}%
\pgfsetfillcolor{currentfill}%
\pgfsetlinewidth{1.003750pt}%
\definecolor{currentstroke}{rgb}{0.000000,0.000000,1.000000}%
\pgfsetstrokecolor{currentstroke}%
\pgfsetdash{}{0pt}%
\pgfsys@defobject{currentmarker}{\pgfqpoint{-0.041667in}{-0.041667in}}{\pgfqpoint{0.041667in}{0.041667in}}{%
\pgfpathmoveto{\pgfqpoint{0.000000in}{-0.041667in}}%
\pgfpathcurveto{\pgfqpoint{0.011050in}{-0.041667in}}{\pgfqpoint{0.021649in}{-0.037276in}}{\pgfqpoint{0.029463in}{-0.029463in}}%
\pgfpathcurveto{\pgfqpoint{0.037276in}{-0.021649in}}{\pgfqpoint{0.041667in}{-0.011050in}}{\pgfqpoint{0.041667in}{0.000000in}}%
\pgfpathcurveto{\pgfqpoint{0.041667in}{0.011050in}}{\pgfqpoint{0.037276in}{0.021649in}}{\pgfqpoint{0.029463in}{0.029463in}}%
\pgfpathcurveto{\pgfqpoint{0.021649in}{0.037276in}}{\pgfqpoint{0.011050in}{0.041667in}}{\pgfqpoint{0.000000in}{0.041667in}}%
\pgfpathcurveto{\pgfqpoint{-0.011050in}{0.041667in}}{\pgfqpoint{-0.021649in}{0.037276in}}{\pgfqpoint{-0.029463in}{0.029463in}}%
\pgfpathcurveto{\pgfqpoint{-0.037276in}{0.021649in}}{\pgfqpoint{-0.041667in}{0.011050in}}{\pgfqpoint{-0.041667in}{0.000000in}}%
\pgfpathcurveto{\pgfqpoint{-0.041667in}{-0.011050in}}{\pgfqpoint{-0.037276in}{-0.021649in}}{\pgfqpoint{-0.029463in}{-0.029463in}}%
\pgfpathcurveto{\pgfqpoint{-0.021649in}{-0.037276in}}{\pgfqpoint{-0.011050in}{-0.041667in}}{\pgfqpoint{0.000000in}{-0.041667in}}%
\pgfpathclose%
\pgfusepath{stroke,fill}%
}%
\begin{pgfscope}%
\pgfsys@transformshift{2.866667in}{0.796524in}%
\pgfsys@useobject{currentmarker}{}%
\end{pgfscope}%
\end{pgfscope}%
\begin{pgfscope}%
\pgfpathrectangle{\pgfqpoint{0.800000in}{0.528000in}}{\pgfqpoint{4.960000in}{3.696000in}}%
\pgfusepath{clip}%
\pgfsetrectcap%
\pgfsetroundjoin%
\pgfsetlinewidth{1.505625pt}%
\definecolor{currentstroke}{rgb}{0.000000,0.000000,1.000000}%
\pgfsetstrokecolor{currentstroke}%
\pgfsetdash{}{0pt}%
\pgfpathmoveto{\pgfqpoint{3.280000in}{1.061164in}}%
\pgfusepath{stroke}%
\end{pgfscope}%
\begin{pgfscope}%
\pgfpathrectangle{\pgfqpoint{0.800000in}{0.528000in}}{\pgfqpoint{4.960000in}{3.696000in}}%
\pgfusepath{clip}%
\pgfsetbuttcap%
\pgfsetmiterjoin%
\definecolor{currentfill}{rgb}{0.000000,0.000000,1.000000}%
\pgfsetfillcolor{currentfill}%
\pgfsetlinewidth{1.003750pt}%
\definecolor{currentstroke}{rgb}{0.000000,0.000000,1.000000}%
\pgfsetstrokecolor{currentstroke}%
\pgfsetdash{}{0pt}%
\pgfsys@defobject{currentmarker}{\pgfqpoint{-0.041667in}{-0.041667in}}{\pgfqpoint{0.041667in}{0.041667in}}{%
\pgfpathmoveto{\pgfqpoint{0.000000in}{0.041667in}}%
\pgfpathlineto{\pgfqpoint{-0.041667in}{-0.041667in}}%
\pgfpathlineto{\pgfqpoint{0.041667in}{-0.041667in}}%
\pgfpathclose%
\pgfusepath{stroke,fill}%
}%
\begin{pgfscope}%
\pgfsys@transformshift{3.280000in}{1.061164in}%
\pgfsys@useobject{currentmarker}{}%
\end{pgfscope}%
\end{pgfscope}%
\begin{pgfscope}%
\pgfpathrectangle{\pgfqpoint{0.800000in}{0.528000in}}{\pgfqpoint{4.960000in}{3.696000in}}%
\pgfusepath{clip}%
\pgfsetrectcap%
\pgfsetroundjoin%
\pgfsetlinewidth{1.505625pt}%
\definecolor{currentstroke}{rgb}{0.000000,0.000000,1.000000}%
\pgfsetstrokecolor{currentstroke}%
\pgfsetdash{}{0pt}%
\pgfpathmoveto{\pgfqpoint{3.693333in}{0.991902in}}%
\pgfusepath{stroke}%
\end{pgfscope}%
\begin{pgfscope}%
\pgfpathrectangle{\pgfqpoint{0.800000in}{0.528000in}}{\pgfqpoint{4.960000in}{3.696000in}}%
\pgfusepath{clip}%
\pgfsetbuttcap%
\pgfsetmiterjoin%
\definecolor{currentfill}{rgb}{0.000000,0.000000,1.000000}%
\pgfsetfillcolor{currentfill}%
\pgfsetlinewidth{1.003750pt}%
\definecolor{currentstroke}{rgb}{0.000000,0.000000,1.000000}%
\pgfsetstrokecolor{currentstroke}%
\pgfsetdash{}{0pt}%
\pgfsys@defobject{currentmarker}{\pgfqpoint{-0.041667in}{-0.041667in}}{\pgfqpoint{0.041667in}{0.041667in}}{%
\pgfpathmoveto{\pgfqpoint{-0.041667in}{-0.041667in}}%
\pgfpathlineto{\pgfqpoint{0.041667in}{-0.041667in}}%
\pgfpathlineto{\pgfqpoint{0.041667in}{0.041667in}}%
\pgfpathlineto{\pgfqpoint{-0.041667in}{0.041667in}}%
\pgfpathclose%
\pgfusepath{stroke,fill}%
}%
\begin{pgfscope}%
\pgfsys@transformshift{3.693333in}{0.991902in}%
\pgfsys@useobject{currentmarker}{}%
\end{pgfscope}%
\end{pgfscope}%
\begin{pgfscope}%
\pgfpathrectangle{\pgfqpoint{0.800000in}{0.528000in}}{\pgfqpoint{4.960000in}{3.696000in}}%
\pgfusepath{clip}%
\pgfsetrectcap%
\pgfsetroundjoin%
\pgfsetlinewidth{1.505625pt}%
\definecolor{currentstroke}{rgb}{0.000000,0.000000,1.000000}%
\pgfsetstrokecolor{currentstroke}%
\pgfsetdash{}{0pt}%
\pgfpathmoveto{\pgfqpoint{4.520000in}{0.993733in}}%
\pgfusepath{stroke}%
\end{pgfscope}%
\begin{pgfscope}%
\pgfpathrectangle{\pgfqpoint{0.800000in}{0.528000in}}{\pgfqpoint{4.960000in}{3.696000in}}%
\pgfusepath{clip}%
\pgfsetbuttcap%
\pgfsetroundjoin%
\definecolor{currentfill}{rgb}{0.000000,0.000000,1.000000}%
\pgfsetfillcolor{currentfill}%
\pgfsetlinewidth{1.003750pt}%
\definecolor{currentstroke}{rgb}{0.000000,0.000000,1.000000}%
\pgfsetstrokecolor{currentstroke}%
\pgfsetdash{}{0pt}%
\pgfsys@defobject{currentmarker}{\pgfqpoint{-0.041667in}{-0.041667in}}{\pgfqpoint{0.041667in}{0.041667in}}{%
\pgfpathmoveto{\pgfqpoint{0.000000in}{-0.041667in}}%
\pgfpathcurveto{\pgfqpoint{0.011050in}{-0.041667in}}{\pgfqpoint{0.021649in}{-0.037276in}}{\pgfqpoint{0.029463in}{-0.029463in}}%
\pgfpathcurveto{\pgfqpoint{0.037276in}{-0.021649in}}{\pgfqpoint{0.041667in}{-0.011050in}}{\pgfqpoint{0.041667in}{0.000000in}}%
\pgfpathcurveto{\pgfqpoint{0.041667in}{0.011050in}}{\pgfqpoint{0.037276in}{0.021649in}}{\pgfqpoint{0.029463in}{0.029463in}}%
\pgfpathcurveto{\pgfqpoint{0.021649in}{0.037276in}}{\pgfqpoint{0.011050in}{0.041667in}}{\pgfqpoint{0.000000in}{0.041667in}}%
\pgfpathcurveto{\pgfqpoint{-0.011050in}{0.041667in}}{\pgfqpoint{-0.021649in}{0.037276in}}{\pgfqpoint{-0.029463in}{0.029463in}}%
\pgfpathcurveto{\pgfqpoint{-0.037276in}{0.021649in}}{\pgfqpoint{-0.041667in}{0.011050in}}{\pgfqpoint{-0.041667in}{0.000000in}}%
\pgfpathcurveto{\pgfqpoint{-0.041667in}{-0.011050in}}{\pgfqpoint{-0.037276in}{-0.021649in}}{\pgfqpoint{-0.029463in}{-0.029463in}}%
\pgfpathcurveto{\pgfqpoint{-0.021649in}{-0.037276in}}{\pgfqpoint{-0.011050in}{-0.041667in}}{\pgfqpoint{0.000000in}{-0.041667in}}%
\pgfpathclose%
\pgfusepath{stroke,fill}%
}%
\begin{pgfscope}%
\pgfsys@transformshift{4.520000in}{0.993733in}%
\pgfsys@useobject{currentmarker}{}%
\end{pgfscope}%
\end{pgfscope}%
\begin{pgfscope}%
\pgfpathrectangle{\pgfqpoint{0.800000in}{0.528000in}}{\pgfqpoint{4.960000in}{3.696000in}}%
\pgfusepath{clip}%
\pgfsetrectcap%
\pgfsetroundjoin%
\pgfsetlinewidth{1.505625pt}%
\definecolor{currentstroke}{rgb}{0.000000,0.000000,1.000000}%
\pgfsetstrokecolor{currentstroke}%
\pgfsetdash{}{0pt}%
\pgfpathmoveto{\pgfqpoint{4.933333in}{1.263792in}}%
\pgfusepath{stroke}%
\end{pgfscope}%
\begin{pgfscope}%
\pgfpathrectangle{\pgfqpoint{0.800000in}{0.528000in}}{\pgfqpoint{4.960000in}{3.696000in}}%
\pgfusepath{clip}%
\pgfsetbuttcap%
\pgfsetmiterjoin%
\definecolor{currentfill}{rgb}{0.000000,0.000000,1.000000}%
\pgfsetfillcolor{currentfill}%
\pgfsetlinewidth{1.003750pt}%
\definecolor{currentstroke}{rgb}{0.000000,0.000000,1.000000}%
\pgfsetstrokecolor{currentstroke}%
\pgfsetdash{}{0pt}%
\pgfsys@defobject{currentmarker}{\pgfqpoint{-0.041667in}{-0.041667in}}{\pgfqpoint{0.041667in}{0.041667in}}{%
\pgfpathmoveto{\pgfqpoint{0.000000in}{0.041667in}}%
\pgfpathlineto{\pgfqpoint{-0.041667in}{-0.041667in}}%
\pgfpathlineto{\pgfqpoint{0.041667in}{-0.041667in}}%
\pgfpathclose%
\pgfusepath{stroke,fill}%
}%
\begin{pgfscope}%
\pgfsys@transformshift{4.933333in}{1.263792in}%
\pgfsys@useobject{currentmarker}{}%
\end{pgfscope}%
\end{pgfscope}%
\begin{pgfscope}%
\pgfpathrectangle{\pgfqpoint{0.800000in}{0.528000in}}{\pgfqpoint{4.960000in}{3.696000in}}%
\pgfusepath{clip}%
\pgfsetrectcap%
\pgfsetroundjoin%
\pgfsetlinewidth{1.505625pt}%
\definecolor{currentstroke}{rgb}{0.000000,0.000000,1.000000}%
\pgfsetstrokecolor{currentstroke}%
\pgfsetdash{}{0pt}%
\pgfpathmoveto{\pgfqpoint{5.346667in}{1.195865in}}%
\pgfusepath{stroke}%
\end{pgfscope}%
\begin{pgfscope}%
\pgfpathrectangle{\pgfqpoint{0.800000in}{0.528000in}}{\pgfqpoint{4.960000in}{3.696000in}}%
\pgfusepath{clip}%
\pgfsetbuttcap%
\pgfsetmiterjoin%
\definecolor{currentfill}{rgb}{0.000000,0.000000,1.000000}%
\pgfsetfillcolor{currentfill}%
\pgfsetlinewidth{1.003750pt}%
\definecolor{currentstroke}{rgb}{0.000000,0.000000,1.000000}%
\pgfsetstrokecolor{currentstroke}%
\pgfsetdash{}{0pt}%
\pgfsys@defobject{currentmarker}{\pgfqpoint{-0.041667in}{-0.041667in}}{\pgfqpoint{0.041667in}{0.041667in}}{%
\pgfpathmoveto{\pgfqpoint{-0.041667in}{-0.041667in}}%
\pgfpathlineto{\pgfqpoint{0.041667in}{-0.041667in}}%
\pgfpathlineto{\pgfqpoint{0.041667in}{0.041667in}}%
\pgfpathlineto{\pgfqpoint{-0.041667in}{0.041667in}}%
\pgfpathclose%
\pgfusepath{stroke,fill}%
}%
\begin{pgfscope}%
\pgfsys@transformshift{5.346667in}{1.195865in}%
\pgfsys@useobject{currentmarker}{}%
\end{pgfscope}%
\end{pgfscope}%
\begin{pgfscope}%
\pgfpathrectangle{\pgfqpoint{0.800000in}{0.528000in}}{\pgfqpoint{4.960000in}{3.696000in}}%
\pgfusepath{clip}%
\pgfsetrectcap%
\pgfsetroundjoin%
\pgfsetlinewidth{1.505625pt}%
\definecolor{currentstroke}{rgb}{0.000000,0.000000,1.000000}%
\pgfsetstrokecolor{currentstroke}%
\pgfsetdash{}{0pt}%
\pgfpathmoveto{\pgfqpoint{1.213333in}{2.067300in}}%
\pgfusepath{stroke}%
\end{pgfscope}%
\begin{pgfscope}%
\pgfpathrectangle{\pgfqpoint{0.800000in}{0.528000in}}{\pgfqpoint{4.960000in}{3.696000in}}%
\pgfusepath{clip}%
\pgfsetrectcap%
\pgfsetroundjoin%
\pgfsetlinewidth{1.505625pt}%
\definecolor{currentstroke}{rgb}{0.000000,0.000000,1.000000}%
\pgfsetstrokecolor{currentstroke}%
\pgfsetdash{}{0pt}%
\pgfpathmoveto{\pgfqpoint{1.626667in}{2.382015in}}%
\pgfusepath{stroke}%
\end{pgfscope}%
\begin{pgfscope}%
\pgfpathrectangle{\pgfqpoint{0.800000in}{0.528000in}}{\pgfqpoint{4.960000in}{3.696000in}}%
\pgfusepath{clip}%
\pgfsetrectcap%
\pgfsetroundjoin%
\pgfsetlinewidth{1.505625pt}%
\definecolor{currentstroke}{rgb}{0.000000,0.000000,1.000000}%
\pgfsetstrokecolor{currentstroke}%
\pgfsetdash{}{0pt}%
\pgfpathmoveto{\pgfqpoint{2.040000in}{2.299652in}}%
\pgfusepath{stroke}%
\end{pgfscope}%
\begin{pgfscope}%
\pgfpathrectangle{\pgfqpoint{0.800000in}{0.528000in}}{\pgfqpoint{4.960000in}{3.696000in}}%
\pgfusepath{clip}%
\pgfsetrectcap%
\pgfsetroundjoin%
\pgfsetlinewidth{1.505625pt}%
\definecolor{currentstroke}{rgb}{0.000000,0.000000,1.000000}%
\pgfsetstrokecolor{currentstroke}%
\pgfsetdash{}{0pt}%
\pgfpathmoveto{\pgfqpoint{2.866667in}{0.796524in}}%
\pgfusepath{stroke}%
\end{pgfscope}%
\begin{pgfscope}%
\pgfpathrectangle{\pgfqpoint{0.800000in}{0.528000in}}{\pgfqpoint{4.960000in}{3.696000in}}%
\pgfusepath{clip}%
\pgfsetrectcap%
\pgfsetroundjoin%
\pgfsetlinewidth{1.505625pt}%
\definecolor{currentstroke}{rgb}{0.000000,0.000000,1.000000}%
\pgfsetstrokecolor{currentstroke}%
\pgfsetdash{}{0pt}%
\pgfpathmoveto{\pgfqpoint{3.280000in}{1.061164in}}%
\pgfusepath{stroke}%
\end{pgfscope}%
\begin{pgfscope}%
\pgfpathrectangle{\pgfqpoint{0.800000in}{0.528000in}}{\pgfqpoint{4.960000in}{3.696000in}}%
\pgfusepath{clip}%
\pgfsetrectcap%
\pgfsetroundjoin%
\pgfsetlinewidth{1.505625pt}%
\definecolor{currentstroke}{rgb}{0.000000,0.000000,1.000000}%
\pgfsetstrokecolor{currentstroke}%
\pgfsetdash{}{0pt}%
\pgfpathmoveto{\pgfqpoint{3.693333in}{0.991902in}}%
\pgfusepath{stroke}%
\end{pgfscope}%
\begin{pgfscope}%
\pgfpathrectangle{\pgfqpoint{0.800000in}{0.528000in}}{\pgfqpoint{4.960000in}{3.696000in}}%
\pgfusepath{clip}%
\pgfsetrectcap%
\pgfsetroundjoin%
\pgfsetlinewidth{1.505625pt}%
\definecolor{currentstroke}{rgb}{0.000000,0.000000,1.000000}%
\pgfsetstrokecolor{currentstroke}%
\pgfsetdash{}{0pt}%
\pgfpathmoveto{\pgfqpoint{4.520000in}{0.993733in}}%
\pgfusepath{stroke}%
\end{pgfscope}%
\begin{pgfscope}%
\pgfpathrectangle{\pgfqpoint{0.800000in}{0.528000in}}{\pgfqpoint{4.960000in}{3.696000in}}%
\pgfusepath{clip}%
\pgfsetrectcap%
\pgfsetroundjoin%
\pgfsetlinewidth{1.505625pt}%
\definecolor{currentstroke}{rgb}{0.000000,0.000000,1.000000}%
\pgfsetstrokecolor{currentstroke}%
\pgfsetdash{}{0pt}%
\pgfpathmoveto{\pgfqpoint{4.933333in}{1.263792in}}%
\pgfusepath{stroke}%
\end{pgfscope}%
\begin{pgfscope}%
\pgfpathrectangle{\pgfqpoint{0.800000in}{0.528000in}}{\pgfqpoint{4.960000in}{3.696000in}}%
\pgfusepath{clip}%
\pgfsetrectcap%
\pgfsetroundjoin%
\pgfsetlinewidth{1.505625pt}%
\definecolor{currentstroke}{rgb}{0.000000,0.000000,1.000000}%
\pgfsetstrokecolor{currentstroke}%
\pgfsetdash{}{0pt}%
\pgfpathmoveto{\pgfqpoint{5.346667in}{1.195865in}}%
\pgfusepath{stroke}%
\end{pgfscope}%
\begin{pgfscope}%
\pgfsetrectcap%
\pgfsetmiterjoin%
\pgfsetlinewidth{0.803000pt}%
\definecolor{currentstroke}{rgb}{0.000000,0.000000,0.000000}%
\pgfsetstrokecolor{currentstroke}%
\pgfsetdash{}{0pt}%
\pgfpathmoveto{\pgfqpoint{0.800000in}{0.528000in}}%
\pgfpathlineto{\pgfqpoint{0.800000in}{4.224000in}}%
\pgfusepath{stroke}%
\end{pgfscope}%
\begin{pgfscope}%
\pgfsetrectcap%
\pgfsetmiterjoin%
\pgfsetlinewidth{0.803000pt}%
\definecolor{currentstroke}{rgb}{0.000000,0.000000,0.000000}%
\pgfsetstrokecolor{currentstroke}%
\pgfsetdash{}{0pt}%
\pgfpathmoveto{\pgfqpoint{5.760000in}{0.528000in}}%
\pgfpathlineto{\pgfqpoint{5.760000in}{4.224000in}}%
\pgfusepath{stroke}%
\end{pgfscope}%
\begin{pgfscope}%
\pgfsetrectcap%
\pgfsetmiterjoin%
\pgfsetlinewidth{0.803000pt}%
\definecolor{currentstroke}{rgb}{0.000000,0.000000,0.000000}%
\pgfsetstrokecolor{currentstroke}%
\pgfsetdash{}{0pt}%
\pgfpathmoveto{\pgfqpoint{0.800000in}{0.528000in}}%
\pgfpathlineto{\pgfqpoint{5.760000in}{0.528000in}}%
\pgfusepath{stroke}%
\end{pgfscope}%
\begin{pgfscope}%
\pgfsetrectcap%
\pgfsetmiterjoin%
\pgfsetlinewidth{0.803000pt}%
\definecolor{currentstroke}{rgb}{0.000000,0.000000,0.000000}%
\pgfsetstrokecolor{currentstroke}%
\pgfsetdash{}{0pt}%
\pgfpathmoveto{\pgfqpoint{0.800000in}{4.224000in}}%
\pgfpathlineto{\pgfqpoint{5.760000in}{4.224000in}}%
\pgfusepath{stroke}%
\end{pgfscope}%
\begin{pgfscope}%
\pgfsetbuttcap%
\pgfsetmiterjoin%
\definecolor{currentfill}{rgb}{1.000000,1.000000,1.000000}%
\pgfsetfillcolor{currentfill}%
\pgfsetfillopacity{0.800000}%
\pgfsetlinewidth{1.003750pt}%
\definecolor{currentstroke}{rgb}{0.800000,0.800000,0.800000}%
\pgfsetstrokecolor{currentstroke}%
\pgfsetstrokeopacity{0.800000}%
\pgfsetdash{}{0pt}%
\pgfpathmoveto{\pgfqpoint{3.851195in}{2.693447in}}%
\pgfpathlineto{\pgfqpoint{5.623889in}{2.693447in}}%
\pgfpathquadraticcurveto{\pgfqpoint{5.662778in}{2.693447in}}{\pgfqpoint{5.662778in}{2.732335in}}%
\pgfpathlineto{\pgfqpoint{5.662778in}{4.087889in}}%
\pgfpathquadraticcurveto{\pgfqpoint{5.662778in}{4.126778in}}{\pgfqpoint{5.623889in}{4.126778in}}%
\pgfpathlineto{\pgfqpoint{3.851195in}{4.126778in}}%
\pgfpathquadraticcurveto{\pgfqpoint{3.812306in}{4.126778in}}{\pgfqpoint{3.812306in}{4.087889in}}%
\pgfpathlineto{\pgfqpoint{3.812306in}{2.732335in}}%
\pgfpathquadraticcurveto{\pgfqpoint{3.812306in}{2.693447in}}{\pgfqpoint{3.851195in}{2.693447in}}%
\pgfpathclose%
\pgfusepath{stroke,fill}%
\end{pgfscope}%
\begin{pgfscope}%
\pgfsetrectcap%
\pgfsetroundjoin%
\pgfsetlinewidth{1.505625pt}%
\definecolor{currentstroke}{rgb}{0.000000,0.000000,1.000000}%
\pgfsetstrokecolor{currentstroke}%
\pgfsetdash{}{0pt}%
\pgfpathmoveto{\pgfqpoint{3.890084in}{3.978167in}}%
\pgfpathlineto{\pgfqpoint{4.278973in}{3.978167in}}%
\pgfusepath{stroke}%
\end{pgfscope}%
\begin{pgfscope}%
\pgfsetbuttcap%
\pgfsetroundjoin%
\definecolor{currentfill}{rgb}{0.000000,0.000000,1.000000}%
\pgfsetfillcolor{currentfill}%
\pgfsetlinewidth{1.003750pt}%
\definecolor{currentstroke}{rgb}{0.000000,0.000000,1.000000}%
\pgfsetstrokecolor{currentstroke}%
\pgfsetdash{}{0pt}%
\pgfsys@defobject{currentmarker}{\pgfqpoint{-0.041667in}{-0.041667in}}{\pgfqpoint{0.041667in}{0.041667in}}{%
\pgfpathmoveto{\pgfqpoint{0.000000in}{-0.041667in}}%
\pgfpathcurveto{\pgfqpoint{0.011050in}{-0.041667in}}{\pgfqpoint{0.021649in}{-0.037276in}}{\pgfqpoint{0.029463in}{-0.029463in}}%
\pgfpathcurveto{\pgfqpoint{0.037276in}{-0.021649in}}{\pgfqpoint{0.041667in}{-0.011050in}}{\pgfqpoint{0.041667in}{0.000000in}}%
\pgfpathcurveto{\pgfqpoint{0.041667in}{0.011050in}}{\pgfqpoint{0.037276in}{0.021649in}}{\pgfqpoint{0.029463in}{0.029463in}}%
\pgfpathcurveto{\pgfqpoint{0.021649in}{0.037276in}}{\pgfqpoint{0.011050in}{0.041667in}}{\pgfqpoint{0.000000in}{0.041667in}}%
\pgfpathcurveto{\pgfqpoint{-0.011050in}{0.041667in}}{\pgfqpoint{-0.021649in}{0.037276in}}{\pgfqpoint{-0.029463in}{0.029463in}}%
\pgfpathcurveto{\pgfqpoint{-0.037276in}{0.021649in}}{\pgfqpoint{-0.041667in}{0.011050in}}{\pgfqpoint{-0.041667in}{0.000000in}}%
\pgfpathcurveto{\pgfqpoint{-0.041667in}{-0.011050in}}{\pgfqpoint{-0.037276in}{-0.021649in}}{\pgfqpoint{-0.029463in}{-0.029463in}}%
\pgfpathcurveto{\pgfqpoint{-0.021649in}{-0.037276in}}{\pgfqpoint{-0.011050in}{-0.041667in}}{\pgfqpoint{0.000000in}{-0.041667in}}%
\pgfpathclose%
\pgfusepath{stroke,fill}%
}%
\begin{pgfscope}%
\pgfsys@transformshift{4.084528in}{3.978167in}%
\pgfsys@useobject{currentmarker}{}%
\end{pgfscope}%
\end{pgfscope}%
\begin{pgfscope}%
\definecolor{textcolor}{rgb}{0.000000,0.000000,0.000000}%
\pgfsetstrokecolor{textcolor}%
\pgfsetfillcolor{textcolor}%
\pgftext[x=4.434528in,y=3.910111in,left,base]{\color{textcolor}\rmfamily\fontsize{14.000000}{16.800000}\selectfont Scenario IV}%
\end{pgfscope}%
\begin{pgfscope}%
\pgfsetrectcap%
\pgfsetroundjoin%
\pgfsetlinewidth{1.505625pt}%
\definecolor{currentstroke}{rgb}{0.000000,0.000000,1.000000}%
\pgfsetstrokecolor{currentstroke}%
\pgfsetdash{}{0pt}%
\pgfpathmoveto{\pgfqpoint{3.890084in}{3.703167in}}%
\pgfpathlineto{\pgfqpoint{4.278973in}{3.703167in}}%
\pgfusepath{stroke}%
\end{pgfscope}%
\begin{pgfscope}%
\pgfsetbuttcap%
\pgfsetmiterjoin%
\definecolor{currentfill}{rgb}{0.000000,0.000000,1.000000}%
\pgfsetfillcolor{currentfill}%
\pgfsetlinewidth{1.003750pt}%
\definecolor{currentstroke}{rgb}{0.000000,0.000000,1.000000}%
\pgfsetstrokecolor{currentstroke}%
\pgfsetdash{}{0pt}%
\pgfsys@defobject{currentmarker}{\pgfqpoint{-0.041667in}{-0.041667in}}{\pgfqpoint{0.041667in}{0.041667in}}{%
\pgfpathmoveto{\pgfqpoint{0.000000in}{0.041667in}}%
\pgfpathlineto{\pgfqpoint{-0.041667in}{-0.041667in}}%
\pgfpathlineto{\pgfqpoint{0.041667in}{-0.041667in}}%
\pgfpathclose%
\pgfusepath{stroke,fill}%
}%
\begin{pgfscope}%
\pgfsys@transformshift{4.084528in}{3.703167in}%
\pgfsys@useobject{currentmarker}{}%
\end{pgfscope}%
\end{pgfscope}%
\begin{pgfscope}%
\definecolor{textcolor}{rgb}{0.000000,0.000000,0.000000}%
\pgfsetstrokecolor{textcolor}%
\pgfsetfillcolor{textcolor}%
\pgftext[x=4.434528in,y=3.635112in,left,base]{\color{textcolor}\rmfamily\fontsize{14.000000}{16.800000}\selectfont Scenario VII}%
\end{pgfscope}%
\begin{pgfscope}%
\pgfsetrectcap%
\pgfsetroundjoin%
\pgfsetlinewidth{1.505625pt}%
\definecolor{currentstroke}{rgb}{0.000000,0.000000,1.000000}%
\pgfsetstrokecolor{currentstroke}%
\pgfsetdash{}{0pt}%
\pgfpathmoveto{\pgfqpoint{3.890084in}{3.428168in}}%
\pgfpathlineto{\pgfqpoint{4.278973in}{3.428168in}}%
\pgfusepath{stroke}%
\end{pgfscope}%
\begin{pgfscope}%
\pgfsetbuttcap%
\pgfsetmiterjoin%
\definecolor{currentfill}{rgb}{0.000000,0.000000,1.000000}%
\pgfsetfillcolor{currentfill}%
\pgfsetlinewidth{1.003750pt}%
\definecolor{currentstroke}{rgb}{0.000000,0.000000,1.000000}%
\pgfsetstrokecolor{currentstroke}%
\pgfsetdash{}{0pt}%
\pgfsys@defobject{currentmarker}{\pgfqpoint{-0.041667in}{-0.041667in}}{\pgfqpoint{0.041667in}{0.041667in}}{%
\pgfpathmoveto{\pgfqpoint{-0.041667in}{-0.041667in}}%
\pgfpathlineto{\pgfqpoint{0.041667in}{-0.041667in}}%
\pgfpathlineto{\pgfqpoint{0.041667in}{0.041667in}}%
\pgfpathlineto{\pgfqpoint{-0.041667in}{0.041667in}}%
\pgfpathclose%
\pgfusepath{stroke,fill}%
}%
\begin{pgfscope}%
\pgfsys@transformshift{4.084528in}{3.428168in}%
\pgfsys@useobject{currentmarker}{}%
\end{pgfscope}%
\end{pgfscope}%
\begin{pgfscope}%
\definecolor{textcolor}{rgb}{0.000000,0.000000,0.000000}%
\pgfsetstrokecolor{textcolor}%
\pgfsetfillcolor{textcolor}%
\pgftext[x=4.434528in,y=3.360112in,left,base]{\color{textcolor}\rmfamily\fontsize{14.000000}{16.800000}\selectfont Scenario IX}%
\end{pgfscope}%
\begin{pgfscope}%
\pgfsetbuttcap%
\pgfsetmiterjoin%
\definecolor{currentfill}{rgb}{1.000000,0.647059,0.000000}%
\pgfsetfillcolor{currentfill}%
\pgfsetfillopacity{0.700000}%
\pgfsetlinewidth{1.003750pt}%
\definecolor{currentstroke}{rgb}{1.000000,0.647059,0.000000}%
\pgfsetstrokecolor{currentstroke}%
\pgfsetstrokeopacity{0.700000}%
\pgfsetdash{}{0pt}%
\pgfpathmoveto{\pgfqpoint{3.890084in}{3.085113in}}%
\pgfpathlineto{\pgfqpoint{4.278973in}{3.085113in}}%
\pgfpathlineto{\pgfqpoint{4.278973in}{3.221224in}}%
\pgfpathlineto{\pgfqpoint{3.890084in}{3.221224in}}%
\pgfpathclose%
\pgfusepath{stroke,fill}%
\end{pgfscope}%
\begin{pgfscope}%
\definecolor{textcolor}{rgb}{0.000000,0.000000,0.000000}%
\pgfsetstrokecolor{textcolor}%
\pgfsetfillcolor{textcolor}%
\pgftext[x=4.434528in,y=3.085113in,left,base]{\color{textcolor}\rmfamily\fontsize{14.000000}{16.800000}\selectfont SM}%
\end{pgfscope}%
\begin{pgfscope}%
\pgfsetbuttcap%
\pgfsetmiterjoin%
\definecolor{currentfill}{rgb}{0.000000,0.501961,0.000000}%
\pgfsetfillcolor{currentfill}%
\pgfsetfillopacity{0.700000}%
\pgfsetlinewidth{1.003750pt}%
\definecolor{currentstroke}{rgb}{0.000000,0.501961,0.000000}%
\pgfsetstrokecolor{currentstroke}%
\pgfsetstrokeopacity{0.700000}%
\pgfsetdash{}{0pt}%
\pgfpathmoveto{\pgfqpoint{3.890084in}{2.810113in}}%
\pgfpathlineto{\pgfqpoint{4.278973in}{2.810113in}}%
\pgfpathlineto{\pgfqpoint{4.278973in}{2.946224in}}%
\pgfpathlineto{\pgfqpoint{3.890084in}{2.946224in}}%
\pgfpathclose%
\pgfusepath{stroke,fill}%
\end{pgfscope}%
\begin{pgfscope}%
\definecolor{textcolor}{rgb}{0.000000,0.000000,0.000000}%
\pgfsetstrokecolor{textcolor}%
\pgfsetfillcolor{textcolor}%
\pgftext[x=4.434528in,y=2.810113in,left,base]{\color{textcolor}\rmfamily\fontsize{14.000000}{16.800000}\selectfont Experimental}%
\end{pgfscope}%
\end{pgfpicture}%
\makeatother%
\endgroup%

%% file: pullsVII_21.tex
\begin{longtable}{|c|c|c|c|c|}\hline
 & Observable &	 NP prediction &	 NP pull & SM pull\endhead\hline
0 &	 $\langle \frac{d\overline{\mathrm{BR}}}{dq^2} \rangle(B_s\to \phi \mu^+\mu^-)^{[1.0,\ 6.0]}$ &	 $4.864\times 10^{-8}$ &	 \cellcolor{green!35} 3.1 $ \sigma$ &	 3.8 $ \sigma$ \\ \hline
1 &	 $a_\mu$ &	 0.0011659 &	 \cellcolor{green!0} 3.5 $ \sigma$ &	 3.5 $ \sigma$ \\ \hline
2 &	 $\langle P_5^\prime\rangle(B^0\to K^{\ast 0}\mu^+\mu^-)^{[4,\ 6]}$ &	 -0.73724 &	 \cellcolor{green!6} 3.2 $ \sigma$ &	 3.3 $ \sigma$ \\ \hline
3 &	 $\epsilon^\prime/\epsilon$ &	 $-2.9466\times 10^{-5}$ &	 \cellcolor{green!0} 2.7 $ \sigma$ &	 2.7 $ \sigma$ \\ \hline
4 &	 $R_{\tau \ell}(B\to D^{\ast}\ell^+\nu)$ &	 0.26431 &	 \cellcolor{green!30} 2.1 $ \sigma$ &	 2.7 $ \sigma$ \\ \hline
5 &	 $\mathrm{BR}(W^\pm\to \tau^\pm\nu)$ &	 0.1082 &	 \cellcolor{red!5} 2.7 $ \sigma$ &	 2.6 $ \sigma$ \\ \hline
6 &	 $\langle R_{\mu e} \rangle(B^0\to K^{\ast 0}\ell^+\ell^-)^{[1.1,\ 6.0]}$ &	 0.80189 &	 \cellcolor{green!50} 1.1 $ \sigma$ &	 2.5 $ \sigma$ \\ \hline
7 &	 $A_\mathrm{FB}^{0, b}$ &	 0.10365 &	 \cellcolor{red!18} 2.8 $ \sigma$ &	 2.4 $ \sigma$ \\ \hline
8 &	 $\langle R_{\mu e} \rangle(B^0\to K^{\ast 0}\ell^+\ell^-)^{[0.045,\ 1.1]}$ &	 0.87107 &	 \cellcolor{green!20} 2 $ \sigma$ &	 2.4 $ \sigma$ \\ \hline
9 &	 $\langle R_{\mu e} \rangle(B^\pm\to K^\pm \ell^+\ell^-)^{[1.1,\ 6.0]}$ &	 0.8 &	 \cellcolor{green!50} 0.8 $ \sigma$ &	 2.3 $ \sigma$ \\ \hline
10 &	 $\frac{\langle \mathrm{BR} \rangle}{\mathrm{BR}}(B\to D^\ast\tau^+\nu)^{[10.4,\ 10.93]}$ &	 0.019509 &	 \cellcolor{green!0} 2.3 $ \sigma$ &	 2.3 $ \sigma$ \\ \hline
11 &	 $\langle \frac{d\mathrm{BR}}{dq^2} \rangle(B^+\to K^{\ast +}\mu^+\mu^-)^{[15.0,\ 19.0]}$ &	 $5.7706\times 10^{-8}$ &	 \cellcolor{green!31} 1.7 $ \sigma$ &	 2.3 $ \sigma$ \\ \hline
12 &	 $\langle P_2\rangle(B^0\to K^{\ast 0}\mu^+\mu^-)^{[4,\ 6]}$ &	 0.26583 &	 \cellcolor{green!11} 2 $ \sigma$ &	 2.3 $ \sigma$ \\ \hline
13 &	 $\overline{\mathrm{BR}}(B_s\to \mu^+\mu^-)$ &	 $3.31\times 10^{-9}$ &	 \cellcolor{green!40} 1.4 $ \sigma$ &	 2.2 $ \sigma$ \\ \hline
14 &	 $A_ e$ &	 0.14785 &	 \cellcolor{green!21} 1.8 $ \sigma$ &	 2.2 $ \sigma$ \\ \hline
15 &	 $\langle \frac{d\mathrm{BR}}{dq^2} \rangle(B^0\to K^{\ast 0}\mu^+\mu^-)^{[15.0,\ 19.0]}$ &	 $5.3257\times 10^{-8}$ &	 \cellcolor{green!42} 1.4 $ \sigma$ &	 2.2 $ \sigma$ \\ \hline
16 &	 $\langle \frac{d\overline{\mathrm{BR}}}{dq^2} \rangle(B_s\to \phi \mu^+\mu^-)^{[15.0,\ 19.0]}$ &	 $4.9957\times 10^{-8}$ &	 \cellcolor{green!41} 1.4 $ \sigma$ &	 2.2 $ \sigma$ \\ \hline
17 &	 $\langle \frac{d\mathrm{BR}}{dq^2} \rangle(B^+\to K^{\ast +}\mu^+\mu^-)^{[4.0,\ 6.0]}$ &	 $4.8554\times 10^{-8}$ &	 \cellcolor{green!21} 1.7 $ \sigma$ &	 2.1 $ \sigma$ \\ \hline
18 &	 $R_{\tau \mu}(B\to D^{\ast}\ell^+\nu)$ &	 0.27198 &	 \cellcolor{green!31} 1.4 $ \sigma$ &	 2.1 $ \sigma$ \\ \hline
19 &	 $\langle A_\mathrm{FB}^{\ell h}\rangle(\Lambda_b\to\Lambda \mu^+\mu^-)^{[15,\ 20]}$ &	 0.16297 &	 \cellcolor{red!0} 2 $ \sigma$ &	 2 $ \sigma$ \\ \hline
20 &	 $\vert\epsilon_K\vert$ &	 0.0018127 &	 \cellcolor{green!0} 2.1 $ \sigma$ &	 2.1 $ \sigma$ \\ \hline
21 &	 $\mathrm{BR}(B^\pm\to K^\pm \tau^+\tau^-)$ &	 $1.9363\times 10^{-7}$ &	 \cellcolor{green!0} 2 $ \sigma$ &	 2 $ \sigma$ \\ \hline
22 &	 $\mathrm{BR}(K_L\to e^+e^-)$ &	 $1.9327\times 10^{-13}$ &	 \cellcolor{red!0} 2.1 $ \sigma$ &	 2.1 $ \sigma$ \\ \hline
23 &	 $\langle \frac{d\mathrm{BR}}{dq^2} \rangle(B^\pm\to K^\pm \mu^+\mu^-)^{[4.0,\ 5.0]}$ &	 $3.1212\times 10^{-8}$ &	 \cellcolor{green!28} 1.5 $ \sigma$ &	 2 $ \sigma$ \\ \hline
24 &	 $\frac{\langle \mathrm{BR} \rangle}{\mathrm{BR}}(B\to D^\ast\tau^+\nu)^{[5.07,\ 5.6]}$ &	 0.059848 &	 \cellcolor{green!0} 2 $ \sigma$ &	 2 $ \sigma$ \\ \hline
25 &	 $\mathrm{BR}(\tau^-\to \mu^- \nu\bar\nu)$ &	 0.17277 &	 \cellcolor{red!12} 2.3 $ \sigma$ &	 2 $ \sigma$ \\ \hline
26 &	 $\langle \frac{d\mathrm{BR}}{dq^2} \rangle(B^0\to K^0\mu^+\mu^-)^{[15.0,\ 22.0]}$ &	 $1.2492\times 10^{-8}$ &	 \cellcolor{green!31} 1.3 $ \sigma$ &	 1.9 $ \sigma$ \\ \hline
27 &	 $\langle \frac{d\mathrm{BR}}{dq^2} \rangle(B^0\to K^0\mu^+\mu^-)^{[4.0,\ 6.0]}$ &	 $2.8846\times 10^{-8}$ &	 \cellcolor{green!21} 1.5 $ \sigma$ &	 1.9 $ \sigma$ \\ \hline
28 &	 $a_e$ &	 0.0011597 &	 1.9 $ \sigma$ &	 1.9 $ \sigma$ \\ \hline
29 &	 $\frac{\langle \mathrm{BR} \rangle}{\mathrm{BR}}(B\to D\tau^+\nu)^{[7.73,\ 8.27]}$ &	 0.091906 &	 \cellcolor{green!0} 1.9 $ \sigma$ &	 1.9 $ \sigma$ \\ \hline
30 &	 $\frac{\langle \mathrm{BR} \rangle}{\mathrm{BR}}(B\to D^\ast\tau^+\nu)^{[7.2,\ 7.73]}$ &	 0.10205 &	 \cellcolor{green!0} 1.9 $ \sigma$ &	 1.9 $ \sigma$ \\ \hline
31 &	 $\langle \frac{d\mathrm{BR}}{dq^2} \rangle(B^\pm\to K^\pm \mu^+\mu^-)^{[1.1,\ 2.0]}$ &	 $3.1704\times 10^{-8}$ &	 \cellcolor{green!27} 1.3 $ \sigma$ &	 1.8 $ \sigma$ \\ \hline
32 &	 $\langle \frac{d\mathrm{BR}}{dq^2} \rangle(B^\pm\to K^\pm \mu^+\mu^-)^{[5.0,\ 6.0]}$ &	 $3.0985\times 10^{-8}$ &	 \cellcolor{green!28} 1.3 $ \sigma$ &	 1.9 $ \sigma$ \\ \hline
33 &	 $\langle P_1\rangle(B^0\to K^{\ast 0}\mu^+\mu^-)^{[4.3,\ 6]}$ &	 -0.17908 &	 \cellcolor{green!0} 1.9 $ \sigma$ &	 1.9 $ \sigma$ \\ \hline
34 &	 $F_L(B^0\to D^{\ast -}\tau^+\nu_\tau)$ &	 0.44235 &	 \cellcolor{red!0} 1.8 $ \sigma$ &	 1.8 $ \sigma$ \\ \hline
35 &	 $\langle \frac{d\mathrm{BR}}{dq^2} \rangle(B^0\to K^{\ast 0}\mu^+\mu^-)^{[1.1,\ 2.5]}$ &	 $4.2757\times 10^{-8}$ &	 \cellcolor{green!24} 1.3 $ \sigma$ &	 1.8 $ \sigma$ \\ \hline
36 &	 $\langle \frac{d\mathrm{BR}}{dq^2} \rangle(B^0\to K^{\ast 0}\mu^+\mu^-)^{[4.0,\ 6.0]}$ &	 $4.4839\times 10^{-8}$ &	 \cellcolor{green!31} 1.1 $ \sigma$ &	 1.7 $ \sigma$ \\ \hline
37 &	 $\langle \frac{d\mathrm{BR}}{dq^2} \rangle(\Lambda_b\to\Lambda \mu^+\mu^-)^{[15,\ 20]}$ &	 $6.3747\times 10^{-8}$ &	 \cellcolor{red!13} 2 $ \sigma$ &	 1.7 $ \sigma$ \\ \hline
38 &	 $\langle \frac{d\mathrm{BR}}{dq^2} \rangle(B^0\to K^{\ast 0}\mu^+\mu^-)^{[4.3,\ 6]}$ &	 $4.5307\times 10^{-8}$ &	 \cellcolor{green!28} 1.2 $ \sigma$ &	 1.7 $ \sigma$ \\ \hline
39 &	 $m_W$ &	 80.365 &	 \cellcolor{green!24} 1.2 $ \sigma$ &	 1.7 $ \sigma$ \\ \hline
40 &	 $\langle \frac{d\mathrm{BR}}{dq^2} \rangle(B^0\to K^0\mu^+\mu^-)^{[2.0,\ 4.0]}$ &	 $2.9205\times 10^{-8}$ &	 \cellcolor{green!21} 1.3 $ \sigma$ &	 1.7 $ \sigma$ \\ \hline
41 &	 $A_{\Delta\Gamma}(B_s\to \phi\gamma)$ &	 0.030556 &	 1.7 $ \sigma$ &	 1.7 $ \sigma$ \\ \hline
42 &	 $\frac{\langle \mathrm{BR} \rangle}{\mathrm{BR}}(B\to D\tau^+\nu)^{[9.0,\ 9.5]}$ &	 0.068292 &	 \cellcolor{green!0} 1.6 $ \sigma$ &	 1.6 $ \sigma$ \\ \hline
43 &	 $\langle \frac{d\mathrm{BR}}{dq^2} \rangle(B^\pm\to K^\pm \mu^+\mu^-)^{[15.0,\ 22.0]}$ &	 $1.3561\times 10^{-8}$ &	 \cellcolor{green!41} 0.81 $ \sigma$ &	 1.6 $ \sigma$ \\ \hline
44 &	 $\langle F_L\rangle(B^0\to K^{\ast 0}\mu^+\mu^-)^{[4,\ 6]}$ &	 0.71389 &	 \cellcolor{red!1} 1.7 $ \sigma$ &	 1.6 $ \sigma$ \\ \hline
45 &	 $R_\mu^0$ &	 20.743 &	 \cellcolor{green!13} 1.3 $ \sigma$ &	 1.5 $ \sigma$ \\ \hline
46 &	 $\langle D_{P_5^\prime}^{\mu e} \rangle(B^0\to K^{\ast 0}\ell^+\ell^-)^{[14.18,\ 19.0]}$ &	 0.0027745 &	 \cellcolor{green!0} 1.5 $ \sigma$ &	 1.5 $ \sigma$ \\ \hline
47 &	 $\sigma_\mathrm{had}^0$ &	 0.00010666 &	 \cellcolor{green!50} 0.28 $ \sigma$ &	 1.5 $ \sigma$ \\ \hline
48 &	 $A_\mathrm{FB}^{0,\tau}$ &	 0.016409 &	 \cellcolor{green!5} 1.4 $ \sigma$ &	 1.5 $ \sigma$ \\ \hline
49 &	 $\langle \frac{d\mathrm{BR}}{dq^2} \rangle(B^\pm\to K^\pm \mu^+\mu^-)^{[3.0,\ 4.0]}$ &	 $3.1401\times 10^{-8}$ &	 \cellcolor{green!28} 0.95 $ \sigma$ &	 1.5 $ \sigma$ \\ \hline
50 &	 $\langle \frac{d\mathrm{BR}}{dq^2} \rangle(B^0\to K^{\ast 0}\mu^+\mu^-)^{[2.5,\ 4.0]}$ &	 $4.0397\times 10^{-8}$ &	 \cellcolor{green!28} 0.9 $ \sigma$ &	 1.5 $ \sigma$ \\ \hline
51 &	 $\mathrm{BR}(B^-\to \pi^- e^+\tau^-)$ &	 0 &	 1.5 $ \sigma$ &	 1.5 $ \sigma$ \\ \hline
52 &	 $\mathrm{BR}(B^+\to K^+\nu\bar\nu)$ &	 $4.3974\times 10^{-6}$ &	 \cellcolor{green!0} 1.4 $ \sigma$ &	 1.4 $ \sigma$ \\ \hline
53 &	 $\langle \overline{S_4}\rangle(B_s\to \phi \mu^+\mu^-)^{[15.0,\ 19.0]}$ &	 -0.3018 &	 \cellcolor{red!0} 1.5 $ \sigma$ &	 1.5 $ \sigma$ \\ \hline
54 &	 $\mathrm{BR}(W^\pm\to \mu^\pm\nu)$ &	 0.1086 &	 \cellcolor{red!6} 1.5 $ \sigma$ &	 1.4 $ \sigma$ \\ \hline
55 &	 $\langle A_9\rangle(B^0\to K^{\ast 0}\mu^+\mu^-)^{[15,\ 19]}$ &	 $6.0621\times 10^{-5}$ &	 \cellcolor{green!0} 1.4 $ \sigma$ &	 1.4 $ \sigma$ \\ \hline
56 &	 $R_ e^0$ &	 20.727 &	 \cellcolor{red!7} 1.5 $ \sigma$ &	 1.4 $ \sigma$ \\ \hline
57 &	 $R_{e\mu}(K^+\to \ell^+\nu)$ &	 $2.4662\times 10^{-5}$ &	 \cellcolor{red!50} 2.4 $ \sigma$ &	 1.4 $ \sigma$ \\ \hline
58 &	 $\langle \mathrm{BR} \rangle(B\to X_se^+e^-)^{[14.2,\ 25.0]}$ &	 $3.3821\times 10^{-7}$ &	 \cellcolor{green!9} 1.2 $ \sigma$ &	 1.4 $ \sigma$ \\ \hline
59 &	 $R_{\tau \ell}(B\to D\ell^+\nu)$ &	 0.31432 &	 \cellcolor{green!18} 1 $ \sigma$ &	 1.4 $ \sigma$ \\ \hline
60 &	 $S_{\phi\gamma}$ &	 -0.00023899 &	 1.3 $ \sigma$ &	 1.3 $ \sigma$ \\ \hline
61 &	 $\langle D_{P_5^\prime}^{\mu e} \rangle(B^0\to K^{\ast 0}\ell^+\ell^-)^{[1.0,\ 6.0]}$ &	 0.084096 &	 \cellcolor{green!9} 1.2 $ \sigma$ &	 1.3 $ \sigma$ \\ \hline
62 &	 $\mathrm{BR}(K_L\to\pi^0\nu\bar\nu)$ &	 $3.3218\times 10^{-11}$ &	 \cellcolor{red!0} 1.3 $ \sigma$ &	 1.3 $ \sigma$ \\ \hline
63 &	 $\mathrm{BR}(B^+\to e^+\nu)$ &	 $9.5326\times 10^{-12}$ &	 \cellcolor{green!0} 1.3 $ \sigma$ &	 1.3 $ \sigma$ \\ \hline
64 &	 $\frac{\langle \mathrm{BR} \rangle}{\mathrm{BR}}(B\to D^\ast\tau^+\nu)^{[6.0,\ 6.5]}$ &	 0.078123 &	 \cellcolor{green!0} 1.3 $ \sigma$ &	 1.3 $ \sigma$ \\ \hline
65 &	 $\langle \frac{d\mathrm{BR}}{dq^2} \rangle(B^\pm\to K^\pm \mu^+\mu^-)^{[0,\ 2]}$ &	 $3.1751\times 10^{-8}$ &	 \cellcolor{green!24} 0.83 $ \sigma$ &	 1.3 $ \sigma$ \\ \hline
66 &	 $\mathrm{BR}(B^0\to \rho^{0}\nu\bar\nu)$ &	 $2.0283\times 10^{-7}$ &	 \cellcolor{green!0} 1.3 $ \sigma$ &	 1.3 $ \sigma$ \\ \hline
67 &	 $\mathrm{BR}(K_S\to \mu^+\mu^-)$ &	 $5.1934\times 10^{-12}$ &	 \cellcolor{green!0} 1.3 $ \sigma$ &	 1.3 $ \sigma$ \\ \hline
68 &	 $\mathrm{BR}(K_S\to e^+e^-)$ &	 $1.6247\times 10^{-16}$ &	 \cellcolor{red!0} 1.3 $ \sigma$ &	 1.3 $ \sigma$ \\ \hline
69 &	 $\mathrm{BR}(B^-\to \pi^- \tau^+e^-)$ &	 0 &	 1.3 $ \sigma$ &	 1.3 $ \sigma$ \\ \hline
70 &	 $\langle P_4^\prime\rangle(B^0\to K^{\ast 0}\mu^+\mu^-)^{[2,\ 4]}$ &	 -0.32702 &	 \cellcolor{red!2} 1.4 $ \sigma$ &	 1.3 $ \sigma$ \\ \hline
71 &	 $\mathrm{BR}(B^0\to K^{*0}\nu\bar\nu)$ &	 $9.5415\times 10^{-6}$ &	 \cellcolor{red!0} 1.3 $ \sigma$ &	 1.3 $ \sigma$ \\ \hline
72 &	 $\frac{\langle \mathrm{BR} \rangle}{\mathrm{BR}}(B\to D^\ast\tau^+\nu)^{[8.27,\ 8.8]}$ &	 0.10572 &	 \cellcolor{green!0} 1.3 $ \sigma$ &	 1.3 $ \sigma$ \\ \hline
73 &	 $\Delta M_d$ &	 $3.9784\times 10^{-13}$ &	 \cellcolor{green!0} 1.3 $ \sigma$ &	 1.3 $ \sigma$ \\ \hline
74 &	 $\mathrm{BR}(\tau^+\to K^+\bar\nu)$ &	 0.0071011 &	 \cellcolor{green!6} 1.1 $ \sigma$ &	 1.2 $ \sigma$ \\ \hline
75 &	 $\langle F_L\rangle(B^0\to K^{\ast 0}\mu^+\mu^-)^{[2,\ 4]}$ &	 0.7937 &	 \cellcolor{green!1} 1.2 $ \sigma$ &	 1.3 $ \sigma$ \\ \hline
76 &	 $\langle P_5^\prime\rangle(B^0\to K^{\ast 0}\mu^+\mu^-)^{[2.5,\ 4]}$ &	 -0.45216 &	 \cellcolor{green!6} 1 $ \sigma$ &	 1.2 $ \sigma$ \\ \hline
77 &	 $\langle P_4^\prime\rangle(B^0\to K^{\ast 0}\mu^+\mu^-)^{[4,\ 6]}$ &	 -0.49808 &	 \cellcolor{green!1} 1.2 $ \sigma$ &	 1.3 $ \sigma$ \\ \hline
78 &	 $\frac{\langle \mathrm{BR} \rangle}{\mathrm{BR}}(B\to D\tau^+\nu)^{[9.86,\ 10.4]}$ &	 0.054658 &	 \cellcolor{red!0} 1.2 $ \sigma$ &	 1.2 $ \sigma$ \\ \hline
79 &	 $a_\tau$ &	 0.0011772 &	 1.1 $ \sigma$ &	 1.1 $ \sigma$ \\ \hline
80 &	 $\langle \frac{d\mathrm{BR}}{dq^2} \rangle(B^0\to K^{\ast 0}\mu^+\mu^-)^{[2,\ 4.3]}$ &	 $4.0581\times 10^{-8}$ &	 \cellcolor{green!26} 0.61 $ \sigma$ &	 1.1 $ \sigma$ \\ \hline
81 &	 $\mathrm{BR}(B^+\to K^{*+}\nu\bar\nu)$ &	 $1.0272\times 10^{-5}$ &	 \cellcolor{green!0} 1.1 $ \sigma$ &	 1.1 $ \sigma$ \\ \hline
82 &	 $\langle P_1\rangle(B^0\to K^{\ast 0}e^+e^-)^{[0.002,\ 1.12]}$ &	 0.035971 &	 \cellcolor{red!0} 1.1 $ \sigma$ &	 1.1 $ \sigma$ \\ \hline
83 &	 $\langle P_1\rangle(B^0\to K^{\ast 0}\mu^+\mu^-)^{[15,\ 19]}$ &	 -0.62376 &	 \cellcolor{red!0} 1.1 $ \sigma$ &	 1.1 $ \sigma$ \\ \hline
84 &	 $\frac{\langle \mathrm{BR} \rangle}{\mathrm{BR}}(B\to D^\ast\tau^+\nu)^{[4.0,\ 4.5]}$ &	 0.024346 &	 \cellcolor{green!0} 1.1 $ \sigma$ &	 1.1 $ \sigma$ \\ \hline
85 &	 $\langle F_L\rangle(B^0\to K^{\ast 0}\mu^+\mu^-)^{[1.1,\ 2.5]}$ &	 0.74205 &	 \cellcolor{green!9} 0.87 $ \sigma$ &	 1.1 $ \sigma$ \\ \hline
86 &	 $\Delta M_s$ &	 $1.2498\times 10^{-11}$ &	 \cellcolor{green!0} 1.1 $ \sigma$ &	 1.1 $ \sigma$ \\ \hline
87 &	 $\langle \frac{d\mathrm{BR}}{dq^2} \rangle(B^\pm\to K^\pm \mu^+\mu^-)^{[2.0,\ 3.0]}$ &	 $3.1564\times 10^{-8}$ &	 \cellcolor{green!27} 0.52 $ \sigma$ &	 1.1 $ \sigma$ \\ \hline
88 &	 $\langle \mathrm{BR} \rangle(B\to X_s\mu^+\mu^-)^{[1.0,\ 6.0]}$ &	 $1.5116\times 10^{-6}$ &	 \cellcolor{green!9} 0.88 $ \sigma$ &	 1.1 $ \sigma$ \\ \hline
89 &	 $\langle \overline{S_3}\rangle(B_s\to \phi \mu^+\mu^-)^{[15.0,\ 19.0]}$ &	 -0.20989 &	 \cellcolor{red!0} 1 $ \sigma$ &	 1 $ \sigma$ \\ \hline
90 &	 $\langle P_1\rangle(B^0\to K^{\ast 0}\mu^+\mu^-)^{[2,\ 4]}$ &	 -0.084115 &	 \cellcolor{red!0} 1.1 $ \sigma$ &	 1.1 $ \sigma$ \\ \hline
91 &	 $\langle P_5^\prime\rangle(B^0\to K^{\ast 0}\mu^+\mu^-)^{[15,\ 19]}$ &	 -0.59215 &	 \cellcolor{red!0} 1 $ \sigma$ &	 1 $ \sigma$ \\ \hline
92 &	 $\frac{\langle \mathrm{BR} \rangle}{\mathrm{BR}}(B\to D^\ast\tau^+\nu)^{[10.5,\ 11.0]}$ &	 0.010417 &	 \cellcolor{green!0} 0.94 $ \sigma$ &	 0.94 $ \sigma$ \\ \hline
93 &	 $\langle A_7\rangle(B^0\to K^{\ast 0}\mu^+\mu^-)^{[1.1,\ 6]}$ &	 0.0024641 &	 \cellcolor{red!0} 0.94 $ \sigma$ &	 0.94 $ \sigma$ \\ \hline
94 &	 $A_\mathrm{CP}(B\to X_{s+d}\gamma)$ &	 $-3.6951\times 10^{-18}$ &	 0.94 $ \sigma$ &	 0.94 $ \sigma$ \\ \hline
95 &	 $\langle P_1\rangle(B^0\to K^{\ast 0}\mu^+\mu^-)^{[4,\ 6]}$ &	 -0.17624 &	 \cellcolor{green!0} 0.96 $ \sigma$ &	 0.97 $ \sigma$ \\ \hline
96 &	 $\frac{\langle \mathrm{BR} \rangle}{\mathrm{BR}}(B\to D^\ast\tau^+\nu)^{[7.73,\ 8.27]}$ &	 0.1077 &	 \cellcolor{green!0} 0.92 $ \sigma$ &	 0.92 $ \sigma$ \\ \hline
97 &	 $\frac{\langle \mathrm{BR} \rangle}{\mathrm{BR}}(B\to D\tau^+\nu)^{[6.67,\ 7.2]}$ &	 0.094855 &	 \cellcolor{green!0} 0.91 $ \sigma$ &	 0.91 $ \sigma$ \\ \hline
98 &	 $\langle D_{P_4^\prime}^{\mu e} \rangle(B^0\to K^{\ast 0}\ell^+\ell^-)^{[1.0,\ 6.0]}$ &	 0.035438 &	 \cellcolor{green!3} 0.84 $ \sigma$ &	 0.91 $ \sigma$ \\ \hline
99 &	 $\langle \mathrm{BR} \rangle(B\to X_s\mu^+\mu^-)^{[14.2,\ 25.0]}$ &	 $3.095\times 10^{-7}$ &	 \cellcolor{red!5} 0.99 $ \sigma$ &	 0.88 $ \sigma$ \\ \hline
100 &	 $A_\tau$ &	 0.14798 &	 \cellcolor{red!11} 1.1 $ \sigma$ &	 0.9 $ \sigma$ \\ \hline
101 &	 $\frac{\langle \mathrm{BR} \rangle}{\mathrm{BR}}(B\to D\tau^+\nu)^{[5.5,\ 6.0]}$ &	 0.079431 &	 \cellcolor{green!0} 0.9 $ \sigma$ &	 0.9 $ \sigma$ \\ \hline
102 &	 $\langle A_7\rangle(B^0\to K^{\ast 0}\mu^+\mu^-)^{[15,\ 19]}$ &	 0.0001042 &	 \cellcolor{red!0} 0.89 $ \sigma$ &	 0.89 $ \sigma$ \\ \hline
103 &	 $\mathrm{BR}(B^0\to \pi^- \tau^+\nu_\tau)$ &	 $8.7524\times 10^{-5}$ &	 \cellcolor{green!2} 0.91 $ \sigma$ &	 0.96 $ \sigma$ \\ \hline
104 &	 $\langle \overline{S_4}\rangle(B_s\to \phi \mu^+\mu^-)^{[2.0,\ 5.0]}$ &	 -0.14259 &	 \cellcolor{red!0} 0.9 $ \sigma$ &	 0.88 $ \sigma$ \\ \hline
105 &	 $\frac{\langle \mathrm{BR} \rangle}{\mathrm{BR}}(B\to D\tau^+\nu)^{[10.93,\ 11.47]}$ &	 0.024307 &	 \cellcolor{red!0} 0.88 $ \sigma$ &	 0.88 $ \sigma$ \\ \hline
106 &	 $\frac{\langle \mathrm{BR} \rangle}{\mathrm{BR}}(B\to D\tau^+\nu)^{[9.5,\ 10.0]}$ &	 0.058778 &	 \cellcolor{green!0} 0.87 $ \sigma$ &	 0.87 $ \sigma$ \\ \hline
107 &	 $\mathrm{BR}(B^-\to K^- \tau^+e^-)$ &	 0 &	 0.87 $ \sigma$ &	 0.87 $ \sigma$ \\ \hline
108 &	 $\langle F_L\rangle(B^0\to K^{\ast 0}\mu^+\mu^-)^{[1,\ 2]}$ &	 0.70254 &	 \cellcolor{green!10} 0.66 $ \sigma$ &	 0.87 $ \sigma$ \\ \hline
109 &	 $\frac{\langle \mathrm{BR} \rangle}{\mathrm{BR}}(B\to D\tau^+\nu)^{[10.4,\ 10.93]}$ &	 0.040011 &	 \cellcolor{green!0} 0.86 $ \sigma$ &	 0.86 $ \sigma$ \\ \hline
110 &	 $\mathrm{BR}(\tau^-\to e^- \nu\bar\nu)$ &	 0.17697 &	 \cellcolor{red!50} 2.4 $ \sigma$ &	 0.82 $ \sigma$ \\ \hline
111 &	 $S_{\psi K_S}$ &	 0.70565 &	 \cellcolor{red!0} 0.83 $ \sigma$ &	 0.83 $ \sigma$ \\ \hline
112 &	 $\frac{\langle \mathrm{BR} \rangle}{\mathrm{BR}}(B\to D\tau^+\nu)^{[8.8,\ 9.33]}$ &	 0.075718 &	 \cellcolor{red!0} 0.84 $ \sigma$ &	 0.84 $ \sigma$ \\ \hline
113 &	 $\frac{\langle \mathrm{BR} \rangle}{\mathrm{BR}}(B\to D\tau^+\nu)^{[7.2,\ 7.73]}$ &	 0.093958 &	 \cellcolor{red!0} 0.84 $ \sigma$ &	 0.84 $ \sigma$ \\ \hline
114 &	 $\frac{\langle \mathrm{BR} \rangle}{\mathrm{BR}}(B\to D\tau^+\nu)^{[10.0,\ 10.5]}$ &	 0.047878 &	 \cellcolor{red!0} 0.84 $ \sigma$ &	 0.84 $ \sigma$ \\ \hline
115 &	 $\langle \overline{F_L}\rangle(B_s\to \phi \mu^+\mu^-)^{[2.0,\ 5.0]}$ &	 0.80894 &	 \cellcolor{green!0} 0.86 $ \sigma$ &	 0.87 $ \sigma$ \\ \hline
116 &	 $A_\mathrm{FB}^{0, c}$ &	 0.074039 &	 \cellcolor{red!6} 0.95 $ \sigma$ &	 0.83 $ \sigma$ \\ \hline
117 &	 $\langle A_8\rangle(B^0\to K^{\ast 0}\mu^+\mu^-)^{[1.1,\ 6]}$ &	 0.0011623 &	 \cellcolor{red!0} 0.83 $ \sigma$ &	 0.83 $ \sigma$ \\ \hline
118 &	 $\mathrm{BR}(W^\pm\to  e^\pm\nu)$ &	 0.10827 &	 \cellcolor{green!4} 0.73 $ \sigma$ &	 0.82 $ \sigma$ \\ \hline
119 &	 $\frac{\langle \mathrm{BR} \rangle}{\mathrm{BR}}(B\to D^\ast\tau^+\nu)^{[6.67,\ 7.2]}$ &	 0.095374 &	 \cellcolor{red!0} 0.82 $ \sigma$ &	 0.82 $ \sigma$ \\ \hline
120 &	 $\frac{\langle \mathrm{BR} \rangle}{\mathrm{BR}}(B\to D\tau^+\nu)^{[6.0,\ 6.5]}$ &	 0.085948 &	 \cellcolor{green!0} 0.82 $ \sigma$ &	 0.82 $ \sigma$ \\ \hline
121 &	 $\langle P_1\rangle(B^0\to K^{\ast 0}\mu^+\mu^-)^{[1.1,\ 2.5]}$ &	 0.02995 &	 \cellcolor{red!0} 0.84 $ \sigma$ &	 0.83 $ \sigma$ \\ \hline
122 &	 $\mathrm{BR}(B^+\to \tau^+\nu)$ &	 $9.1574\times 10^{-5}$ &	 \cellcolor{green!7} 0.69 $ \sigma$ &	 0.83 $ \sigma$ \\ \hline
123 &	 $\langle A_9\rangle(B^0\to K^{\ast 0}\mu^+\mu^-)^{[1.1,\ 6]}$ &	 0.00013215 &	 \cellcolor{red!0} 0.8 $ \sigma$ &	 0.8 $ \sigma$ \\ \hline
124 &	 $\mathrm{BR}(K^+\to\pi^+\nu\bar\nu)$ &	 $9.2404\times 10^{-11}$ &	 \cellcolor{red!0} 0.83 $ \sigma$ &	 0.83 $ \sigma$ \\ \hline
125 &	 $\frac{\langle \mathrm{BR} \rangle}{\mathrm{BR}}(B\to D^\ast\tau^+\nu)^{[6.13,\ 6.67]}$ &	 0.087527 &	 \cellcolor{green!0} 0.79 $ \sigma$ &	 0.79 $ \sigma$ \\ \hline
126 &	 $\frac{\langle \mathrm{BR} \rangle}{\mathrm{BR}}(B\to D\tau^+\nu)^{[6.13,\ 6.67]}$ &	 0.094177 &	 \cellcolor{red!0} 0.79 $ \sigma$ &	 0.79 $ \sigma$ \\ \hline
127 &	 $\langle F_L\rangle(B^0\to K^{\ast 0}\mu^+\mu^-)^{[0,\ 2]}$ &	 0.36291 &	 \cellcolor{green!8} 0.65 $ \sigma$ &	 0.83 $ \sigma$ \\ \hline
128 &	 $\overline{\mathrm{BR}}(B_s\to \phi\gamma)$ &	 $4.0151\times 10^{-5}$ &	 0.8 $ \sigma$ &	 0.8 $ \sigma$ \\ \hline
129 &	 $\langle A_\mathrm{FB}^\ell\rangle(\Lambda_b\to\Lambda \mu^+\mu^-)^{[15,\ 20]}$ &	 -0.35201 &	 \cellcolor{red!1} 0.8 $ \sigma$ &	 0.77 $ \sigma$ \\ \hline
130 &	 $\frac{\langle \mathrm{BR} \rangle}{\mathrm{BR}}(B\to D^\ast\tau^+\nu)^{[8.8,\ 9.33]}$ &	 0.10126 &	 \cellcolor{green!0} 0.77 $ \sigma$ &	 0.77 $ \sigma$ \\ \hline
131 &	 $\langle A_\mathrm{FB}\rangle(B^0\to K^{\ast 0}\mu^+\mu^-)^{[4.3,\ 6]}$ &	 0.12076 &	 \cellcolor{green!3} 0.69 $ \sigma$ &	 0.76 $ \sigma$ \\ \hline
132 &	 $\frac{\langle \mathrm{BR} \rangle}{\mathrm{BR}}(B\to D\tau^+\nu)^{[7.5,\ 8.0]}$ &	 0.087076 &	 \cellcolor{green!0} 0.75 $ \sigma$ &	 0.75 $ \sigma$ \\ \hline
133 &	 $\langle \overline{F_L}\rangle(B_s\to \phi \mu^+\mu^-)^{[15.0,\ 19.0]}$ &	 0.34168 &	 \cellcolor{red!0} 0.74 $ \sigma$ &	 0.73 $ \sigma$ \\ \hline
134 &	 $\langle P_1\rangle(B^0\to K^{\ast 0}\mu^+\mu^-)^{[2,\ 4.3]}$ &	 -0.095443 &	 \cellcolor{red!1} 0.76 $ \sigma$ &	 0.74 $ \sigma$ \\ \hline
135 &	 $\langle F_L\rangle(B^0\to K^{\ast 0}\mu^+\mu^-)^{[2.5,\ 4]}$ &	 0.79337 &	 \cellcolor{red!1} 0.78 $ \sigma$ &	 0.75 $ \sigma$ \\ \hline
136 &	 $\langle A_\mathrm{FB}\rangle(B^0\to K^{\ast 0}\mu^+\mu^-)^{[1,\ 2]}$ &	 -0.16588 &	 \cellcolor{green!2} 0.66 $ \sigma$ &	 0.72 $ \sigma$ \\ \hline
137 &	 $R_ b^0$ &	 0.21583 &	 \cellcolor{green!1} 0.7 $ \sigma$ &	 0.73 $ \sigma$ \\ \hline
138 &	 $\frac{\langle \mathrm{BR} \rangle}{\mathrm{BR}}(B\to D^\ast\tau^+\nu)^{[5.5,\ 6.0]}$ &	 0.067058 &	 \cellcolor{green!0} 0.72 $ \sigma$ &	 0.72 $ \sigma$ \\ \hline
139 &	 $\frac{\langle \mathrm{BR} \rangle}{\mathrm{BR}}(B\to D\tau^+\nu)^{[10.5,\ 11.0]}$ &	 0.035542 &	 \cellcolor{green!0} 0.71 $ \sigma$ &	 0.71 $ \sigma$ \\ \hline
140 &	 $\frac{\langle \mathrm{BR} \rangle}{\mathrm{BR}}(B\to D\tau^+\nu)^{[8.5,\ 9.0]}$ &	 0.076305 &	 \cellcolor{green!0} 0.71 $ \sigma$ &	 0.71 $ \sigma$ \\ \hline
141 &	 $\frac{\langle \mathrm{BR} \rangle}{\mathrm{BR}}(B\to D\tau^+\nu)^{[4.0,\ 4.53]}$ &	 0.038684 &	 \cellcolor{green!0} 0.69 $ \sigma$ &	 0.69 $ \sigma$ \\ \hline
142 &	 $A_\mathrm{FB}^{0, e}$ &	 0.016394 &	 \cellcolor{red!3} 0.76 $ \sigma$ &	 0.69 $ \sigma$ \\ \hline
143 &	 $\mathrm{BR}(B^+\to \pi^+\nu\bar\nu)$ &	 $1.2672\times 10^{-7}$ &	 \cellcolor{green!0} 0.68 $ \sigma$ &	 0.68 $ \sigma$ \\ \hline
144 &	 $\langle P_5^\prime\rangle(B^0\to K^{\ast 0}\mu^+\mu^-)^{[4.3,\ 6]}$ &	 -0.75087 &	 \cellcolor{red!3} 0.71 $ \sigma$ &	 0.65 $ \sigma$ \\ \hline
145 &	 $\mathrm{BR}(B^+\to \rho^{+}\nu\bar\nu)$ &	 $4.3699\times 10^{-7}$ &	 \cellcolor{green!0} 0.67 $ \sigma$ &	 0.67 $ \sigma$ \\ \hline
146 &	 $\langle P_5^\prime\rangle(B^0\to K^{\ast 0}\mu^+\mu^-)^{[1.1,\ 2.5]}$ &	 0.1884 &	 \cellcolor{green!10} 0.44 $ \sigma$ &	 0.66 $ \sigma$ \\ \hline
147 &	 $\langle A_\mathrm{FB}\rangle(B^0\to K^{\ast 0}\mu^+\mu^-)^{[2,\ 4.3]}$ &	 -0.041339 &	 \cellcolor{green!5} 0.53 $ \sigma$ &	 0.65 $ \sigma$ \\ \hline
148 &	 $\frac{\langle \mathrm{BR} \rangle}{\mathrm{BR}}(B\to D\tau^+\nu)^{[4.0,\ 4.5]}$ &	 0.035905 &	 \cellcolor{green!0} 0.65 $ \sigma$ &	 0.65 $ \sigma$ \\ \hline
149 &	 $\frac{\langle \mathrm{BR} \rangle}{\mathrm{BR}}(B\to D^\ast\tau^+\nu)^{[7.5,\ 8.0]}$ &	 0.098515 &	 \cellcolor{green!0} 0.65 $ \sigma$ &	 0.65 $ \sigma$ \\ \hline
150 &	 $\langle P_2\rangle(B^0\to K^{\ast 0}\mu^+\mu^-)^{[2.5,\ 4]}$ &	 -0.11542 &	 \cellcolor{green!3} 0.6 $ \sigma$ &	 0.66 $ \sigma$ \\ \hline
151 &	 $S_{K^{*}\gamma}$ &	 -0.022785 &	 \cellcolor{green!0} 0.63 $ \sigma$ &	 0.63 $ \sigma$ \\ \hline
152 &	 $\mathrm{BR}(B^0\to \pi^0\nu\bar\nu)$ &	 $5.898\times 10^{-8}$ &	 \cellcolor{green!0} 0.63 $ \sigma$ &	 0.63 $ \sigma$ \\ \hline
153 &	 $\langle A_T^\mathrm{Im}\rangle(B^0\to K^{\ast 0}e^+e^-)^{[0.002,\ 1.12]}$ &	 0.00032829 &	 \cellcolor{green!0} 0.64 $ \sigma$ &	 0.64 $ \sigma$ \\ \hline
154 &	 $\langle R_{\mu e} \rangle(B^+\to K^{\ast +}\ell^+\ell^-)^{[15.0,\ 19.0]}$ &	 0.79446 &	 \cellcolor{red!14} 0.89 $ \sigma$ &	 0.59 $ \sigma$ \\ \hline
155 &	 $\langle F_L\rangle(B^0\to K^{\ast 0}\mu^+\mu^-)^{[4.3,\ 6]}$ &	 0.70627 &	 \cellcolor{red!1} 0.6 $ \sigma$ &	 0.58 $ \sigma$ \\ \hline
156 &	 $\frac{\langle \mathrm{BR} \rangle}{\mathrm{BR}}(B\to D^\ast\tau^+\nu)^{[4.5,\ 5.0]}$ &	 0.039696 &	 \cellcolor{red!0} 0.59 $ \sigma$ &	 0.59 $ \sigma$ \\ \hline
157 &	 $A_ b$ &	 0.93475 &	 \cellcolor{red!0} 0.59 $ \sigma$ &	 0.59 $ \sigma$ \\ \hline
158 &	 $\langle A_\mathrm{FB}\rangle(B^0\to K^{\ast 0}\mu^+\mu^-)^{[0,\ 2]}$ &	 -0.10469 &	 \cellcolor{red!0} 0.62 $ \sigma$ &	 0.62 $ \sigma$ \\ \hline
159 &	 $\mathrm{BR}(\tau^-\to e^-\mu^+e^-)$ &	 0 &	 0.58 $ \sigma$ &	 0.58 $ \sigma$ \\ \hline
160 &	 $\frac{\langle \mathrm{BR} \rangle}{\mathrm{BR}}(B\to D\tau^+\nu)^{[8.27,\ 8.8]}$ &	 0.083993 &	 \cellcolor{green!0} 0.58 $ \sigma$ &	 0.58 $ \sigma$ \\ \hline
161 &	 $\frac{\langle \mathrm{BR} \rangle}{\mathrm{BR}}(B\to D^\ast\tau^+\nu)^{[10.0,\ 10.5]}$ &	 0.059012 &	 \cellcolor{green!0} 0.57 $ \sigma$ &	 0.57 $ \sigma$ \\ \hline
162 &	 $\mathrm{BR}(B^-\to K^- \mu^+\tau^-)$ &	 0 &	 0.57 $ \sigma$ &	 0.57 $ \sigma$ \\ \hline
163 &	 $\frac{\langle \mathrm{BR} \rangle}{\mathrm{BR}}(B\to D\tau^+\nu)^{[4.53,\ 5.07]}$ &	 0.060577 &	 \cellcolor{green!0} 0.56 $ \sigma$ &	 0.56 $ \sigma$ \\ \hline
164 &	 $R_{\mu e}(B\to D^{\ast}\ell^+\nu)$ &	 0.94513 &	 \cellcolor{red!39} 1.4 $ \sigma$ &	 0.56 $ \sigma$ \\ \hline
165 &	 $\langle P_5^\prime\rangle(B^0\to K^{\ast 0}\mu^+\mu^-)^{[0.04,\ 2]}$ &	 0.53161 &	 \cellcolor{green!3} 0.49 $ \sigma$ &	 0.55 $ \sigma$ \\ \hline
166 &	 $\frac{\langle \mathrm{BR} \rangle}{\mathrm{BR}}(B\to D^\ast\tau^+\nu)^{[4.53,\ 5.07]}$ &	 0.044486 &	 \cellcolor{green!0} 0.54 $ \sigma$ &	 0.54 $ \sigma$ \\ \hline
167 &	 $\mathrm{BR}(\tau^+\to \pi^+\bar\nu)$ &	 0.10871 &	 \cellcolor{green!9} 0.35 $ \sigma$ &	 0.54 $ \sigma$ \\ \hline
168 &	 $A_\mathrm{FB}^{0,\mu}$ &	 0.016318 &	 \cellcolor{green!4} 0.45 $ \sigma$ &	 0.53 $ \sigma$ \\ \hline
169 &	 $\langle A_8\rangle(B^0\to K^{\ast 0}\mu^+\mu^-)^{[15,\ 19]}$ &	 $7.7484\times 10^{-5}$ &	 \cellcolor{green!0} 0.52 $ \sigma$ &	 0.52 $ \sigma$ \\ \hline
170 &	 $\langle P_5^\prime\rangle(B^0\to K^{\ast 0}\mu^+\mu^-)^{[1,\ 2]}$ &	 0.32831 &	 \cellcolor{red!6} 0.66 $ \sigma$ &	 0.54 $ \sigma$ \\ \hline
171 &	 $\frac{\langle \mathrm{BR} \rangle}{\mathrm{BR}}(B\to D\tau^+\nu)^{[11.5,\ 12.0]}$ &	 0.0020025 &	 \cellcolor{red!0} 0.51 $ \sigma$ &	 0.51 $ \sigma$ \\ \hline
172 &	 $\mathrm{BR}(\tau^-\to \mu^-e^+\mu^-)$ &	 0 &	 0.51 $ \sigma$ &	 0.51 $ \sigma$ \\ \hline
173 &	 $\mathrm{BR}(\pi^+\to e^+\nu)$ &	 0.00012294 &	 \cellcolor{red!45} 1.4 $ \sigma$ &	 0.51 $ \sigma$ \\ \hline
174 &	 $\langle \frac{d\mathrm{BR}}{dq^2} \rangle(B^0\to K^{\ast 0}\mu^+\mu^-)^{[0,\ 2]}$ &	 $7.9334\times 10^{-8}$ &	 \cellcolor{red!7} 0.69 $ \sigma$ &	 0.54 $ \sigma$ \\ \hline
175 &	 $\mathrm{BR}(K_L\to \mu^+\mu^-)$ &	 $7.4841\times 10^{-9}$ &	 \cellcolor{red!0} 0.5 $ \sigma$ &	 0.5 $ \sigma$ \\ \hline
176 &	 $\mathrm{BR}(\tau^-\to \mu^-e^+e^-)$ &	 0 &	 0.49 $ \sigma$ &	 0.49 $ \sigma$ \\ \hline
177 &	 $\langle \frac{d\mathrm{BR}}{dq^2} \rangle(B^0\to K^0\mu^+\mu^-)^{[0,\ 2]}$ &	 $2.9458\times 10^{-8}$ &	 \cellcolor{green!9} 0.3 $ \sigma$ &	 0.49 $ \sigma$ \\ \hline
178 &	 $\langle \frac{d\mathrm{BR}}{dq^2} \rangle(B^+\to K^{\ast +}\mu^+\mu^-)^{[2.0,\ 4.0]}$ &	 $4.3931\times 10^{-8}$ &	 \cellcolor{red!13} 0.78 $ \sigma$ &	 0.5 $ \sigma$ \\ \hline
179 &	 $\mathrm{BR}(B^0\to K^0\nu\bar\nu)$ &	 $4.0717\times 10^{-6}$ &	 \cellcolor{green!0} 0.48 $ \sigma$ &	 0.48 $ \sigma$ \\ \hline
180 &	 $\Gamma_Z$ &	 2.4939 &	 \cellcolor{red!5} 0.58 $ \sigma$ &	 0.47 $ \sigma$ \\ \hline
181 &	 $\mathrm{BR}(B_c\to \tau^+\nu)$ &	 0.024743 &	 \cellcolor{red!0} 0.48 $ \sigma$ &	 0.47 $ \sigma$ \\ \hline
182 &	 $\frac{\langle \mathrm{BR} \rangle}{\mathrm{BR}}(B\to D^\ast\tau^+\nu)^{[7.0,\ 7.5]}$ &	 0.094054 &	 \cellcolor{red!0} 0.47 $ \sigma$ &	 0.47 $ \sigma$ \\ \hline
183 &	 $\frac{\langle \mathrm{BR} \rangle}{\mathrm{BR}}(B\to D\tau^+\nu)^{[11.0,\ 11.5]}$ &	 0.020875 &	 \cellcolor{green!0} 0.46 $ \sigma$ &	 0.46 $ \sigma$ \\ \hline
184 &	 $\langle P_2\rangle(B^0\to K^{\ast 0}\mu^+\mu^-)^{[1.1,\ 2.5]}$ &	 -0.45133 &	 \cellcolor{green!0} 0.47 $ \sigma$ &	 0.47 $ \sigma$ \\ \hline
185 &	 $\mathrm{BR}(B^-\to K^{*-} e^+\mu^-)$ &	 0 &	 0.45 $ \sigma$ &	 0.45 $ \sigma$ \\ \hline
186 &	 $\langle \mathrm{BR} \rangle(B\to X_se^+e^-)^{[1.0,\ 6.0]}$ &	 $1.9486\times 10^{-6}$ &	 \cellcolor{green!18} 0.049 $ \sigma$ &	 0.42 $ \sigma$ \\ \hline
187 &	 $\langle P_4^\prime\rangle(B^0\to K^{\ast 0}\mu^+\mu^-)^{[15,\ 19]}$ &	 -0.63521 &	 \cellcolor{red!0} 0.44 $ \sigma$ &	 0.44 $ \sigma$ \\ \hline
188 &	 $\langle P_4^\prime\rangle(B^0\to K^{\ast 0}\mu^+\mu^-)^{[2.5,\ 4]}$ &	 -0.3742 &	 \cellcolor{red!1} 0.47 $ \sigma$ &	 0.45 $ \sigma$ \\ \hline
189 &	 $\langle P_4^\prime\rangle(B^0\to K^{\ast 0}\mu^+\mu^-)^{[0.04,\ 2]}$ &	 0.15774 &	 \cellcolor{green!0} 0.41 $ \sigma$ &	 0.42 $ \sigma$ \\ \hline
190 &	 $\langle P_4^\prime\rangle(B^0\to K^{\ast 0}\mu^+\mu^-)^{[1.1,\ 2.5]}$ &	 -0.040805 &	 \cellcolor{red!4} 0.51 $ \sigma$ &	 0.41 $ \sigma$ \\ \hline
191 &	 $\langle \frac{d\mathrm{BR}}{dq^2} \rangle(B^\pm\to K^\pm \mu^+\mu^-)^{[2,\ 4.3]}$ &	 $3.1456\times 10^{-8}$ &	 \cellcolor{green!20} 0.0094 $ \sigma$ &	 0.42 $ \sigma$ \\ \hline
192 &	 $\langle F_L\rangle(B^0\to K^{\ast 0}\mu^+\mu^-)^{[0.04,\ 2]}$ &	 0.36291 &	 \cellcolor{red!10} 0.65 $ \sigma$ &	 0.44 $ \sigma$ \\ \hline
193 &	 $\langle F_L\rangle(B^0\to K^{\ast 0}\mu^+\mu^-)^{[2,\ 4.3]}$ &	 0.78916 &	 \cellcolor{green!2} 0.37 $ \sigma$ &	 0.42 $ \sigma$ \\ \hline
194 &	 $R_\tau^0$ &	 20.77 &	 \cellcolor{green!12} 0.14 $ \sigma$ &	 0.38 $ \sigma$ \\ \hline
195 &	 $\langle \frac{d\mathrm{BR}}{dq^2} \rangle(B^0\to K^0\mu^+\mu^-)^{[2,\ 4.3]}$ &	 $2.9181\times 10^{-8}$ &	 \cellcolor{green!9} 0.21 $ \sigma$ &	 0.39 $ \sigma$ \\ \hline
196 &	 $\langle R_{\mu e} \rangle(B^0\to K^{\ast 0}\ell^+\ell^-)^{[0.1,\ 8.0]}$ &	 0.82242 &	 \cellcolor{green!8} 0.19 $ \sigma$ &	 0.37 $ \sigma$ \\ \hline
197 &	 $\langle R_{\mu e} \rangle(B^0\to K^{\ast 0}\ell^+\ell^-)^{[15.0,\ 19.0]}$ &	 0.79447 &	 \cellcolor{red!26} 0.89 $ \sigma$ &	 0.36 $ \sigma$ \\ \hline
198 &	 $\langle P_2\rangle(B^0\to K^{\ast 0}\mu^+\mu^-)^{[15,\ 19]}$ &	 0.37143 &	 \cellcolor{green!1} 0.32 $ \sigma$ &	 0.36 $ \sigma$ \\ \hline
199 &	 $\langle P_1\rangle(B^0\to K^{\ast 0}\mu^+\mu^-)^{[2.5,\ 4]}$ &	 -0.10663 &	 \cellcolor{green!0} 0.33 $ \sigma$ &	 0.34 $ \sigma$ \\ \hline
200 &	 $\langle A_\mathrm{FB}^h\rangle(\Lambda_b\to\Lambda \mu^+\mu^-)^{[15,\ 20]}$ &	 -0.31822 &	 \cellcolor{green!0} 0.31 $ \sigma$ &	 0.31 $ \sigma$ \\ \hline
201 &	 $\frac{\langle \mathrm{BR} \rangle}{\mathrm{BR}}(B\to D\tau^+\nu)^{[6.5,\ 7.0]}$ &	 0.089095 &	 \cellcolor{green!0} 0.34 $ \sigma$ &	 0.34 $ \sigma$ \\ \hline
202 &	 $A_\mu$ &	 0.14716 &	 \cellcolor{red!0} 0.34 $ \sigma$ &	 0.34 $ \sigma$ \\ \hline
203 &	 $\frac{\langle \mathrm{BR} \rangle}{\mathrm{BR}}(B\to D^\ast\tau^+\nu)^{[9.86,\ 10.4]}$ &	 0.071012 &	 \cellcolor{green!0} 0.33 $ \sigma$ &	 0.33 $ \sigma$ \\ \hline
204 &	 $\overline{\mathrm{BR}}(B_s\to \tau^+\tau^-)$ &	 $9.0721\times 10^{-7}$ &	 \cellcolor{red!0} 0.33 $ \sigma$ &	 0.33 $ \sigma$ \\ \hline
205 &	 $\langle P_1\rangle(B^0\to K^{\ast 0}\mu^+\mu^-)^{[0.04,\ 2]}$ &	 0.043716 &	 \cellcolor{red!0} 0.32 $ \sigma$ &	 0.32 $ \sigma$ \\ \hline
206 &	 $\mathrm{BR}(\bar B^0\to \bar K^{*0} \mu^+e^-)$ &	 0 &	 0.3 $ \sigma$ &	 0.3 $ \sigma$ \\ \hline
207 &	 $\frac{\langle \mathrm{BR} \rangle}{\mathrm{BR}}(B\to D^\ast\tau^+\nu)^{[8.5,\ 9.0]}$ &	 0.098623 &	 \cellcolor{red!0} 0.3 $ \sigma$ &	 0.3 $ \sigma$ \\ \hline
208 &	 $\mathrm{BR}(B^-\to K^- \tau^+\mu^-)$ &	 0 &	 0.29 $ \sigma$ &	 0.29 $ \sigma$ \\ \hline
209 &	 $\frac{\langle \mathrm{BR} \rangle}{\mathrm{BR}}(B\to D\tau^+\nu)^{[4.5,\ 5.0]}$ &	 0.054469 &	 \cellcolor{red!0} 0.27 $ \sigma$ &	 0.27 $ \sigma$ \\ \hline
210 &	 $\langle \frac{d\mathrm{BR}}{dq^2} \rangle(B^+\to K^{\ast +}\mu^+\mu^-)^{[0,\ 2]}$ &	 $8.3078\times 10^{-8}$ &	 \cellcolor{green!3} 0.14 $ \sigma$ &	 0.21 $ \sigma$ \\ \hline
211 &	 $\mathrm{BR}(B^-\to K^{*-} \mu^+e^-)$ &	 0 &	 0.25 $ \sigma$ &	 0.25 $ \sigma$ \\ \hline
212 &	 $\langle \overline{S_3}\rangle(B_s\to \phi \mu^+\mu^-)^{[2.0,\ 5.0]}$ &	 -0.0078565 &	 \cellcolor{red!0} 0.27 $ \sigma$ &	 0.27 $ \sigma$ \\ \hline
213 &	 $\mathrm{BR}(B^+\to \mu^+\nu)$ &	 $3.8612\times 10^{-7}$ &	 \cellcolor{red!1} 0.26 $ \sigma$ &	 0.24 $ \sigma$ \\ \hline
214 &	 $\langle P_5^\prime\rangle(B^0\to K^{\ast 0}\mu^+\mu^-)^{[2,\ 4.3]}$ &	 -0.39945 &	 \cellcolor{red!6} 0.37 $ \sigma$ &	 0.24 $ \sigma$ \\ \hline
215 &	 $x_{12}^{\mathrm{Im},D}$ &	 $-3.4139\times 10^{-20}$ &	 0.25 $ \sigma$ &	 0.25 $ \sigma$ \\ \hline
216 &	 $\frac{\langle \mathrm{BR} \rangle}{\mathrm{BR}}(B\to D^\ast\tau^+\nu)^{[6.5,\ 7.0]}$ &	 0.087179 &	 \cellcolor{red!0} 0.22 $ \sigma$ &	 0.22 $ \sigma$ \\ \hline
217 &	 $\frac{\langle \mathrm{BR} \rangle}{\mathrm{BR}}(B\to D^\ast\tau^+\nu)^{[9.33,\ 9.86]}$ &	 0.0907 &	 \cellcolor{red!0} 0.22 $ \sigma$ &	 0.22 $ \sigma$ \\ \hline
218 &	 $\frac{\langle \mathrm{BR} \rangle}{\mathrm{BR}}(B\to D\tau^+\nu)^{[11.47,\ 12.0]}$ &	 0.0026758 &	 \cellcolor{red!0} 0.22 $ \sigma$ &	 0.22 $ \sigma$ \\ \hline
219 &	 $\mathrm{BR}(B^-\to K^- e^+\tau^-)$ &	 0 &	 0.2 $ \sigma$ &	 0.2 $ \sigma$ \\ \hline
220 &	 $\langle P_5^\prime\rangle(B^0\to K^{\ast 0}\mu^+\mu^-)^{[2,\ 4]}$ &	 -0.35661 &	 \cellcolor{green!7} 0.11 $ \sigma$ &	 0.26 $ \sigma$ \\ \hline
221 &	 $\mathrm{BR}(B^-\to \pi^- \mu^+\tau^-)$ &	 0 &	 0.18 $ \sigma$ &	 0.18 $ \sigma$ \\ \hline
222 &	 $\mathrm{BR}(B^0\to K^{*0}\gamma)$ &	 $4.1772\times 10^{-5}$ &	 0.18 $ \sigma$ &	 0.18 $ \sigma$ \\ \hline
223 &	 $S_{\psi\phi}$ &	 0.03873 &	 \cellcolor{red!0} 0.18 $ \sigma$ &	 0.18 $ \sigma$ \\ \hline
224 &	 $\mathrm{BR}(B^0\to \mu^+\mu^-)$ &	 $1.0315\times 10^{-10}$ &	 \cellcolor{green!6} 0.034 $ \sigma$ &	 0.17 $ \sigma$ \\ \hline
225 &	 $\frac{\langle \mathrm{BR} \rangle}{\mathrm{BR}}(B\to D^\ast\tau^+\nu)^{[5.6,\ 6.13]}$ &	 0.073949 &	 \cellcolor{green!0} 0.16 $ \sigma$ &	 0.16 $ \sigma$ \\ \hline
226 &	 $\Gamma_W$ &	 2.0917 &	 \cellcolor{green!0} 0.16 $ \sigma$ &	 0.16 $ \sigma$ \\ \hline
227 &	 $\langle \frac{d\mathrm{BR}}{dq^2} \rangle(B^0\to K^{\ast 0}\mu^+\mu^-)^{[1,\ 2]}$ &	 $4.4936\times 10^{-8}$ &	 \cellcolor{red!1} 0.2 $ \sigma$ &	 0.17 $ \sigma$ \\ \hline
228 &	 $\frac{\langle \mathrm{BR} \rangle}{\mathrm{BR}}(B\to D^\ast\tau^+\nu)^{[9.5,\ 10.0]}$ &	 0.081195 &	 \cellcolor{red!0} 0.16 $ \sigma$ &	 0.16 $ \sigma$ \\ \hline
229 &	 $\frac{\langle \mathrm{BR} \rangle}{\mathrm{BR}}(B\to D\tau^+\nu)^{[7.0,\ 7.5]}$ &	 0.089338 &	 \cellcolor{red!0} 0.16 $ \sigma$ &	 0.16 $ \sigma$ \\ \hline
230 &	 $\langle P_1\rangle(B^0\to K^{\ast 0}\mu^+\mu^-)^{[1,\ 2]}$ &	 0.047436 &	 \cellcolor{green!0} 0.13 $ \sigma$ &	 0.14 $ \sigma$ \\ \hline
231 &	 $\frac{\langle \mathrm{BR} \rangle}{\mathrm{BR}}(B\to D^\ast\tau^+\nu)^{[8.0,\ 8.5]}$ &	 0.10022 &	 \cellcolor{red!0} 0.15 $ \sigma$ &	 0.15 $ \sigma$ \\ \hline
232 &	 $\frac{\langle \mathrm{BR} \rangle}{\mathrm{BR}}(B\to D\tau^+\nu)^{[5.6,\ 6.13]}$ &	 0.086109 &	 \cellcolor{green!0} 0.14 $ \sigma$ &	 0.14 $ \sigma$ \\ \hline
233 &	 $\sigma_\mathrm{trident}/\sigma_\mathrm{trident}^\mathrm{SM}$ &	 1.0035 &	 \cellcolor{red!0} 0.14 $ \sigma$ &	 0.13 $ \sigma$ \\ \hline
234 &	 $\frac{\langle \mathrm{BR} \rangle}{\mathrm{BR}}(B\to D\tau^+\nu)^{[8.0,\ 8.5]}$ &	 0.082641 &	 \cellcolor{red!0} 0.12 $ \sigma$ &	 0.12 $ \sigma$ \\ \hline
235 &	 $\frac{\langle \mathrm{BR} \rangle}{\mathrm{BR}}(B\to D\tau^+\nu)^{[5.0,\ 5.5]}$ &	 0.069057 &	 \cellcolor{green!0} 0.12 $ \sigma$ &	 0.12 $ \sigma$ \\ \hline
236 &	 $\langle F_L\rangle(B^0\to K^{\ast 0}\mu^+\mu^-)^{[15,\ 19]}$ &	 0.3406 &	 \cellcolor{green!0} 0.098 $ \sigma$ &	 0.11 $ \sigma$ \\ \hline
237 &	 $\langle R_{\mu e} \rangle(B^+\to K^{\ast +}\ell^+\ell^-)^{[0.1,\ 8.0]}$ &	 0.82185 &	 \cellcolor{red!15} 0.41 $ \sigma$ &	 0.1 $ \sigma$ \\ \hline
238 &	 $\mathrm{BR}(\tau^-\to e^-e^+e^-)$ &	 0 &	 0.1 $ \sigma$ &	 0.1 $ \sigma$ \\ \hline
239 &	 $A_ c$ &	 0.6677 &	 \cellcolor{green!0} 0.085 $ \sigma$ &	 0.092 $ \sigma$ \\ \hline
240 &	 $\mathrm{BR}(B\to X_s\gamma)$ &	 0.0003291 &	 0.086 $ \sigma$ &	 0.086 $ \sigma$ \\ \hline
241 &	 $\langle \frac{d\mathrm{BR}}{dq^2} \rangle(\Lambda_b\to\Lambda \mu^+\mu^-)^{[1.1,\ 6]}$ &	 $9.362\times 10^{-9}$ &	 \cellcolor{green!0} 0.056 $ \sigma$ &	 0.065 $ \sigma$ \\ \hline
242 &	 $\langle D_{P_4^\prime}^{\mu e} \rangle(B^0\to K^{\ast 0}\ell^+\ell^-)^{[14.18,\ 19.0]}$ &	 -0.00016124 &	 \cellcolor{red!0} 0.072 $ \sigma$ &	 0.072 $ \sigma$ \\ \hline
243 &	 $\frac{\langle \mathrm{BR} \rangle}{\mathrm{BR}}(B\to D^\ast\tau^+\nu)^{[4.0,\ 4.53]}$ &	 0.026298 &	 \cellcolor{red!0} 0.068 $ \sigma$ &	 0.068 $ \sigma$ \\ \hline
244 &	 $\frac{\langle \mathrm{BR} \rangle}{\mathrm{BR}}(B\to D\tau^+\nu)^{[5.07,\ 5.6]}$ &	 0.075356 &	 \cellcolor{red!0} 0.067 $ \sigma$ &	 0.067 $ \sigma$ \\ \hline
245 &	 $\frac{\langle \mathrm{BR} \rangle}{\mathrm{BR}}(B\to D\tau^+\nu)^{[9.33,\ 9.86]}$ &	 0.065584 &	 \cellcolor{green!0} 0.048 $ \sigma$ &	 0.048 $ \sigma$ \\ \hline
246 &	 $R_ c^0$ &	 0.17222 &	 \cellcolor{green!0} 0.04 $ \sigma$ &	 0.041 $ \sigma$ \\ \hline
247 &	 $\mathrm{BR}(B^+\to K^{*+}\gamma)$ &	 $4.245\times 10^{-5}$ &	 0.04 $ \sigma$ &	 0.04 $ \sigma$ \\ \hline
248 &	 $\frac{\langle \mathrm{BR} \rangle}{\mathrm{BR}}(B\to D^\ast\tau^+\nu)^{[9.0,\ 9.5]}$ &	 0.09285 &	 \cellcolor{red!0} 0.028 $ \sigma$ &	 0.028 $ \sigma$ \\ \hline
249 &	 $\langle \frac{d\mathrm{BR}}{dq^2} \rangle(B^+\to K^{\ast +}\mu^+\mu^-)^{[2,\ 4.3]}$ &	 $4.4175\times 10^{-8}$ &	 \cellcolor{red!6} 0.13 $ \sigma$ &	 0.0019 $ \sigma$ \\ \hline
250 &	 $\frac{\langle \mathrm{BR} \rangle}{\mathrm{BR}}(B\to D^\ast\tau^+\nu)^{[5.0,\ 5.5]}$ &	 0.054155 &	 \cellcolor{green!0} 0.0066 $ \sigma$ &	 0.0066 $ \sigma$ \\ \hline
251 &	 $\mathrm{BR}(B^0\to \tau^+\tau^-)$ &	 $2.7877\times 10^{-8}$ &	 \cellcolor{red!0} 0.0051 $ \sigma$ &	 0.0047 $ \sigma$ \\ \hline
\end{longtable}